\documentclass[twocolumn,traditabstract]{aa}  

\usepackage{fixltx2e}
\usepackage[english]{babel}
\usepackage{graphicx,amsmath}
\usepackage{epstopdf}
\usepackage{epsf,color}
\usepackage[mathscr]{eucal}
\usepackage{amsmath}
\usepackage{amssymb,amsfonts}
\usepackage{natbib}
\usepackage{graphicx}
\usepackage{txfonts}
\usepackage{dsfont}
\definecolor{Mygreen}{rgb}{0.00, 0.5, 0.5}
\definecolor{Mypink}{rgb}{1.0, 0.0, 0.5}
\definecolor{Myblue}{rgb}{0.00, 0.2, 0.8}
\definecolor{Myred}{rgb}{0.80, 0.2, 0.0}
\usepackage[breaklinks, citecolor=Myblue, linkcolor=Mygreen, urlcolor=Mygreen, colorlinks=true, debug, baseurl=' ']{hyperref}
\usepackage{float}
\usepackage{color}

\def\simlt{\lower.5ex\hbox{$\; \buildrel < \over \sim \;$}}
\def\simgt{\lower.5ex\hbox{$\; \buildrel > \over \sim \;$}}

\bibpunct{(}{)}{ and}{a}{}{,}
\newfont{\gwpfont}{cmssq8 scaled 1000}

\begin{document}

\def\aj{AJ}%
\def\araa{ARA\&A}%
\def\apj{ApJ}%
\def\apjl{ApJ}%
\def\apjs{ApJS}%
\def\aap{A\&A}%
 \def\aapr{A\&A~Rev.}%
\def\aaps{A\&AS}%
\def\mnras{MNRAS}
\def\ssr{SSRv}
\def\nat{Nature}
\def\jcap{JCAP}

\def\Mgv{M_{\rm g,500}}
\def\Mg{M_{\rm g}}
\def\YX {Y_{\rm X}}
\def\LXv {L_{\rm X,500}}
\def\TX {T_{\rm X}}
\def\fgv {f_{\rm g,500}}
\def\fg  {f_{\rm g}}
\def\kT {{\rm k}T}
\def\ne {n_{\rm e}}
\def\Mv {M_{\rm 500}}
\def \Rv {R_{500}}
\def\keV {\rm keV}
\def\Yv{Y_{500}}

\def\MT {$M$--$T_{\rm X}$}
\def\MYX {$M$--$Y_{\rm X}$}
\def\MMg {$M_{500}$--$M_{\rm g,500}$}
\def\MgT {$M_{\rm g,500}$--$T_{\rm X}$}
\def\MgY {$M_{\rm g,500}$--$Y_{\rm X}$}

\def\msol {{\rm M_{\odot}}}

\def\lesssim{\mathrel{\hbox{\rlap{\hbox{\lower4pt\hbox{$\sim$}}}\hbox{$<$}}}}
\def\gtrsim{\mathrel{\hbox{\rlap{\hbox{\lower4pt\hbox{$\sim$}}}\hbox{$>$}}}}

\def\psz{PSZ2\,G144.83$+$25.11}

\def\xmm{XMM-{\it Newton}}
\def\planck{{\it Planck}} 
\def\chandra{{\it Chandra}}
\def \rosat {\hbox{\it ROSAT}}
\newcommand{\excpres}{{\gwpfont EXCPRES}}
\newcommand{\ma}[1]{\textcolor{red}{{ #1}}}
\title{First Sunyaev-Zel'dovich mapping with the NIKA2 camera: Implication of cluster substructures for the pressure profile and mass estimate}

\author{version 2.1}

\author{F.~Ruppin \inst{\ref{LPSC}}
\and  F.~Mayet \inst{\ref{LPSC}}
\and  G.W.~Pratt \inst{\ref{CEA1},\ref{CEA2}}
\and  R.~Adam \inst{\ref{OCA},\ref{Teruel}}
\and  P.~Ade \inst{\ref{Cardiff}}
\and  P.~Andr\'e \inst{\ref{CEA1},\ref{CEA2}}
\and  M.~Arnaud \inst{\ref{CEA1},\ref{CEA2}}
\and  H.~Aussel \inst{\ref{CEA1},\ref{CEA2}}
\and  I.~Bartalucci \inst{\ref{CEA1},\ref{CEA2}}
\and  A.~Beelen \inst{\ref{IAS}}
\and  A.~Beno\^it \inst{\ref{Neel}}
\and  A.~Bideaud \inst{\ref{Neel}}
\and  O.~Bourrion \inst{\ref{LPSC}}
\and  M.~Calvo \inst{\ref{Neel}}
\and  A.~Catalano \inst{\ref{LPSC}}
\and  B.~Comis \inst{\ref{LPSC}}
\and  M.~De~Petris \inst{\ref{Roma}}
\and  F.-X.~D\'esert \inst{\ref{IPAG}}
\and  S.~Doyle \inst{\ref{Cardiff}}
\and  E.~F.~C.~Driessen \inst{\ref{IRAMF}}
\and  J.~Goupy \inst{\ref{Neel}}
\and  C.~Kramer \inst{\ref{IRAME}}
\and  G.~Lagache \inst{\ref{LAM}}
\and  S.~Leclercq \inst{\ref{IRAMF}}
\and  J.-F.~Lestrade \inst{\ref{LERMA}}
\and  J.F.~Mac\'ias-P\'erez \inst{\ref{LPSC}}
\and  P.~Mauskopf \inst{\ref{Cardiff},\ref{Arizona}}
\and  A.~Monfardini \inst{\ref{Neel}}
\and  L.~Perotto \inst{\ref{LPSC}}
\and  G.~Pisano \inst{\ref{Cardiff}}
\and  E.~Pointecouteau \inst{\ref{IRAP},\ref{IRAP2}}
\and  N.~Ponthieu \inst{\ref{IPAG}}
\and  V.~Rev\'eret \inst{\ref{CEA1},\ref{CEA2}}
\and  A.~Ritacco \inst{\ref{IRAME}}
\and  C.~Romero \inst{\ref{IRAMF}}
\and  H.~Roussel \inst{\ref{IAP}}
\and  K.~Schuster \inst{\ref{IRAMF}}
\and  A.~Sievers \inst{\ref{IRAME}}
\and  C.~Tucker \inst{\ref{Cardiff}}
\and  R.~Zylka \inst{\ref{IRAMF}}}

\institute{
Laboratoire de Physique Subatomique et de Cosmologie, Universit\'e Grenoble Alpes, CNRS/IN2P3, 53, avenue des Martyrs, Grenoble, France
  \label{LPSC}
\and
IRFU, CEA, Universit\'e Paris-Saclay, F-91191 Gif-sur-Yvette, France
\label{CEA1}
\and
Universit\'e Paris Diderot, AIM, Sorbonne Paris Cit\'e, CEA, CNRS, F-91191 Gif-sur-Yvette, France
\label{CEA2}
\and
Laboratoire Lagrange, Universit\'e C\^ote d'Azur, Observatoire de la C\^ote d'Azur, CNRS, Blvd de l'Observatoire, CS 34229, 06304 Nice cedex 4, France
  \label{OCA}  
\and    
Centro de Estudios de F\'isica del Cosmos de Arag\'on (CEFCA), Plaza San Juan, 1, planta 2, E-44001, Teruel, Spain
\label{Teruel}   
\and
Astronomy Instrumentation Group, University of Cardiff, UK
  \label{Cardiff}  
\and
Institut d'Astrophysique Spatiale (IAS), CNRS and Universit\'e Paris Sud, Orsay, France
  \label{IAS}
\and
Institut N\'eel, CNRS and Universit\'e Grenoble Alpes, France
  \label{Neel}
\and
Dipartimento di Fisica, Sapienza Universit\`a di Roma, Piazzale Aldo Moro 5, I-00185 Roma, Italy
  \label{Roma}
\and
Institut de Plan\'etologie et d'Astrophysique de Grenoble, Univ. Grenoble Alpes, CNRS, IPAG, 38000 Grenoble, France 
  \label{IPAG}
\and
Institut de RadioAstronomie Millim\'etrique (IRAM), Grenoble, France
  \label{IRAMF}
\and
Institut de RadioAstronomie Millim\'etrique (IRAM), Granada, Spain
  \label{IRAME}
\and
Aix Marseille Universit\'e, CNRS, LAM (Laboratoire d'Astrophysique de Marseille) UMR 7326, 13388, Marseille, France
  \label{LAM}
\and 
LERMA, Observatoire de Paris, PSL Research University, CNRS, Sorbonne Universit\'es, UPMC Univ., 75014, Paris, France
  \label{LERMA}
\and
School of Earth and Space Exploration and Department of Physics, Arizona State University, Tempe, AZ 85287
  \label{Arizona}
\and 
Universit\'e de Toulouse, UPS-OMP, Institut de Recherche en Astrophysique et Plan\'etologie (IRAP), Toulouse, France
  \label{IRAP}
\and
CNRS, IRAP, 9 Av. colonel Roche, BP 44346, F-31028 Toulouse cedex 4, France 
  \label{IRAP2}
\and
Institut d'Astrophysique de Paris, CNRS (UMR7095), 98 bis boulevard Arago, F-75014, Paris, France
  \label{IAP}
}

\abstract {The complete characterization of the pressure profile of high-redshift galaxy clusters, from their core to their outskirts, is a major issue for the study of the formation of large-scale structures. It is essential to constrain a potential redshift evolution of both the slope and scatter of the mass-observable scaling relations used in cosmology studies based on cluster statistics. In this paper, we present the first thermal Sunyaev-Zel'dovich (tSZ) mapping of a cluster from the sample of the New IRAM Kids Arrays (NIKA2) SZ large program that aims at constraining the redshift evolution of cluster pressure profiles and the tSZ-mass scaling relation. We observed the galaxy cluster \psz\ at redshift $z=0.58$ with the NIKA2 camera, a dual-band (150 and 260 GHz) instrument operated at the Institut de Radioastronomie Millimétrique (IRAM) 30-meter telescope. We identify a thermal pressure excess in the south-west region of \psz\ and a high-redshift sub-millimeter point source that affect the intracluster medium (ICM) morphology of the cluster. The NIKA2 data are used jointly with tSZ data acquired by the Multiplexed SQUID/TES Array at Ninety Gigahertz (MUSTANG), Bolocam, and \planck\ experiments in order to non-parametrically set the best constraints on the electronic pressure distribution from the cluster core ($\rm{R} \sim 0.02 \rm{R_{500}}$) to its outskirts ($\rm{R} \sim 3 \rm{R_{500}} $). We investigate the impact of the over-pressure region on the shape of the pressure profile and on the constraints on the integrated Compton parameter $\rm{Y_{500}}$. A hydrostatic mass analysis is also performed by combining the tSZ-constrained pressure profile with the deprojected electronic density profile from XMM-\emph{Newton}. This allows us to conclude that the estimates of $\rm{Y_{500}}$ and $\rm{M_{500}}$ obtained from the analysis with and without masking the disturbed ICM region differ by 65\%\ and 79\% respectively. This work highlights that NIKA2 will have a crucial impact on the characterization of the scatter of the $\rm{Y_{500}}-\rm{M_{500}}$ scaling relation due to its high potential to constrain the thermodynamic and morphological properties of the ICM when used in synergy with X-ray observations of similar angular resolution. This study also presents the typical products that will be delivered to the community for all clusters included in the NIKA2 tSZ Large Program.}

\titlerunning{tSZ observation of \psz with NIKA2}
\authorrunning{F. Ruppin \emph{et al.}}
\maketitle

\section{Introduction}\label{sec:Introduction}

\indent The abundance of galaxy clusters as a function of mass and redshift is a powerful probe to study large-scale structure formation processes and to constrain cosmological models \citep[e.g.][]{pla16a,deh16}. Significant improvements on the estimation of cosmological parameters may be obtained from the combination of constraints based on cluster statistics and geometrical probes, due to the fact that their parameter degeneracies are different. Taken individually, the comparison of constraints driven by a cluster-based analysis and other probes, such as the power spectrum of the temperature anisotropies of the cosmic microwave background (CMB) \citep{pla16b} or the baryon acoustic oscillations \citep{and14}, can also provide information on possible deviations from the standard $\Lambda$-cold dark matter ($\Lambda$CDM) paradigm to describe the evolution of density perturbations from the recombination epoch until today \citep[e.g.][]{pla16a}.\\
\indent The main limiting factor of current cosmological studies based on galaxy cluster surveys comes from the use of indirect estimates of the mass of galaxy clusters to constrain the cluster mass function \citep[e.g.][]{tin08}. Furthermore, while the hierarchical model of structure formation predicts the existence of mass-observable scaling relations \citep{kai86}, non-gravitational processes may induce significant departures from the theoretical expectations and a possible variation of the scaling relation properties such as their slope and scatter with redshift \citep[e.g.][]{evr91}.\\
\indent The calibration of mass-observable scaling relations is a challenging task as 85\% of the total mass of galaxy clusters is due to the dark matter halo that escapes direct detection. Thus, the total mass of galaxy clusters is inferred from observational probes such as the X-ray emission \citep[e.g.][]{vik06,pra09}, which is due to the bremsstrahlung of hot electrons within the intracluster medium (ICM), the thermal Sunyaev-Zel'dovich effect \citep[tSZ;][]{sun72}, the velocity dispersion of the galaxies of the cluster \citep[e.g.][]{biv06,sif16}, or the lensing distortions of background galaxies (e.g. \citealt{app14}, \citealt{ume14}, and \citealt{hoe15}). However, using such mass proxies requires several hypotheses such as the hydrostatic equilibrium of the ICM and a specific model of the dark matter halo. These assumptions may lead to significant bias on the total mass estimates due to the presence of non-thermal pressure support \citep[e.g., gas turbulence due to merging events and feedback from active galactic nuclei, ][]{mar16} or projection effects \citep[e.g.][]{clo04} due to unidentified substructures along the line of sight.\\
\indent The tSZ effect has been shown to be an excellent cosmological probe as it enables us to establish nearly mass-limited samples of galaxy clusters up to very high redshift \citep[e.g.][]{pla16c} thanks to its redshift independence. Furthermore, its integrated flux provides a very low-scatter mass proxy \citep[e.g.][]{pla11}. In the past years the \planck\ satellite, the South Pole Telescope (SPT), and the Atacama Cosmology Telescope (ACT) surveys have discovered and/or characterized galaxy clusters from tSZ observations (e.g. \citealt{pla16c}, \citealt{ble15}, and \citealt{has13}). While perfectly adapted to establish large cluster samples for cosmology analyses, these instruments have a relatively low angular resolution (full width at half maximum (FWHM) $>$ 1 arcmin), which prevents them from mapping the ICM at intermediate and high redshift. However, the full characterization of the ICM of galaxy clusters over a wide range of redshifts is essential to identify and constrain all the biases and systematics affecting the calibration of the tSZ-mass scaling relation. Furthermore, joint analyses combining data sets of equivalent quality from different probes are now required to constrain potential deviations from the standard assumptions used to extract ICM information from single-probe analyses \citep[e.g.][]{sie16}. It is therefore necessary to use high angular resolution tSZ observations over a large field of view (FOV) in combination with current X-ray and optical data to describe the ICM thermodynamic properties up to high redshift and study potential variations of the tSZ-mass scaling relation characteristics with redshift.\\
\indent The New IRAM KIDs Array 2 (NIKA2) is a $6.5$ arcmin field of view camera designed to observe the sky at 150 and 260 GHz, with an angular resolution of 17.7 and 11.2 arcsec respectively \citep[see][]{ada18}. The NIKA2 camera has successfully been installed and commissioned at the Institut de Radio Astronomie Millimetrique (IRAM) 30-m telescope. The NIKA2 characteristics enable us to map the ICM of high-redshift galaxy clusters with a quality similar to the one reached by current X-ray observatories such as \xmm\ \citep[e.g.][]{boe07} in terms of substructure resolution and ICM spatial extension. Furthermore, its dual band capability enables a limiting of the systematics affecting single frequency tSZ mapping. For instance, it may be used to study contamination of the tSZ signal by, for example,\emph{} dusty and radio point sources or kinetic SZ effect \citep[see][]{ada16a,ada17a}. As a pilot study for NIKA2, several galaxy clusters have been successfully observed at the IRAM 30-m telescope with the NIKA pathfinder \citep{mon10,cat14} to cover the various configurations expected for NIKA2 (see \citealt{ada14,ada15,ada16a,ada17a,ada17b}, \citealt{rup17}, and \citealt{rom17}.\\
\indent The NIKA2 tSZ large program is a follow-up of \planck\ and ACT clusters and is part of the NIKA2 Guaranteed time. It will allow us to characterize a tSZ-selected sample of 50 clusters with a redshift ranging from $0.5$ to $0.9$ \citep{com16}. The tSZ mapping of these clusters will enable us to study the redshift evolution of the electronic pressure distribution within the ICM, in order to calibrate the tSZ-mass scaling relation in this redshift range. It will also allow us to investigate the impact of the ICM dynamical state on the cluster mass estimation. The NIKA2 tSZ large program therefore intends to provide the community with the first tSZ-mass scaling relation calibrated at high redshift.\\
\indent In this paper, we present the first observation of a galaxy cluster of the NIKA2 tSZ large program with the NIKA2 camera, namely \psz. This cluster, also known as {MACS\,J0647$+$7015} \citep{ebe01}, is located at a redshift $z=0.58$. We perform a non-parametric extraction of the ICM pressure profile from a joint multi-instrument analysis based on the tSZ observations of this cluster made by MUSTANG \citep{you15}, NIKA2, Bolocam \citep{say13,cza15}, the Arcminute Microkelvin Imager (AMI) \citep{per15}, and \planck\ \citep{pla16c}. This analysis enables us to obtain the most stringent constraints on the ICM pressure distribution from the cluster center to its outskirts. Furthermore, the potential of NIKA2 to identify both unrelaxed gas substructures and point source contaminants is highlighted by the discovery of a thermal pressure excess in the south-west region of the ICM and a high-redshift sub-millimeter point source at $1.3$ arcmin to the west of the cluster center. A key result of this study is the characterization of the impact of the identified over-pressure region on the estimate of the pressure profile of \psz. The synergy between high angular resolution tSZ and X-ray observations is also exploited by combining the deprojected gas density profile from \xmm\ with the pressure profile constrained by NIKA2 to reconstruct the thermodynamic properties of the ICM without making use of X-ray spectroscopy.\\
\begin{figure*}[h!]
\centering
\includegraphics[height=4.5cm]{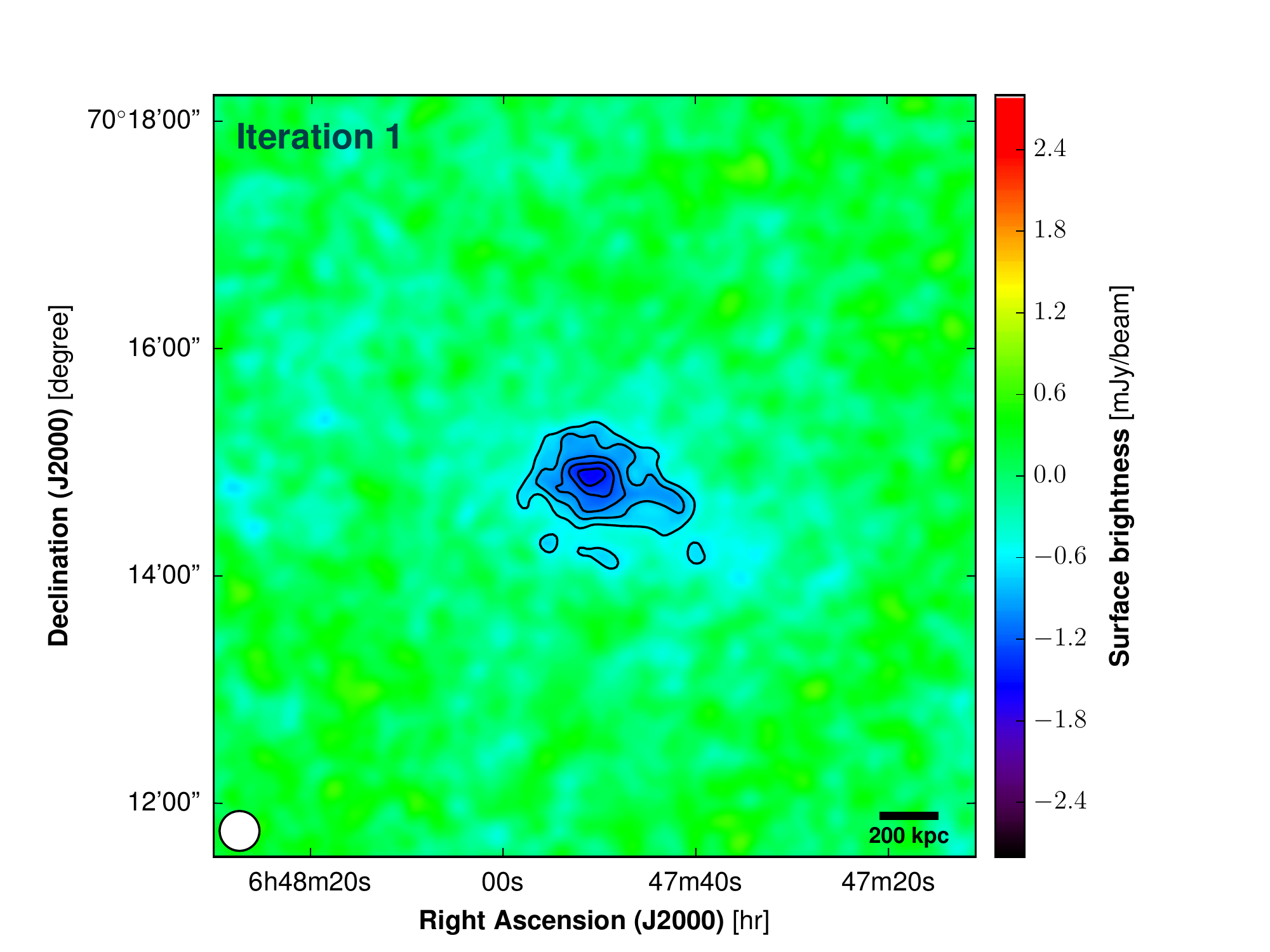}
\includegraphics[height=4.5cm]{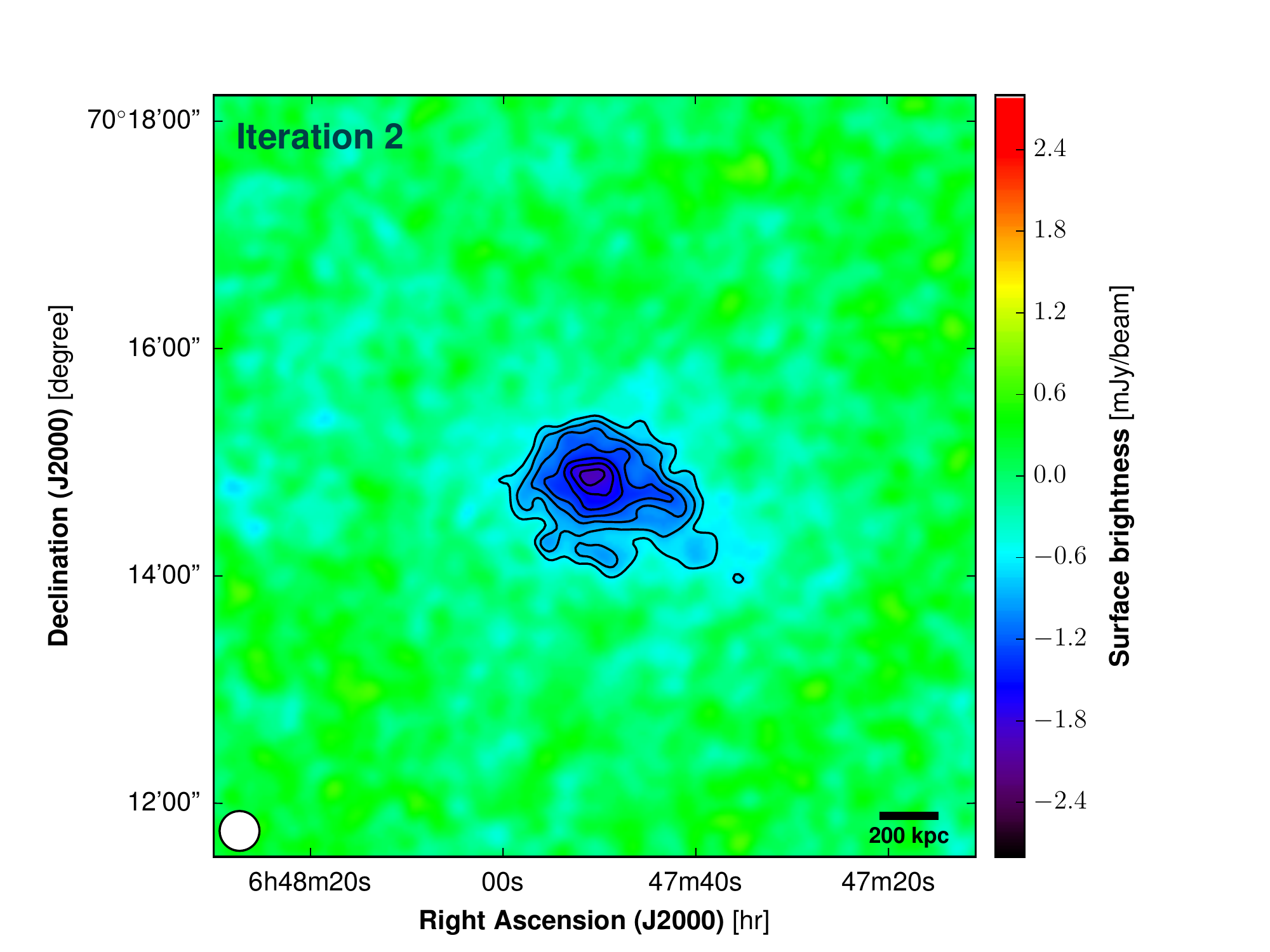}
\includegraphics[height=4.5cm]{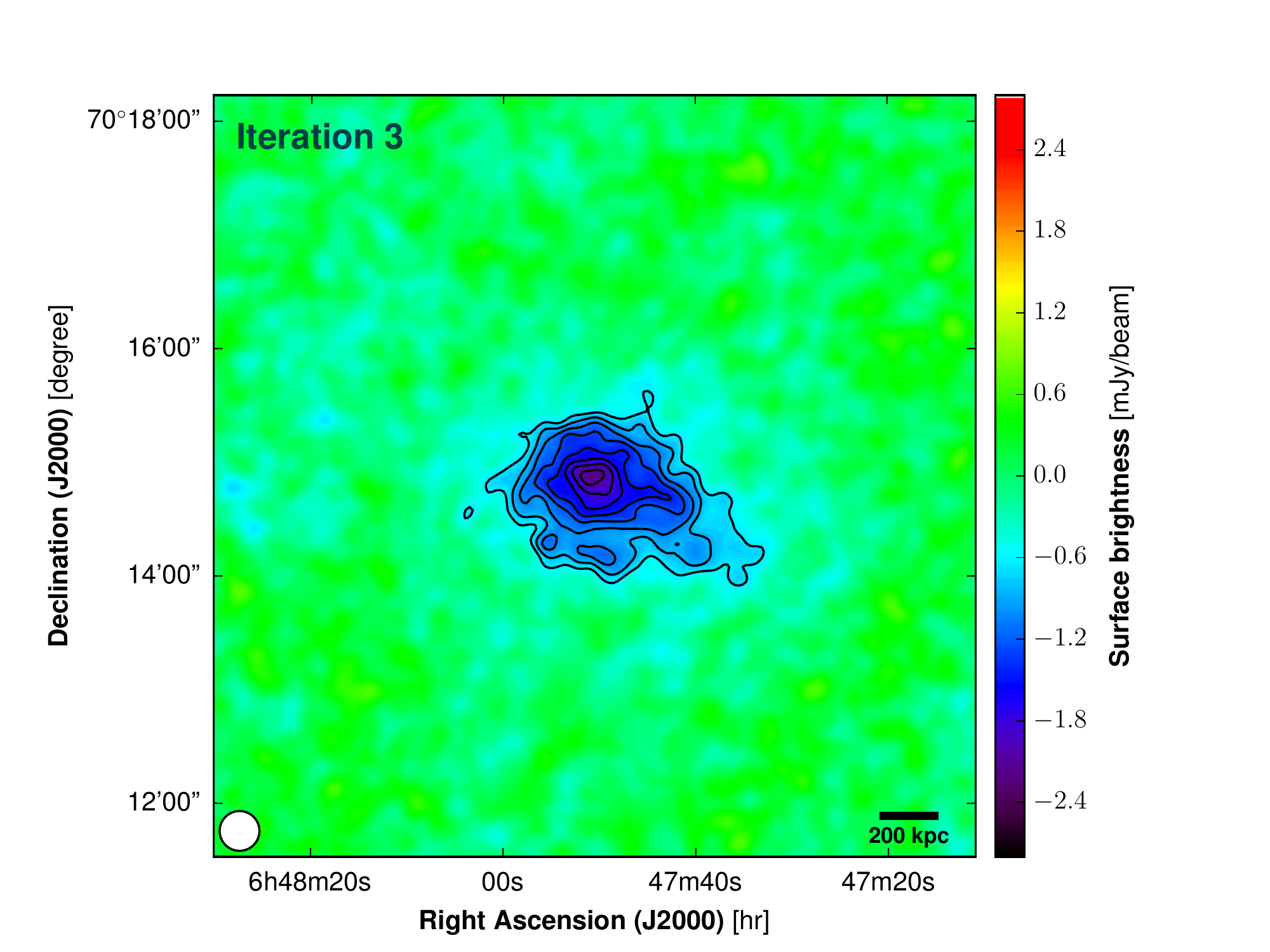}
\caption{{\footnotesize NIKA2 tSZ surface brightness maps at 150 GHz at the end of the first three iterations of the map-making. The black contours give the significance of the measured signal starting at $3\sigma$ with $1\sigma$ spacing. The color range is the same as the one used in Fig. \ref{fig:NIKA2_maps} top left panel.}}
\label{fig:NIKA2_maps_iterative}
\end{figure*}
\indent This paper is organized as follows. The NIKA2 observations of \psz\ at the IRAM 30-meter telescope and the tSZ map-making are described in Sect. \ref{sec:Observations}. Previous tSZ and X-ray observations of this cluster are presented in Sect. \ref{sec:ancillary}. The characterization of the point source contamination is explained in Sect. \ref{sec:point_source}. The identification of the south-west over-pressure region in the ICM is discussed in Sect. \ref{sec:overp}. In Sect. \ref{sec:MCMC} a non-parametric multi-instrument analysis is performed to extract the radial pressure profile and constrain the ICM thermodynamic properties. The perspectives for the NIKA2 tSZ large program and future cluster cosmology studies are discussed in Sect. \ref{sec:discussion}. Throughout this study we assume a flat $\Lambda$CDM cosmology following the latest \planck\ results \citep{pla16b}: H$_0 = 67.8$ km s$^{-1}$ Mpc$^{-1}$, $\Omega_{\rm m} = 0.308$, and $\Omega_\Lambda = 0.692$. Within this framework, one arcsecond corresponds to a distance of 6.66 kpc at the cluster redshift.
\begin{table}[t]
\begin{center}
\begin{tabular}{ccc}
\hline
\hline
Observing band & 150 GHz & 260 GHz \\
\hline
Gaussian beam model FWHM (arcsec) & 17.7 & 11.2 \\
Field of view (arcmin) & 6.5 & 6.5 \\
Number of used detectors & 531 & 1823 \\
Conversion factor $y$-Jy/beam & $-11.9\pm 0.9$ & $3.7\pm 0.4$ \\
Pointing errors (arcsec) & $<3$ & $<3$ \\
Calibration uncertainties & 7\% & 9\% \\
\hline
\hline
\end{tabular}
\end{center}
\caption{{\footnotesize Instrumental characteristics and performance of NIKA2 at the IRAM 30-m telescope in April 2017. The uncertainty on the conversion factor $y$-Jy/beam includes calibration error.}}
\label{tab:NIKA2_instru}
\end{table}
\begin{figure*}
\centering
\includegraphics[height=6.8cm]{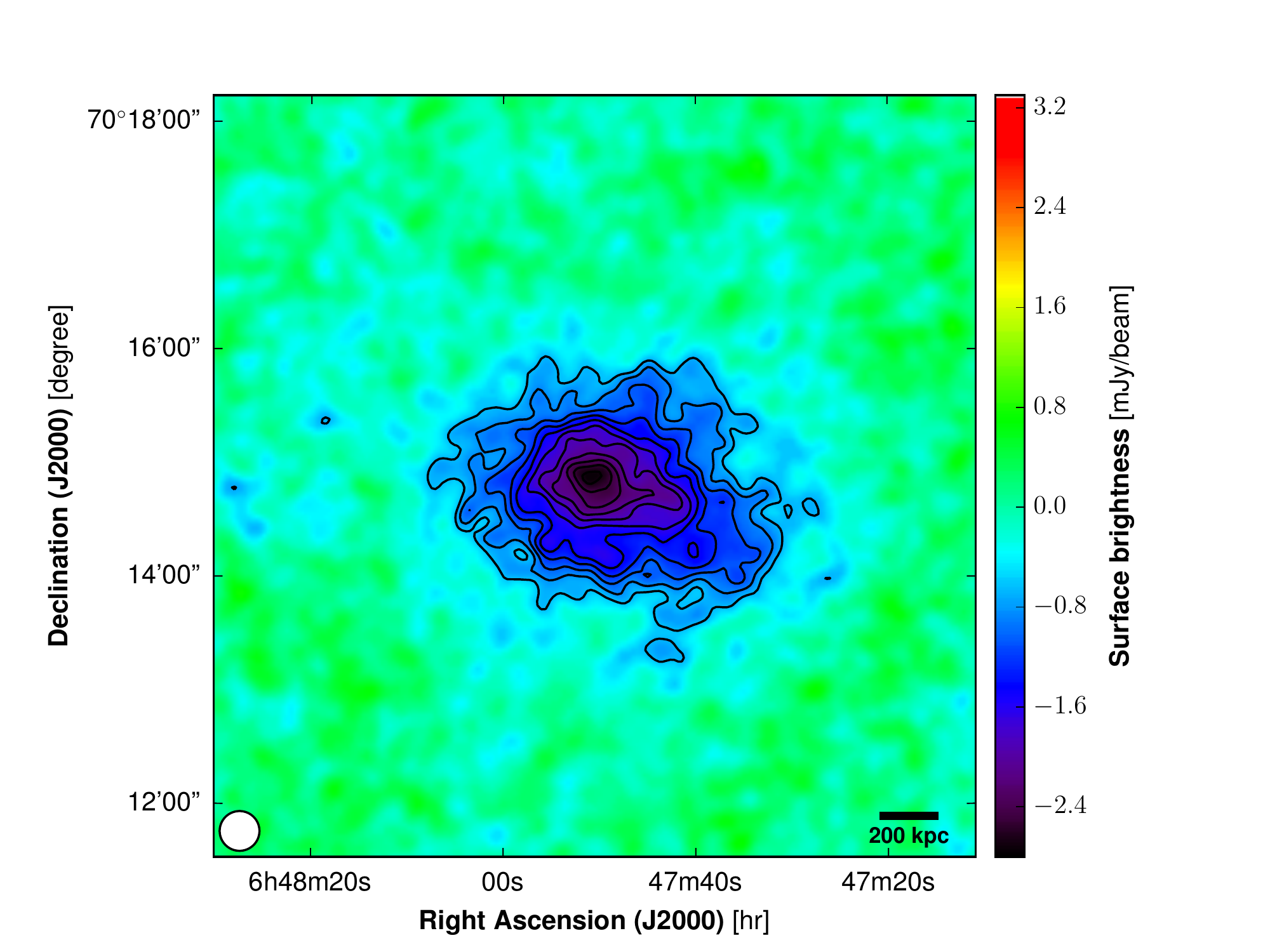}
\includegraphics[height=6.8cm]{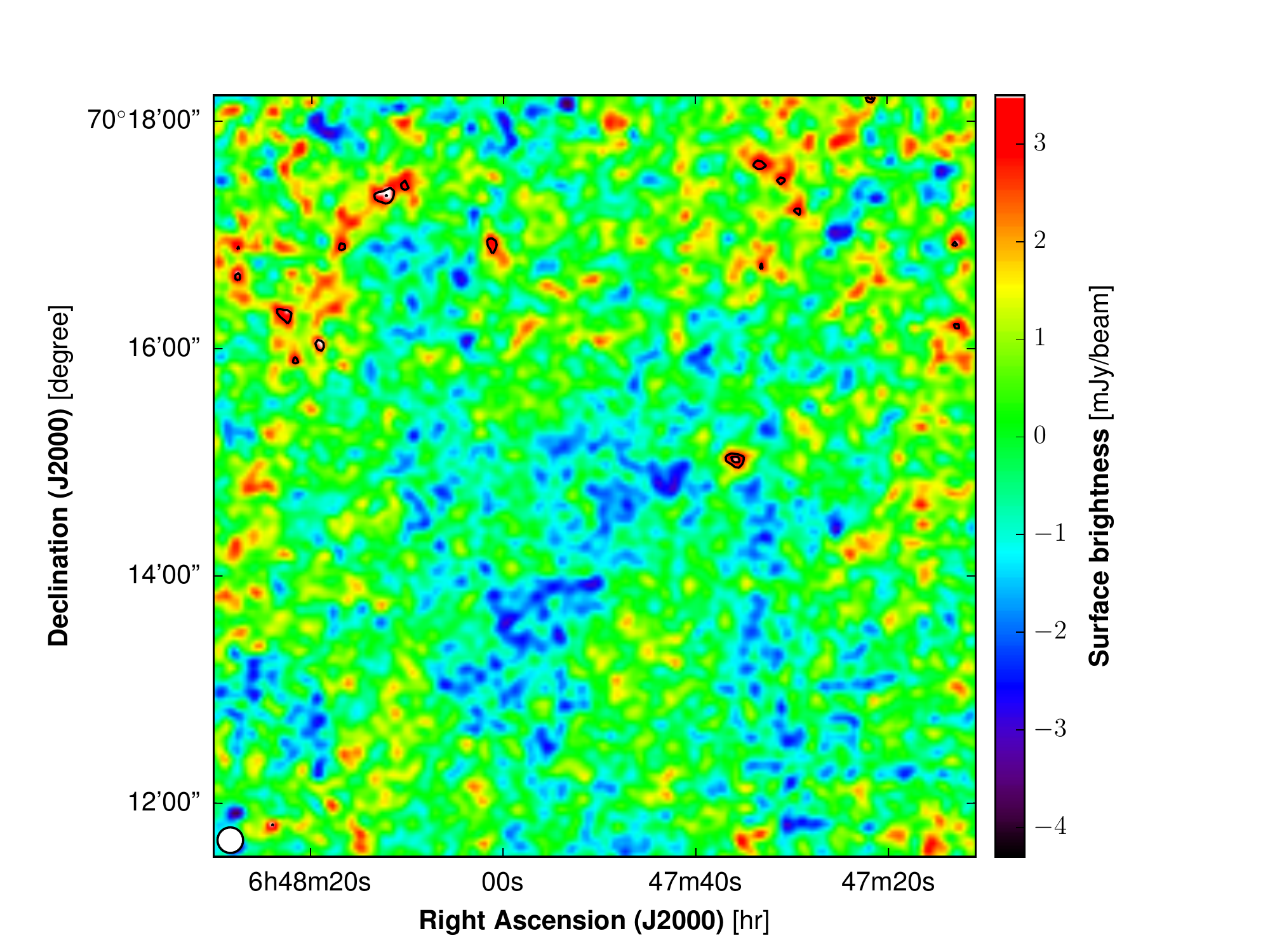}
\includegraphics[height=6.8cm]{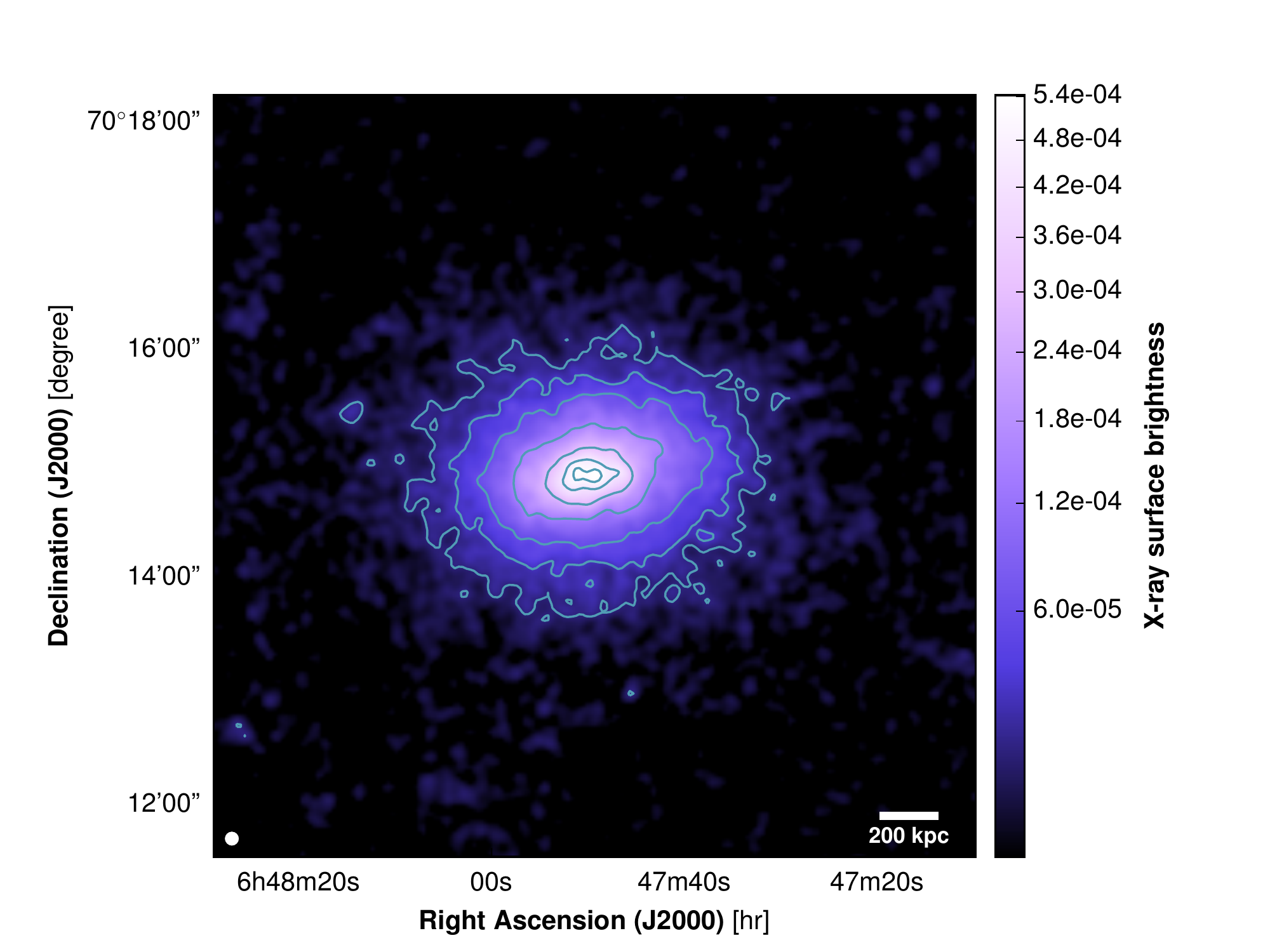}
\includegraphics[height=6.8cm]{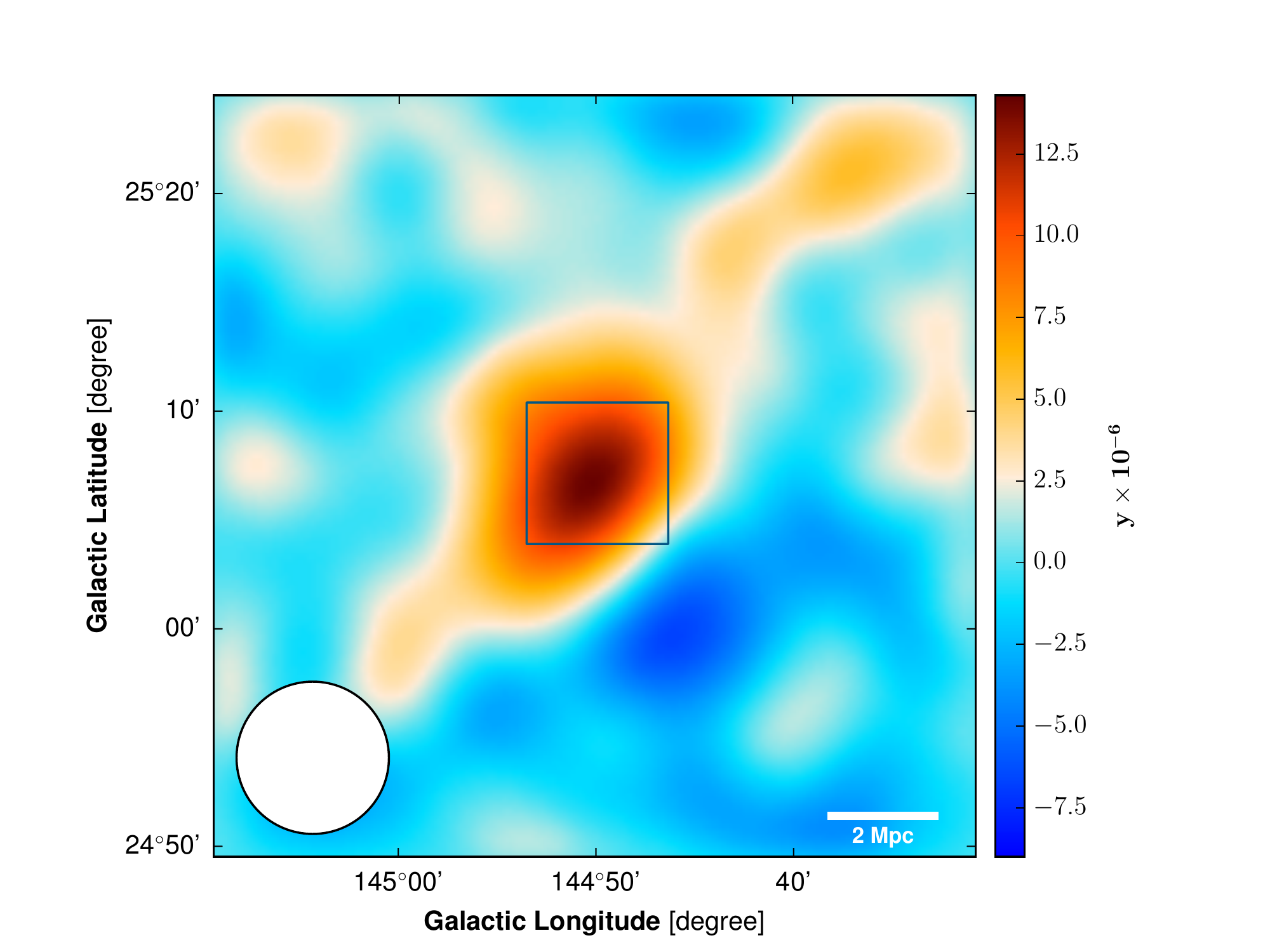}
\includegraphics[height=6.8cm]{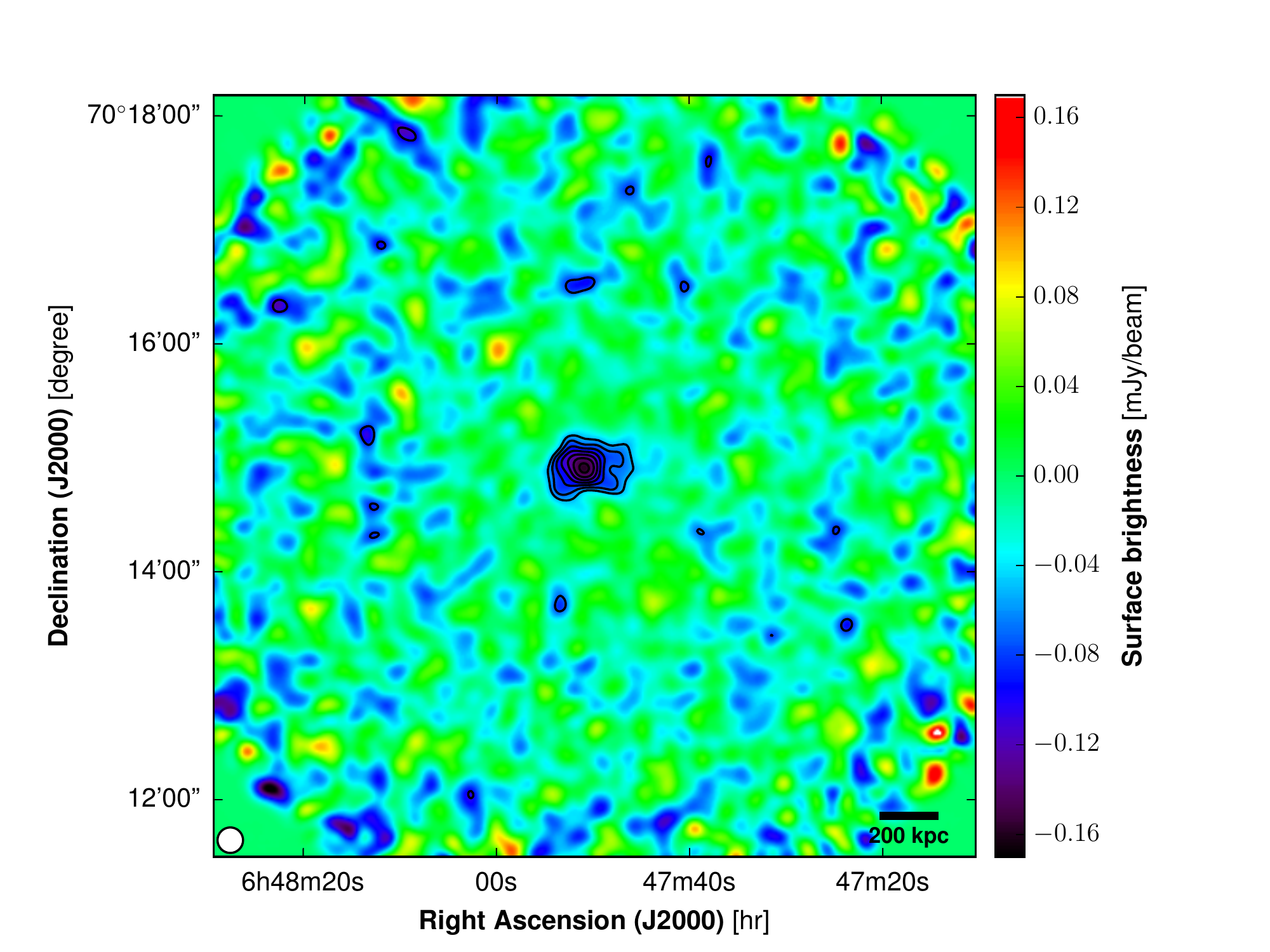}
\includegraphics[height=6.8cm]{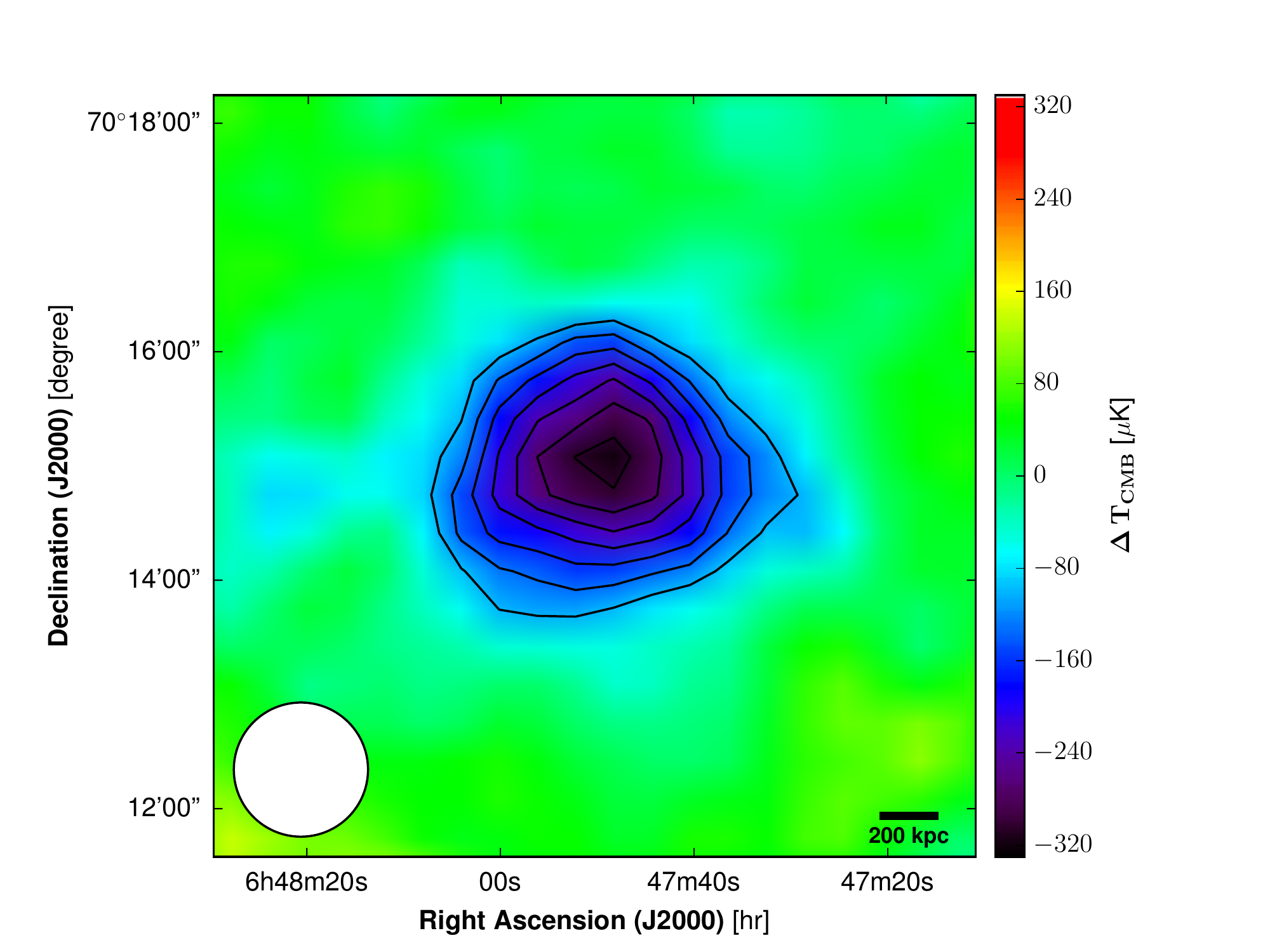}
\caption{{\footnotesize \textbf{Top:} NIKA2 surface brightness maps at 150 GHz (left) and 260 GHz (right). For display purposes, the maps are smoothed with an additional 10 and 6~arcsec FWHM Gaussian filter at 150 and 260 GHz respectively. The NIKA2 effective beam FWHMs are represented as white disks in the bottom left-hand corner of the maps. The considered FOV is 6.7 arcmin wide. \textbf{Middle:} \xmm\ X-ray photon count map (left) of \psz\ obtained after subtracting both the background and point sources and correcting for vignetting. The contours are given for X-ray counts of 1, 2, 4, 10, 20, 30, and 35. The map has been smoothed with an additional 6~arcsec FWHM Gaussian filter and colors are displayed using a square-root scale for display purposes. \planck\ map (right) of the Compton parameter of \psz\ in a wider FOV of 35 arcmin \citep{pla16d}. The dark blue square gives the size of the region displayed for all the other maps in the figure. \textbf{Bottom:} MUSTANG map (left) of the \psz\ surface brightness at 90~GHz smoothed by an additional 9 arcsec FWHM Gaussian filter for display purposes \citep{you15}. Bolocam  map (right) of the \psz\ surface brightness at 140~GHz smoothed by an additional 40 arcsec FWHM Gaussian filter \citep{say13}. For each tSZ map, the black contours give the significance of the measured signal starting at $3\sigma$ with $1\sigma$ spacing. The FWHMs of each instrument are shown as white disks in the bottom left-hand corner of the maps.}}
\label{fig:NIKA2_maps}
\end{figure*}
\begin{figure*}[h!]
\centering
\includegraphics[height=6.6cm]{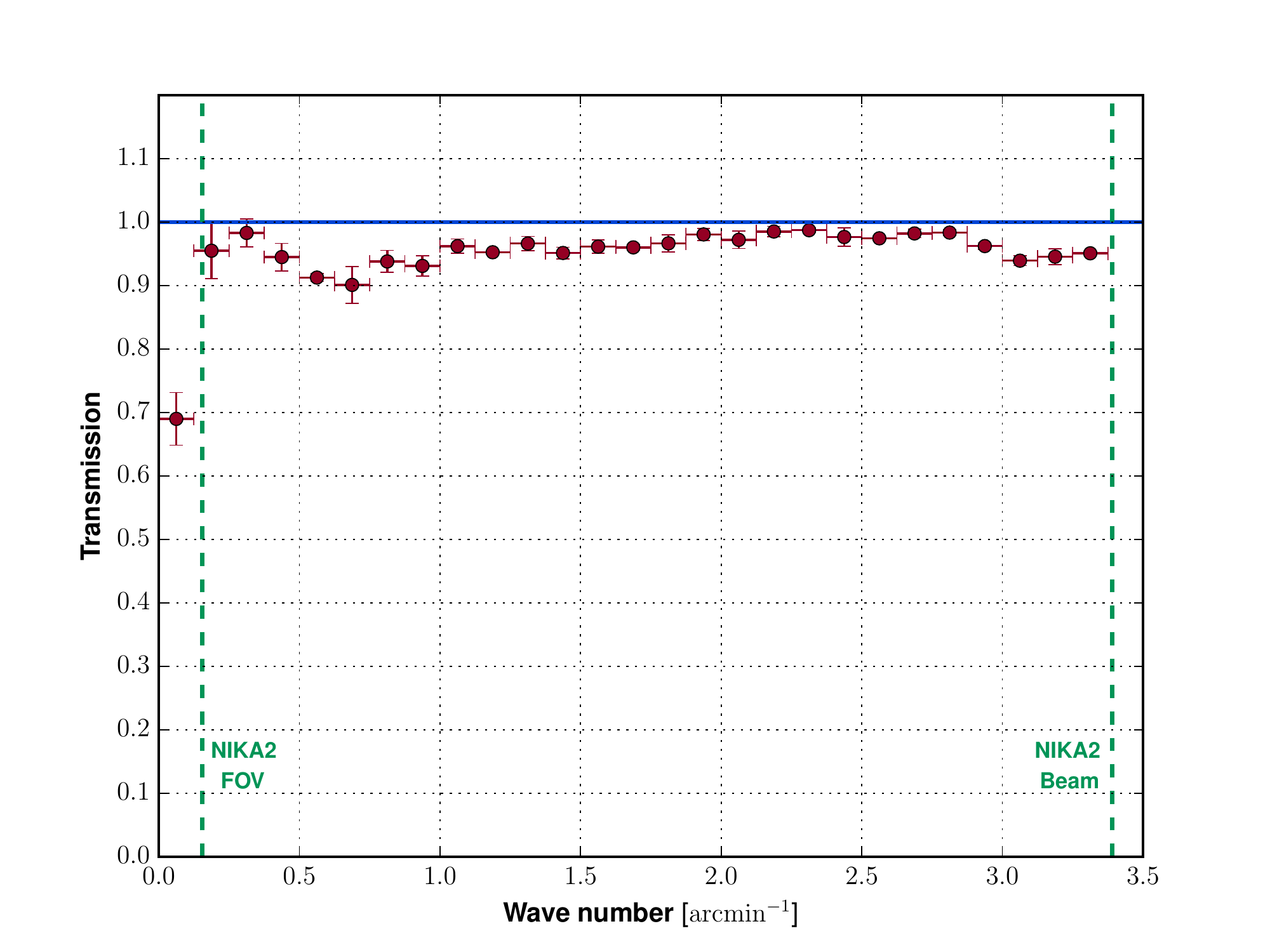}
\includegraphics[height=6.6cm]{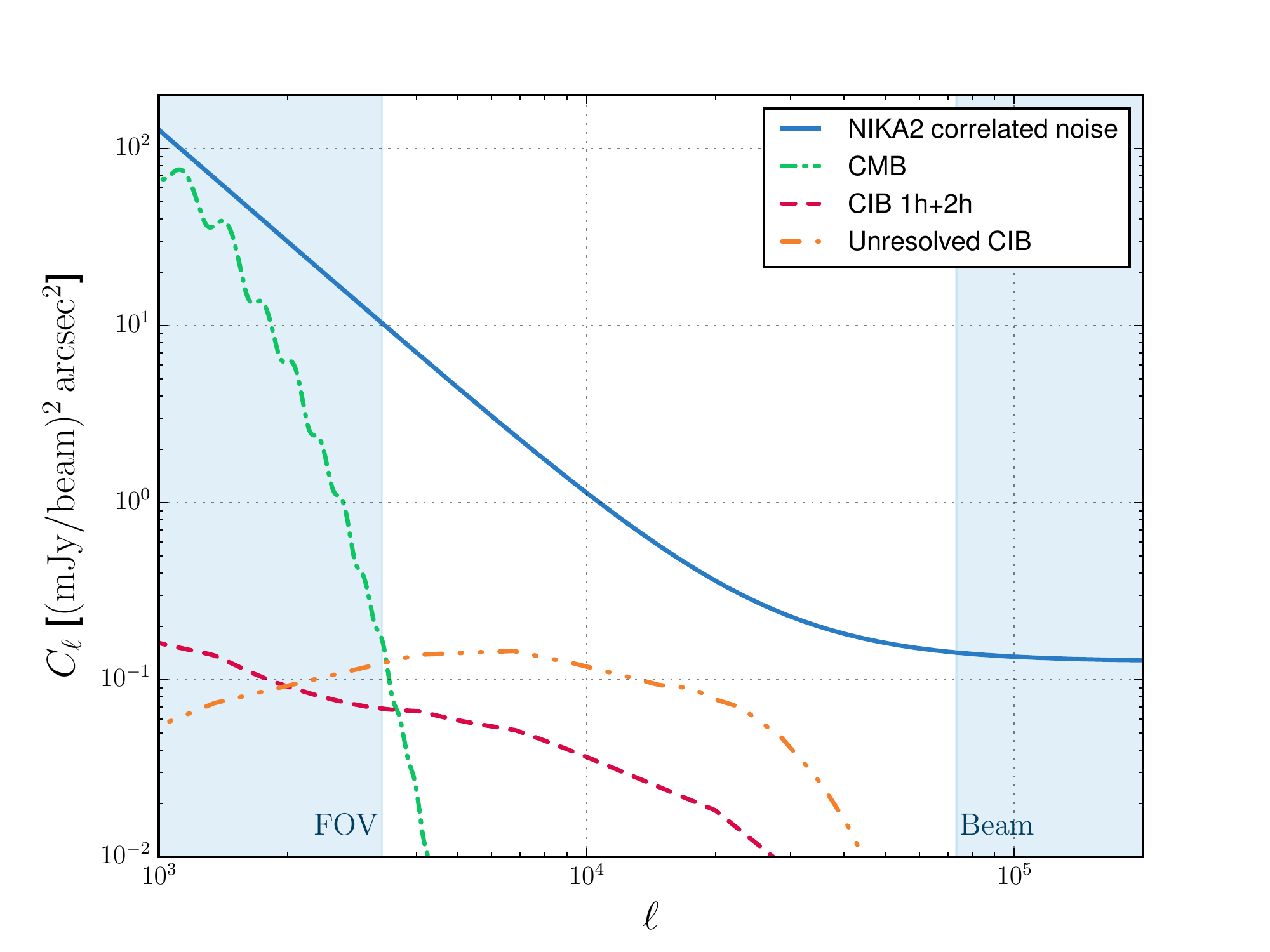}
\caption{{\footnotesize \textbf{Left:} NIKA2 150~GHz data reduction transfer function as a function of the angular frequency. The uncertainties are computed using the dispersion of the transfer functions computed from different simulations. The 150~GHz beam cutoff and the size of the NIKA2 FOV are also represented by green dashed lines. \textbf{Right:} Angular power spectrum of the NIKA2 residual correlated noise (blue), CMB primary anisotropies (green), CIB (red), and unresolved CIB (orange) at 150~GHz. The NIKA2 beam and transfer function filtering has been applied to the contaminant power spectra. The blue regions show the angular scales that are larger and smaller than the NIKA2 FOV and beam respectively.}}
\label{fig:TF_NIKA2_2mm}
\end{figure*}

\section{Observation at the IRAM 30-meter telescope with NIKA2}\label{sec:Observations}

This section is dedicated to the description of the NIKA2 observations of \psz\ made in April 2017 and to the raw data analysis performed to produce the tSZ surface brightness maps at 150~GHz and 260~GHz. We first recall some key elements of the thermal SZ effect for the reader's convenience.

\subsection{The thermal Sunyaev-Zel'dovich effect}
The thermal Sunyaev-Zel'dovich effect \citep{sun72,sun80} is a distortion of the CMB blackbody spectrum due to the Compton scattering of CMB photons on high-energy ICM electrons. The amplitude of the spectral distortion is given by the Compton parameter, which is related to the line-of-sight integral of the electronic pressure $P_e$ for a given sky position,
\begin{equation}
        y = \frac{\sigma_{\mathrm{T}}}{m_{e} c^2} \int P_{e} \, dl,
        \label{eq:y_compton}
\end{equation}
where $m_{e}$ is the electron mass, $\sigma_{\mathrm{T}}$ the Thomson scattering cross section, and $c$ the speed of light. The relative difference in CMB intensity is given by
\begin{equation}
        \frac{\Delta I_{tSZ}}{I_0} = y \, f(\nu, T_e),
\label{eq:deltaI}
\end{equation}
where $f(\nu, T_e)$ gives the frequency dependence of the tSZ spectrum \citep{bir99,car02}, and $T_e$ is the ICM electronic temperature. The temperature dependence of this spectrum is due to relativistic corrections for which we use the results of \cite{ito98,poi98} in the analysis developed in Sect. \ref{sec:MCMC}. The integrated Compton parameter $\rm{Y_{tot}}$ is computed by the aperture photometry performed on the cluster Compton parameter map and is expected to provide a low-scatter mass proxy for galaxy clusters. A robust calibration of the scaling relation between $\rm{Y_{500}}$\footnote{Spherically integrated Compton parameter up to a cluster radius $\rm{R_{500}}$, for which the cluster density is $500$ times the Universe critical density.} and $\rm{M_{500}}$ from tSZ observations of a representative sample of galaxy clusters in a large range of redshift is needed to constrain cosmological models from the analysis of tSZ surveys.

\subsection{The NIKA2 camera}
The NIKA2 camera is a new generation continuum instrument installed at the IRAM 30-m single-dish telescope. It consists of three monolithic arrays of kinetic inductance detectors (KIDs). One of them is made of 616 detectors with a maximum transmission at 150 GHz and two of them are arrays of 1140 detectors with a maximum transmission at about 260 GHz. The detector design is based on a Hilbert fractal geometry for the LEKID (Lumped Element Kinetic Inductance Detector) proposed by \cite{roe12}. The high number of frequency-multiplexed detectors gives the NIKA2 instrument the ability to map the sky in a field of view of 6.5 arcminutes simultaneously at 150 and 260~GHz. The measured angular resolutions (main beam FWHM) are respectively 17.7 and 11.2~arcsec at 150 and 260~GHz, and are close to the IRAM 30-m telescope diffraction limits of 16.8 and 9.7~arcsec at 150 and 260~GHz respectively. 
The combination of a large field of view and high-sensitivity sensors provides the NIKA2 camera with a high mapping speed, which is a key parameter for tSZ science. For instance, a four-hour tSZ mapping of a galaxy cluster at redshift $z = 0.7$ with a total mass of $5 \times 10^{14}~\mathrm{M_{\odot}}$ using $7\times 5~\rm{arcmin}^2$ scans at an elevation of $60^{\circ}$ under an opacity of 0.2 at 150~GHz would lead to a tSZ surface brightness peak detected at 9$\sigma$ for an effective resolution of 21~arcsec. Furthermore, the signal-to-noise ratio would be higher than three on the tSZ surface brightness profile up to $\theta_{500}$. More details on the NIKA2 instrument can be found in \cite{ada18}, \cite{cal16}, and \cite{bou16}.

\subsection{Observing conditions, scanning strategy, calibration, and data reduction}\label{subsec:raw_data_analysis}
The NIKA2 tSZ observations of \psz\ were made in April 2017. This is the very first tSZ observation of a galaxy cluster conducted with the NIKA2 instrument. This section aims at describing the observation conditions, scanning strategy, calibration method, and map making procedure.\\
\indent The weather conditions at the time of the observations were quite poor with a mean opacity of 0.30 at 150~GHz, 0.48 at 260~GHz, and a rather unstable atmosphere. The cluster field was observed for an effective observing time of 11.3 hours at a mean elevation of $41.4^{\circ}$. The pointing center was chosen to be $(\mathrm{R.A., Dec.})_{\mathrm{J2000}}$ = (06:47:50.5, +70:14:53.0), which corresponds to the position of the tSZ surface brightness peak of the MUSTANG observations of \psz \citep{you15}. We applied the same scanning strategy as the one presented in \cite{ada15,rup17} and used on-the-fly raster scans of $10\times 7$~arcmin with $10$~arcsec steps between each subscan in four different directions in equatorial coordinates (J2000). We follow the calibration procedure described in \cite{ada18} and get absolute calibration uncertainties of 7\% and 9\% at 150 and 260~GHz respectively. The root mean square pointing error has been measured to be lower than 3~arcsec. The main instrumental characteristics of the NIKA2 camera during this science verification run are summarized in Table \ref{tab:NIKA2_instru}.\\
\indent We follow the pre-processing method detailed in \cite{ada15} to select valid detectors and remove cosmic ray glitches and cryogenic vibrations from the raw data. The removal of both atmospheric and electronic correlated noise has been done independently for the two bands by using an iterative procedure. This method is different from the one used in previous NIKA analyses and aims at recovering extended signal far from the tSZ peak without using a mask in the raw data analysis. At the end of each iteration, the signal detected with a significance higher than $3\sigma$ on the map is removed from the raw data before the common mode estimation in the next iteration. The latter is computed by using blocks of highly correlated detectors to minimize the amount of residual spatially correlated noise in the final map. As shown in Fig. \ref{fig:NIKA2_maps_iterative} left panel, the recovered tSZ signal and cluster extension at the end of the first iteration are rather weak because the correlated noise template computed as the mean of the time ordered information (TOI) of all detectors includes the cluster signal. This induces an important spatial filtering of the signal at the cluster location. The removal of the significant signal identified on the map before the common mode estimation at each iteration enables the reduction of the bias induced by the cluster signal on the correlated noise template computation.\\
\indent The filtering of the signal is therefore reduced at each iteration as shown in Fig. \ref{fig:NIKA2_maps_iterative}, middle and right panels. The procedure is stopped when the signal-to-noise ratio at the tSZ surface brightness peak does not increase significantly. The processed time-ordered information is projected on a pixelized grid and all the scans are then coadded with an inverse variance weighting to get the final maps shown in the top panels of Fig. \ref{fig:NIKA2_maps}. This method allows us to get a fairly uniform amount of residual correlated noise on the final map as the correlated noise template is estimated by considering all the observed field. Furthermore, this iterative procedure allows us to avoid ringing in the final tSZ map as the amount of cluster signal biasing the correlated noise template in the final iteration is negligible.\\
\indent Removing correlated noise from the raw data induces a filtering of the tSZ signal at scales that are larger than the NIKA2 field of view (6.5~arcmin). The NIKA2 processing circular transfer function at 150~GHz has been computed using simulations as described in \cite{ada15}. As can be seen on the left panel of Fig. \ref{fig:TF_NIKA2_2mm}, it is fairly flat and close to unity at scales smaller than the NIKA2 FOV and vanishes rapidly for larger scales.\\
\begin{table*}[h]
\begin{center}
\begin{tabular}{ccccccccc}
\hline
\hline
Source & $250~\mu$m source position & $100~\mu$m & $160~\mu$m & $250~\mu$m & $350~\mu$m & $500~\mu$m & 1.15 mm & 2.05 mm \\
 &  & [mJy] & [mJy] & [mJy] & [mJy] & [mJy] & [mJy] &  expected [mJy] \\
\hline
SMG1 & 6:47:15.62, +70:14:00.60 & $3.8\pm 0.4$ & $5.3\pm 0.8$ & $30.6\pm 1.1$ & $20.6\pm 1.1$ & $4.6\pm  1.4$& $2.0\pm  0.6$ & $0.34\pm  0.14$\\
SMG2 & 6:47:18.36, +70:13:42.60 & $5.8\pm 0.9$ & $7.8\pm 1.6$ & $28.2\pm 0.9$ & $24.6\pm 1.1$ & $15.9\pm  1.4$& $0.47\pm  0.1$ & $0.21\pm  0.12$\\
SMG3 & 6:47:18.79, +70:17:01.32 & $4.5\pm 0.9$ & $27.1\pm 1.0$ & $60.4\pm 0.9$ & $50.1\pm 0.9$ & $35.9\pm  1.0$& $0.25\pm  0.08$ & $0.32\pm  0.18$\\
SMG4 & 6:47:19.80, +70:14:38.76 & $9.0\pm 0.6$ & $20.5\pm 0.6$ & $31.5\pm 1.1$ & $18.7\pm 1.4$ & $11.2\pm  1.7$& $1.3\pm  0.6$ & $0.42\pm  0.16$\\
SMG5 & 6:47:34.85, +70:15:42.84 & $3.2\pm 0.3$ & $9.5\pm 0.5$ & $20.7\pm 1.0$ & $16.4\pm 1.1$ & $6.3\pm  1.4$& $0.9\pm  0.5$ & $0.29\pm  0.13$\\
SMG6 & 6:47:34.85, +70:11:41.28 & $15.2\pm 0.8$ & $31.5\pm 1.4$ & $88.0\pm 1.1$ & $68.3\pm 1.5$ & $42.2\pm  1.5$& $3.0\pm  0.7$ & $1.12\pm  0.21$\\
SMG7 & 6:47:36.60, +70:16:32.52 & $3.7\pm 0.4$ & $15.1\pm 0.6$ & $34.3\pm 1.0$ & $30.6\pm 1.1$ & $11.6\pm  1.4$& $0.4\pm  0.2$ & $0.21\pm  0.09$\\
SMG8 & 6:47:38.47, +70:13:23.52 & $2.7\pm 0.5$ & $7.4\pm 0.8$ & $13.4\pm 1.0$ & $4.4\pm 0.9$ & $0.4\pm  0.6$& $0.1\pm  0.1$ & $0.03\pm  0.02$\\
SMG9 & 6:47:41.26, +70:16:13.08 & $12.9\pm 0.3$ & $21.9\pm 0.5$ & $44.0\pm 1.1$ & $24.3\pm 1.0$ & $11.3\pm  1.6$& $0.2\pm  0.1$ & $0.13\pm  0.08$\\
SMG10 & 6:47:48.50, +70:13:48.00 & $0.6\pm 0.1$ & $2.4\pm 0.3$ & $18.6\pm 1.1$ & $17.5\pm 1.0$ & $17.2\pm  1.4$& $0.2\pm  0.2$ & $0.30\pm  0.16$\\
SMG11 & 6:47:50.04, +70:15:30.60 & $10.0\pm 0.3$ & $17.8\pm 0.6$ & $27.3\pm 1.0$ & $8.9\pm 1.1$ & $2.6\pm  0.9$& $0.1\pm  0.3$ & $0.08\pm  0.09$\\
SMG12 & 6:47:53.90, +70:14:31.20 & $7.2\pm 0.4$ & $14.7\pm 0.9$ & $22.0\pm 1.0$ & $12.8\pm 1.0$ & $6.6\pm  1.3$& $0.1\pm  0.1$ & $0.07\pm  0.05$\\
SMG13 & 6:47:56.38, +70:17:21.48 & $7.9\pm 0.4$ & $15.9\pm 1.6$ & $26.1\pm 1.3$ & $15.9\pm 1.0$ & $2.2\pm  1.0$& $1.2\pm  0.6$ & $0.15\pm  0.12$\\
SMG14 & 6:48:04.37, +70:17:00.60 & $31.5\pm 0.5$ & $57.9\pm 0.9$ & $79.3\pm 1.1$ & $54.8\pm 1.1$ & $37.6\pm  1.4$& $0.8\pm  0.5$ & $0.72\pm  0.21$\\
SMG15 & 6:48:08.64, +70:14:25.08 & $9.7\pm 0.5$ & $19.3\pm 1.0$ & $36.6\pm 1.0$ & $20.4\pm 1.0$ & $4.1\pm  0.9$& $0.1\pm  0.3$ & $0.13\pm  0.15$\\
SMG16 & 6:48:14.59, +70:14:38.40 & $0.5\pm 0.1$ & $8.1\pm 0.6$ & $31.0\pm 1.0$ & $31.6\pm 1.0$ & $19.5\pm  1.2$& $0.7\pm  0.1$ & $0.34\pm  0.18$\\
SMG17 & 6:48:15.43, +70:17:24.72 & $0.1\pm 0.1$ & $5.6\pm 0.9$ & $23.0\pm 1.1$ & $31.2\pm 1.1$ & $21.4\pm  1.5$& $3.4\pm  0.7$ & $0.88\pm  0.14$\\
SMG18 & 6:48:21.55, +70:16:06.24 & $0.1\pm 0.2$ & $9.6\pm 1.1$ & $27.9\pm 1.1$ & $21.7\pm 1.1$ & $0.6\pm  1.1$& $0.1\pm  0.1$ & $0.12\pm  0.17$\\
SMG19 & 6:48:21.89, +70:16:34.68 & $0.1\pm 0.1$ & $18.3\pm 1.2$ & $34.3\pm 1.0$ & $27.8\pm 1.0$ & $13.9\pm  1.4$& $0.2\pm  0.1$ & $0.04\pm  0.12$\\
SMG20 & 6:48:23.54, +70:15:51.12 & $6.7\pm 0.8$ & $8.1\pm 1.4$ & $32.2\pm 1.0$ & $27.2\pm 1.3$ & $17.1\pm  1.4$& $0.4\pm  0.2$ & $0.19\pm  0.08$\\
SMG21 & 6:48:11.43, +70:16:09.69 & $31.7\pm 0.4$ & $34.0\pm 0.8$ & $28.7\pm 1.0$ & $9.6\pm 1.0$ & $2.5\pm  0.8$& $0.1\pm  0.1$ & $0.07\pm  0.06$\\
SMG22 & 6:48:16.47, +70:15:47.23 & $9.3\pm 0.5$ & $15.7\pm 0.8$ & $23.0\pm 1.1$ & $11.6\pm 1.2$ & $3.0\pm  1.0$& $0.1\pm  0.1$ & $0.05\pm  0.06$\\
SMG23 & 6:48:00.86, +70:15:29.63 & $6.4\pm 0.3$ & $8.3\pm 0.6$ & $1.7\pm 1.0$ & $0.3\pm 0.4$ & $0.1\pm  0.4$& $1.2\pm  0.8$ & $0.04\pm  0.03$\\
SMG24 & 6:47:45.02, +70:15:47.73 & $19.4\pm 0.3$ & $19.6\pm 0.5$ & $14.3\pm 1.1$ & $6.9\pm 1.0$ & $2.1\pm  0.8$& $0.1\pm  0.2$ & $0.03\pm  0.04$\\
SMG25 & 6:48:02.47, +70:13:37.52 & $7.6\pm 0.4$ & $13.4\pm 0.6$ & $7.8\pm 0.9$ & $3.6\pm 0.9$ & $6.3\pm  1.4$& $0.3\pm  0.3$ & $0.07\pm  0.05$\\
SMG26 & 6:47:43.71, +70:14:05.07 & $4.4\pm 0.3$ & $8.3\pm 0.6$ & $11.9\pm 1.0$ & $1.9\pm 1.2$ & $0.7\pm  0.3$& $0.1\pm  0.1$ & $0.03\pm  0.02$\\
SMG27 & 6:47:21.06, +70:14:23.02 & $7.3\pm 0.4$ & $16.2\pm 1.2$ & $20.6\pm 1.2$ & $9.5\pm 1.2$ & $1.0\pm  1.5$& $0.1\pm  0.1$ & $0.05\pm  0.06$\\
SMG28 & 6:47:35.04, +70:15:03.96 & $2.1\pm 0.4$ & $1.5\pm 0.3$ & $6.6\pm 0.9$ & $14.1\pm 1.0$ & $12.5\pm  1.1$& $3.2\pm  0.6$ & $0.87\pm  0.20$\\
\hline
\hline
\end{tabular}
\end{center}
\caption{{\footnotesize Positions and fluxes of the 28 submillimeter sources identified in the $6.7 \times 6.7$ $\mathrm{arcmin^2}$ field around \psz, measured by fitting Gaussian models to the {\it Herschel} maps at each wavelength as described in Sect. \ref{sec:point_source}. The 260~GHz NIKA2 map is also used to constrain each source SED at low frequency. The expected fluxes at 150~GHz are computed by integrating the estimated SED in the corresponding NIKA2 bandpass.}}
\label{tab:Submm_ps_flux}
\end{table*}
\indent The residual correlated noise in the final maps has been characterized following the procedure described in \cite{ada16a}. Null maps are computed by the semi-difference of the coadded maps from two equivalent subsamples of processed scans. Homogeneous noise maps are obtained by normalizing the null maps by the square root of the integration time spent per pixel. These maps are then used to compute the residual noise power spectrum of the NIKA2 150 and 260~GHz maps using the POKER software \citep{pon11}. We eventually fit the noise power spectrum by a model made of a white component and a power law to account for spatial correlations. This model enables the simulation of residual noise map realizations that are used to infer the root mean square (RMS) noise in each pixel of the final maps. The noise map realizations are also used to compute the noise covariance matrix at 150~GHz. The latter is used in the ICM analysis developed in Sect. \ref{sec:MCMC} to account for noise inhomogeneity and spatial correlations in the likelihood function estimation.

\subsection{NIKA2 maps of \psz}\label{sec:Raw_NIKA_observations}
\begin{figure*}[h!]
\centering
\includegraphics[height=4.5cm]{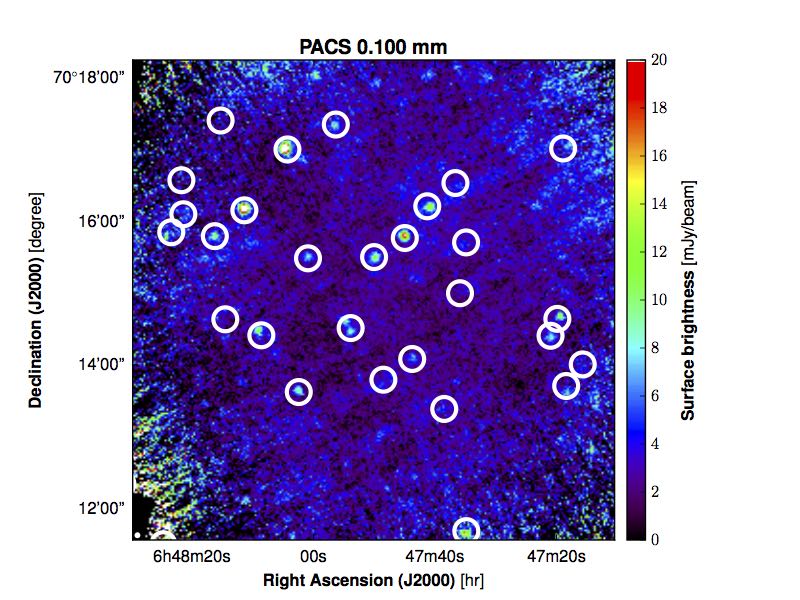}
\includegraphics[height=4.5cm]{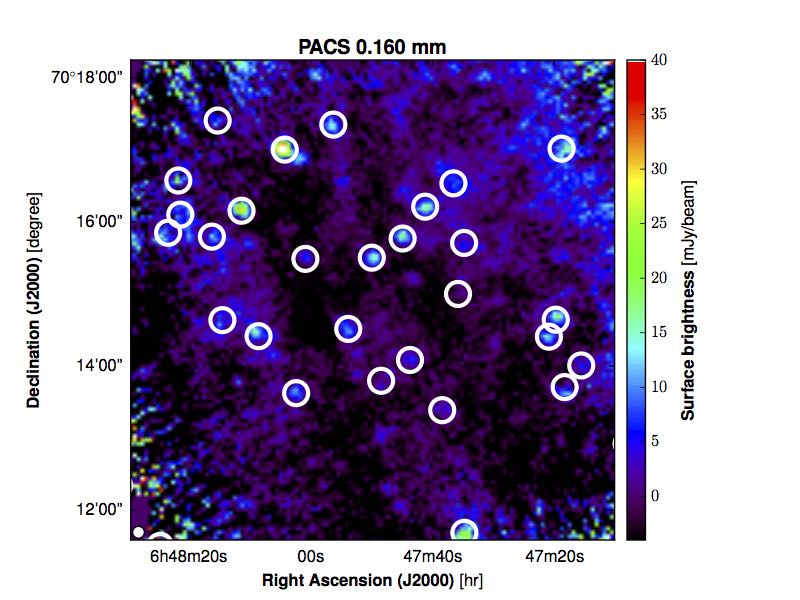}
\includegraphics[height=4.5cm]{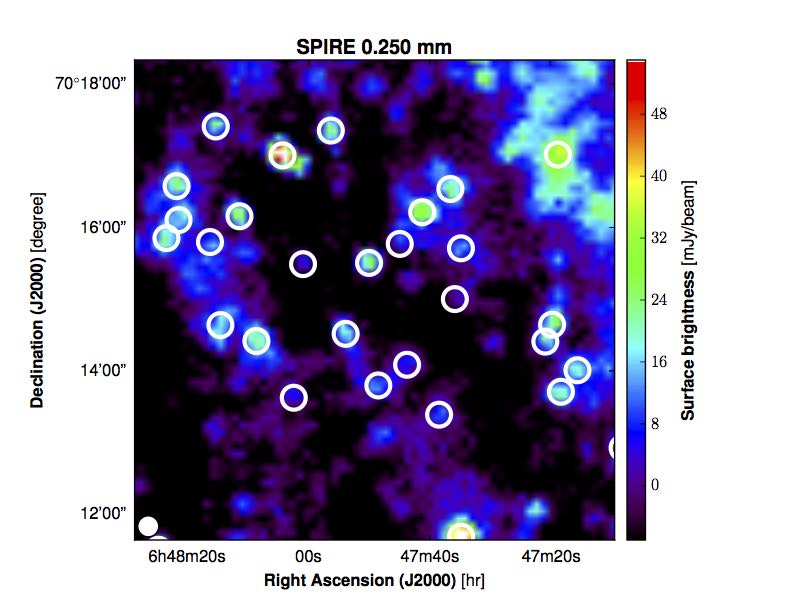}
\includegraphics[height=4.5cm]{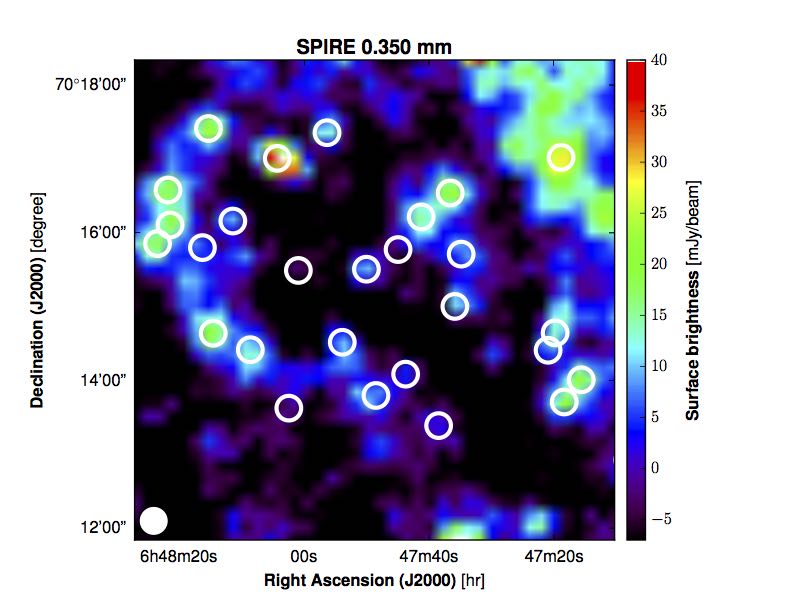}
\includegraphics[height=4.5cm]{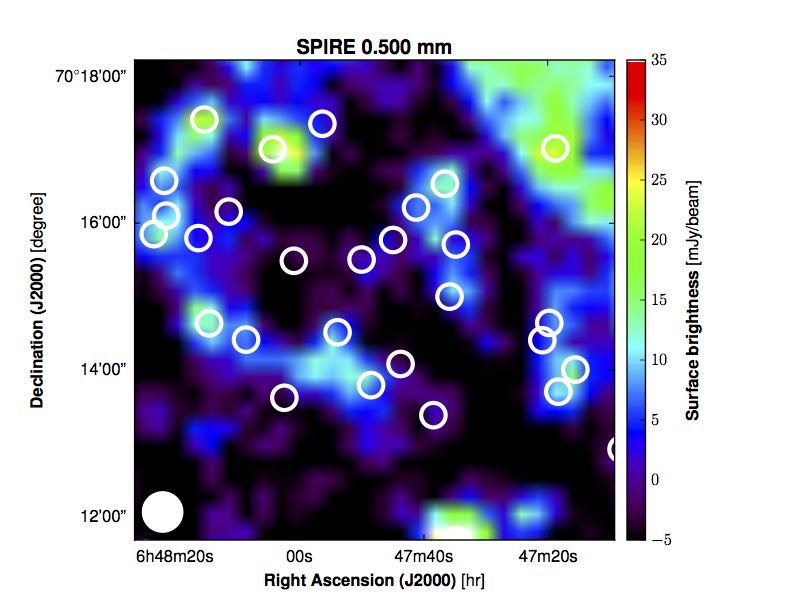}
\includegraphics[height=4.5cm]{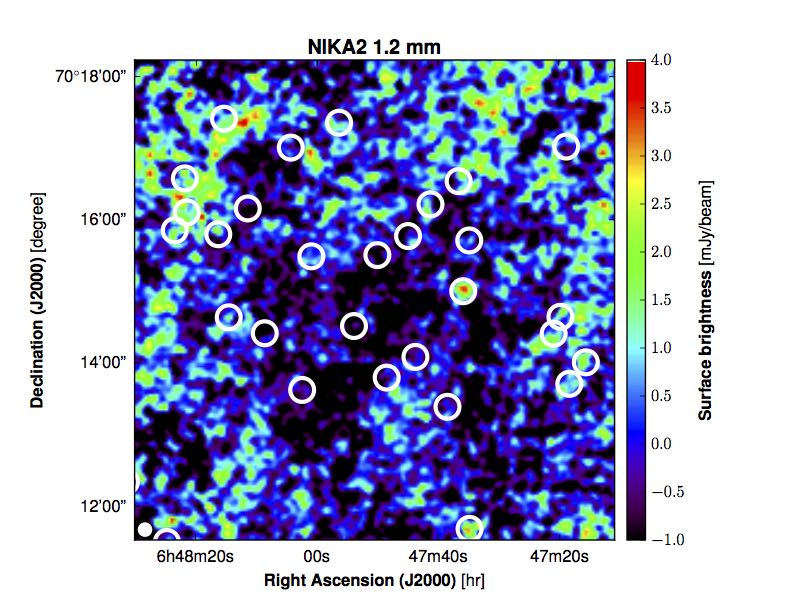}
\caption{{\footnotesize Herschel multiwavelength data set of \psz\ in the sky region considered for NIKA2. The corresponding instrument is indicated at the top of each map \citep{pog10, gri10}. The maps are smoothed and the color range has been optimized for visualization purposes. The submillimeter point source locations are indicated by 10 arcsec radius white circles (Table \ref{tab:Submm_ps_flux}).}}
\label{fig:point_sources}
\end{figure*}
The NIKA2 surface brightness maps of \psz\ are shown in Fig. \ref{fig:NIKA2_maps}, top panels. They are both centered on the tSZ peak coordinates measured by MUSTANG \citep{you15}. The maps have been smoothed with a 10 and 6 arcsec FWHM Gaussian filter at respectively 150 and 260~GHz for visual purposes. At the effective resolution of 21 and 13~arcsec, the RMS noise at the center of the maps is $203~\mathrm{\mu Jy/beam}$ and $933~\mathrm{\mu Jy/beam}$ at 150 and 260~GHz respectively. It has been computed by simulating noise maps from the residual noise power spectrum in the 150 and 260~GHz maps, and from astrophysical contaminant models (\citealt{pla14b}, \citealt{bet12}, and \citealt{tuc11}). Both the contributions induced by the CIB and CMB temperature anisotropies in the residual noise are taken into account following the procedure developed for the NIKA camera and described in \cite{ada17a}. We find that the RMS noise caused by these astrophysical contaminants is $77~\mathrm{\mu Jy/beam}$ at 150~GHz and $280~\mathrm{\mu Jy/beam}$ at 260~GHz at our effective angular resolution of 21 and 13~arcsec respectively. This corresponds to an increase of the RMS noise otherwise caused by both the instrumental and atmospheric noise contributions of 7.9\% and 4.8\% at 150 and 260~GHz respectively.\\
Given that the NIKA2 FOV is about four times larger than the NIKA one, we also investigated the variations of the amplitude of the astrophysical contaminants as a function of the angular scale and compared them to the angular power spectrum of the NIKA2 residual noise at 150~GHz (see Sect. \ref{subsec:raw_data_analysis}). The right panel of Fig. \ref{fig:TF_NIKA2_2mm} shows the angular power spectrum of the NIKA2 residual correlated noise (blue), CMB primary anisotropies (green), CIB (red), and unresolved CIB (orange). The angular power spectrum of the CMB primary anisotropy has been computed using the Cosmic Linear Anisotropy Solving System software \citep{les11} up to a multipole of $10^4$ using the \planck\ cosmology \citep{pla16b}. The CIB power spectrum measured by \cite{pla14b} at 143 and 217 GHz takes into account the one halo (1h) and two halo (2h) terms of the clustering of dusty star-forming galaxies and has been extrapolated to the NIKA2 frequencies. The unresolved CIB power spectrum is modeled as a shot noise component using the models from \cite{bet12,tuc11} for the dusty star-forming galaxies and the radio sources respectively. The three power spectra are convolved with the NIKA2 beam and transfer function at 150~GHz. Most of the astrophysical background at small scales comes from the shot noise induced by the unresolved CIB. The CMB primary anisotropies are largely dominant over the CIB background for angular scales that are larger than the NIKA2 FOV. However, all the astrophysical contaminants are far smaller than the NIKA2 residual correlated noise for all the scales that can be recovered by the NIKA2 instrument.\\

\indent The NIKA2 map at 150~GHz is shown in the top left panel of Fig. \ref{fig:NIKA2_maps}. There is a clear negative tSZ signal that reaches a peak signal-to-noise ratio of $13.5\sigma$ less than 3 arcsec away from the map center. Significant tSZ signal is recovered up to an angular distance from the center of ${\sim}1.4$~arcmin. This is comparable to the ICM extension mapped in X-ray by the \xmm\ observatory as shown in the middle left panel of Fig. \ref{fig:NIKA2_maps}. The diffuse tSZ signal exhibits an elliptical morphology with an E-W orientation, and some evidence for an extension towards the S-W. The RMS noise is relatively constant across the field, which is slightly larger than the NIKA2 instantaneous FOV. As shown in Sect. \ref{sec:MCMC}, the tSZ surface brightness profile extracted from this map is detected at signal-to-noise greater than three out to beyond $\rm{R_{500}}$.\\
\indent The 260~GHz map (top right panel in Fig. \ref{fig:NIKA2_maps}) does not show any significant tSZ signal, as expected given the level of residual noise. Indeed, knowing the tSZ surface brightness peak at 150~GHz, the tSZ spectrum analytic expression, the NIKA2 bandpasses, and the mean opacities for each NIKA2 channel, we can estimate the expected tSZ surface brightness peak at 260 GHz. The computed value of ${\sim}1.4$~mJy/beam is of the order of the RMS noise at 260~GHz. A point source is detected at 260~GHz with a significance of $4\sigma$ at $1.3$~arcmin to the west of the cluster center. The source signal compensates the negative tSZ signal at the same position in the NIKA2 map at 150~GHz. As shown in \cite{ada16a}, removing such a contaminant is essential to obtain unbiased ICM thermodynamic profiles. The removal of submillimeter point sources in the NIKA2 field is explained in Sect. \ref{sec:point_source}.


\section{\psz\ ancillary data}\label{sec:ancillary}

The galaxy cluster \psz\ is one of the 124 spectroscopically confirmed clusters of the MAssive Cluster Survey \citep[MACS,][]{ebe01}. It has been widely observed at different wavelengths. This section aims at describing the previous tSZ and X-ray observations of this cluster and highlights the complementarity between the existing tSZ data sets and the NIKA2 data.

\subsection{Previous tSZ observations of \psz}\label{sec:previous_sz}
\emph{MUSTANG:}\\
The Multiplexed Squid/TES Array at Ninety Gigahertz \citep[MUSTANG, ][]{dic08} instrument operating on the 100-meter Green Bank Telescope observed \psz\ during 16.4 hours between February 2011 and January 2013. Details on the MUSTANG observations and data reduction can be found in \cite{mas10,kor11}. As shown in Fig. \ref{fig:NIKA2_maps} bottom left panel, the cluster tSZ signal at 90~GHz has been mapped with a peak signal-to-noise ratio of 8.1 at an angular resolution of 9~arcsec. The removal of atmospheric noise induces the filtering of the astronomical signals on angular scales larger than the instantaneous FOV of 42~arcsec. The high angular resolution of MUSTANG enables the constraining of angular scales that are two times smaller than the ones constrained by NIKA2. The MUSTANG data brings therefore complementary information to that of NIKA2 to constrain the core dynamics of \psz.\\

\noindent\emph{Bolocam:}\\
Bolocam has been used to obtain a high significance tSZ map of \psz\ at 140~GHz \citep{say13,cza15}. The Bolocam instrument is a 144-pixel bolometer array at the Caltech Submillimeter Observatory \citep[see][for more details on the Bolocam instrument]{hai04}. Its angular resolution of 58~arcsec at 140~GHz does not enable the mapping of the inner structures of the cluster. However, thanks to its instantaneous FOV of 8~arcmin diameter, Bolocam provides valuable information on the pressure distribution in the cluster outskirts. The tSZ signal is significant up to an angular distance of ${\sim}1.7~\mathrm{arcmin}$ from the center as shown in Fig. \ref{fig:NIKA2_maps}, bottom right panel. Used in combination with the NIKA2 observations, the Bolocam data can therefore help constrain the cluster external region of the pressure profile where the NIKA2 signal starts to be significantly filtered \citep[see, e.g.][]{rom17}. Both MUSTANG and Bolocam data have been used by \cite{you15} to constrain the pressure distribution of \psz\ by using a spherical parametric model (see Sect. \ref{sec:MCMC}).\\

\noindent\emph{Planck:}\\
The galaxy cluster \psz\ has been identified by \planck\ with a signal-to-noise ratio of 5.28. Its integrated Compton parameter estimated at $5R_{500}$ is given in the early \planck\ tSZ catalog by $\mathrm{Y_{5R500}} = (1.38 \pm 0.29) \times 10^{-3}~\mathrm{arcmin}^2$ and in the second catalog release by $\mathrm{Y_{5R500}} = (1.34 \pm 0.29) \times 10^{-3}~\mathrm{arcmin}^2$. The \planck\ $y$-map of \psz\ is shown in the middle right panel of Fig. \ref{fig:NIKA2_maps}. It has been extracted from the \planck\ full sky $y$-map \citep{pla16d} using a Gnomonic projection. We compute the cylindrical integrated Compton parameter by aperture photometry on this map and find a value of $\mathrm{Y_{5R500}} = (1.00 \pm 0.44) \times 10^{-3}~\mathrm{arcmin}^2$, which is compatible with the \planck\ catalog value but slightly lower due to negative correlated noise at the cluster location. The \planck\ map of \psz\ will help constrain the outer slope of the pressure profile in the ICM characterization presented in Sect. \ref{sec:MCMC} by providing information on the total integrated Compton parameter.\\

\begin{figure*}[h!]
\centering
\includegraphics[height=6.5cm]{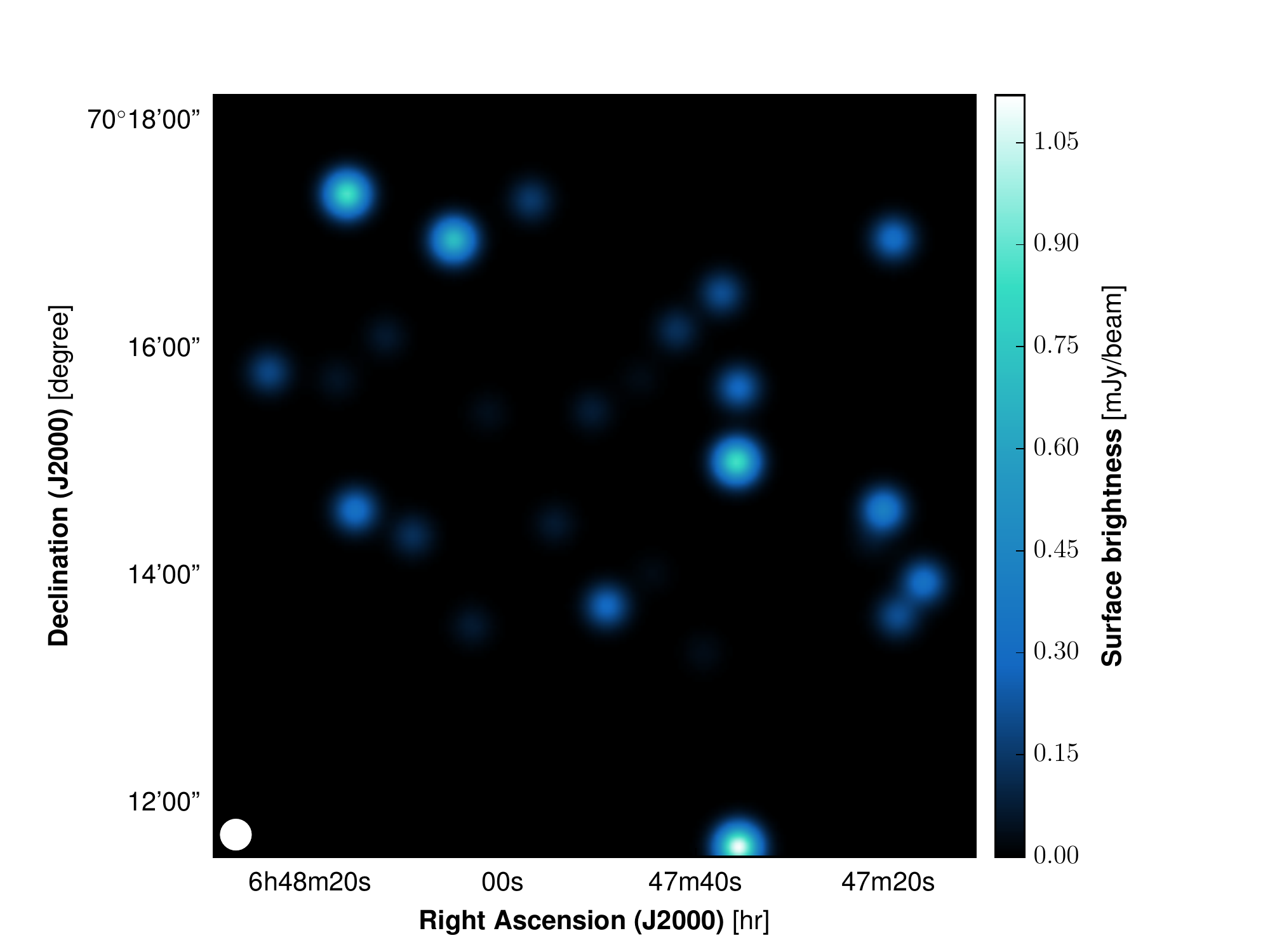}
\includegraphics[height=6.5cm]{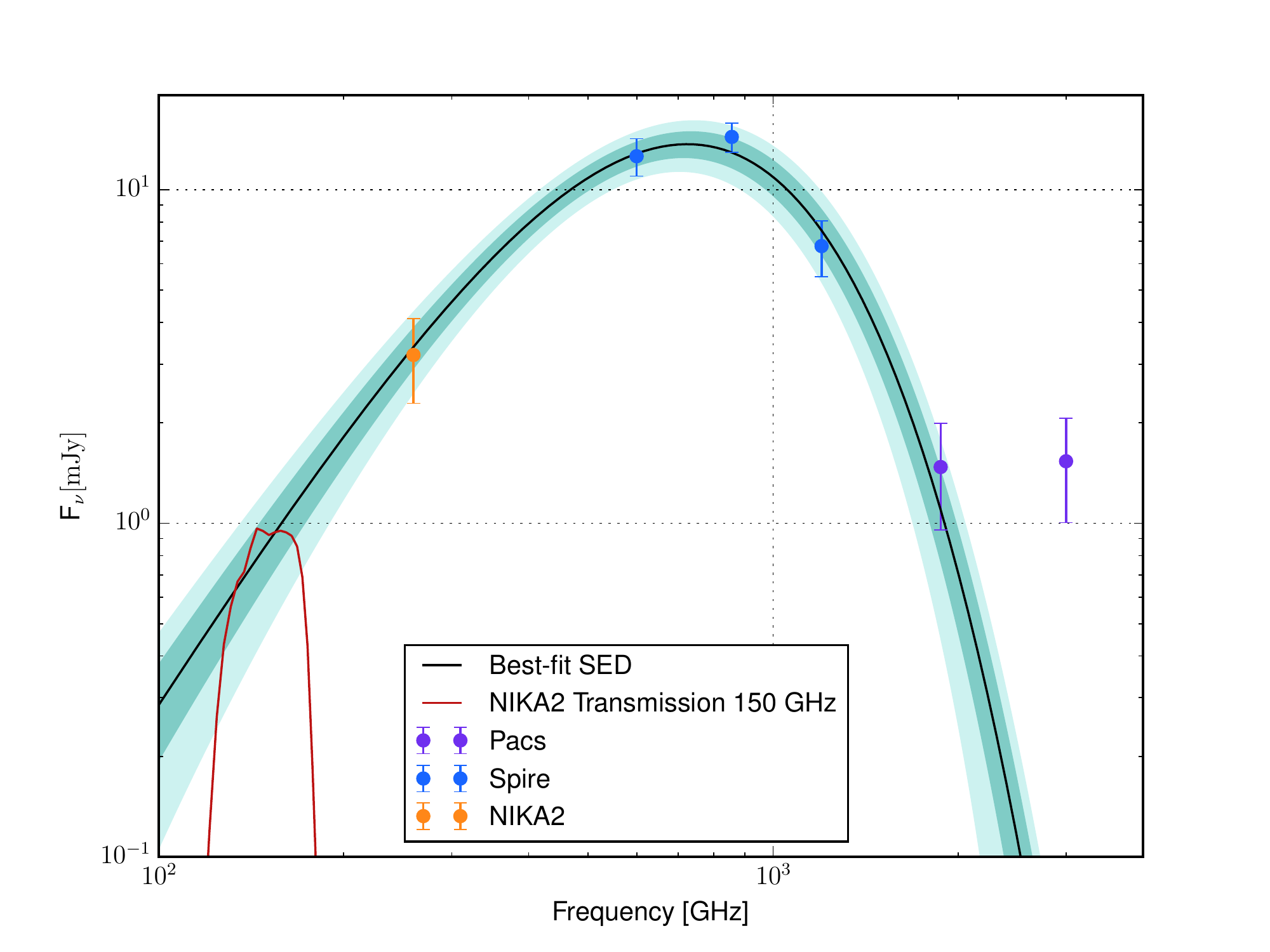}
\caption{{\footnotesize \textbf{Left:} Map of the expected flux at 150~GHz for each submillimeter source identified in the Herschel and NIKA2 channels. Each point source is modeled by a 2D Gaussian function of 17.7~arcsec FWHM and with an amplitude given by the last column of Table \ref{tab:Submm_ps_flux}. \textbf{Right:} SED of the submillimeter source SMG28 (see Table \ref{tab:Submm_ps_flux}) identified in the NIKA2 map at 260~GHz. The fluxes measured in the Herschel maps are color corrected using both PACS \citep{pog10} and SPIRE calibration files. The best-fit SED model (black solid line) is integrated in the NIKA2 bandpass at 150~GHz (red solid line) to estimate the expected flux at this frequency. The 1 and 2 sigma uncertainties on the best-fit SED are shown as dark and light blue-green shaded areas.}}
\label{fig:Expected_pts_NIKA2}
\end{figure*}

\noindent\emph{AMI:}\\
The Arcminute Microkelvin Imager (AMI) also provides an independent measurement of the integrated Compton parameter of \psz\ \citep{per15} and partially alleviates the degeneracy between the cluster characteristic size ($\theta_s$) and flux ($\mathrm{Y_{5R500}}$). The combined {\planck}/AMI constraint on the integrated Compton parameter is given by $\mathrm{Y_{5R500}} = (1.72\pm 0.47) \times 10^{-3}~\mathrm{arcmin}^2$, which is compatible with all the previously quoted estimates. This constraint is used to give a first estimation of the radial pressure profile of \psz\ (see Sect. \ref{sec:MCMC}).

\subsection{XMM-{\it Newton} observations}\label{sec:XMM}
The galaxy cluster \psz\ has been observed by both \xmm\ and \chandra. Here we focus on the former observations, which yield enough counts for a spatially-resolved spectroscopic analysis. Data were retrieved from the \xmm\ archive and the standard procedures \citep[see, e.g.][]{bar17} have been followed for the production of cleaned, calibrated event files, vignetting correction, point source detection and exclusion, and background subtraction. The total exposure time after cleaning for flares is ${\sim}68~\rm{ks}$. The \xmm\ X-ray surface brightness map of \psz\ is shown in the middle left panel of Fig. \ref{fig:NIKA2_maps}. The overall ICM morphology is comparable to the one mapped by NIKA2 at 150 GHz (see Sect. \ref{sec:Raw_NIKA_observations}). However, no significant ICM extension is identified in the S-W region of the cluster. The implication of such an observation is discussed in Sect. \ref{sec:overp}.


\section{Point source contamination}\label{sec:point_source}

The tSZ signal of the cluster can be affected by foreground, background, and cluster contaminants. Such contaminants have to be characterized and carefully removed for each cluster of the NIKA2 tSZ large program as they can bias the pressure profile estimation and therefore the calibration of the $\rm{Y_{500}}-\rm{M_{500}}$ scaling relation \citep[see, e.g.][]{ada16a}. The cosmological backgrounds have already been shown to be negligible with respect to the cluster signal in Sect. \ref{sec:Raw_NIKA_observations}. In this section, we consider the impact of point source contamination by constraining the flux at 150~GHz for all the sources found in the considered field shown in Fig. \ref{fig:NIKA2_maps}.

\subsection{Radio sources}\label{sec:radio}

High resolution radio observations have not been performed for \psz. The map of the cluster field obtained by the Giant Metrewave Radio Telescope (GMRT) at 150~MHz for the Tata Institute of Fundamental Research (TIFR) GMRT Sky Survey \citep[TGSS,][]{int17} does not show evidence of any radio source. The Very Large Array of the National Radio Astronomy Observatory (NRAO VLA) Sky Survey \citep[NVSS,][]{con98} at 1.4~GHz enabled the identification of one radio source in the $6.7 \times 6.7~\mathrm{arcmin}^2$ field that we take into account. It is located to the east of \psz\ at $(\mathrm{R.A., Dec.})_{\mathrm{J2000}}$ = (06:48:15.9, +70:15:7.2). The flux of this source given in the NVSS catalog is $(3.09\pm 0.51)~\mathrm{mJy}$ at 1.4~GHz. No significant signal is seen either at 150 or 260~GHz with NIKA2 at this position. If we model the radio source spectral energy distributions (SED) by a power law, $F_{\nu} = F_{\rm{1~GHz}}\left(\frac{\nu}{\rm{1~GHz}}\right)^{\alpha_{\rm{radio}}}$ and assume a mean spectral index of -0.7 \citep{wit79,pla13}, the expected fluxes at 150 and 260~GHz are $(0.09\pm 0.02)~\mathrm{mJy}$ and $(0.06\pm 0.01)~\mathrm{mJy}$ respectively. As these fluxes are well below the RMS noise for both NIKA2 frequencies, the radio source contamination is therefore considered negligible in the following.
\begin{figure*}[h!]
\centering
\includegraphics[height=6.8cm]{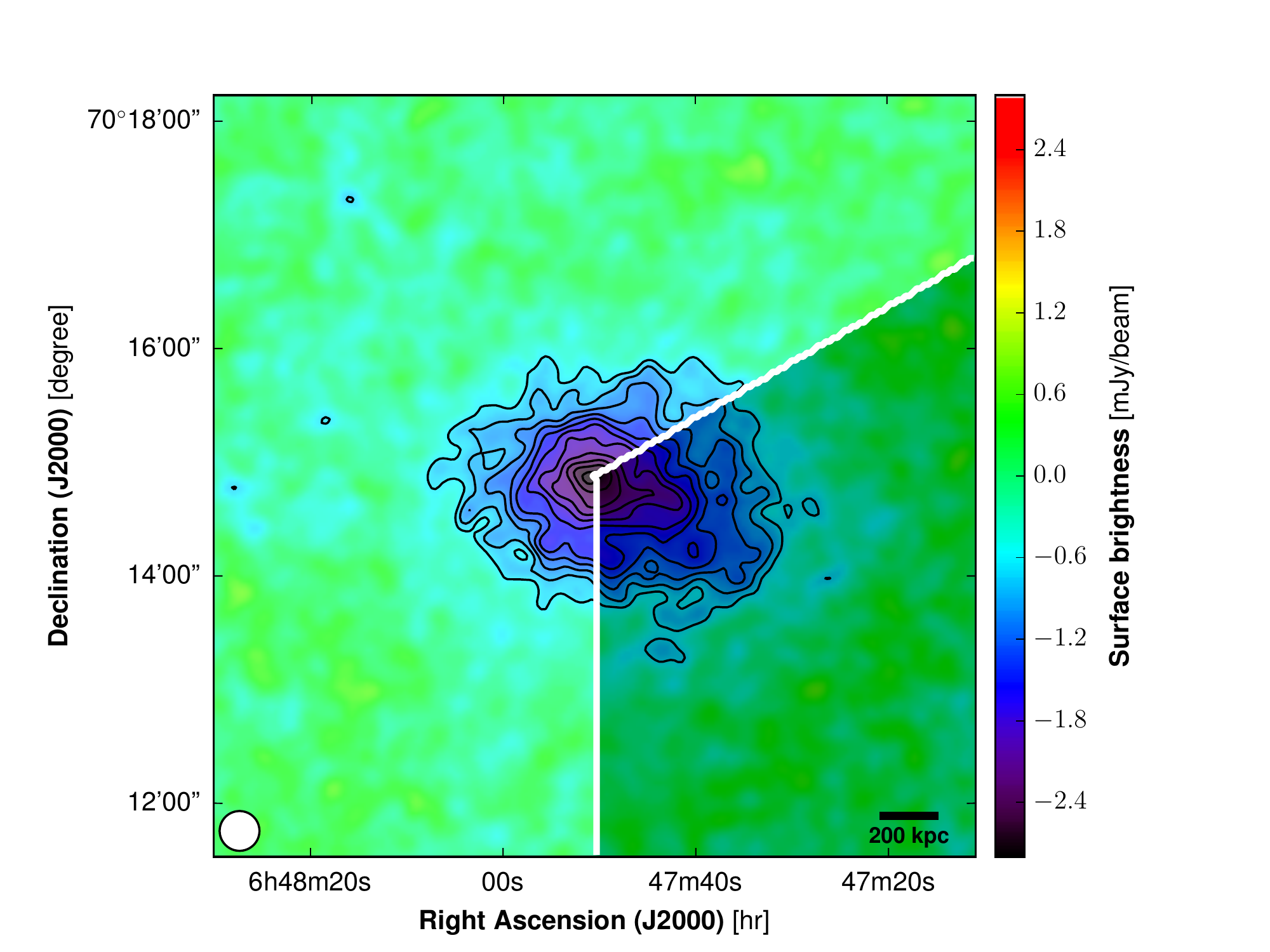}
\includegraphics[height=6.8cm]{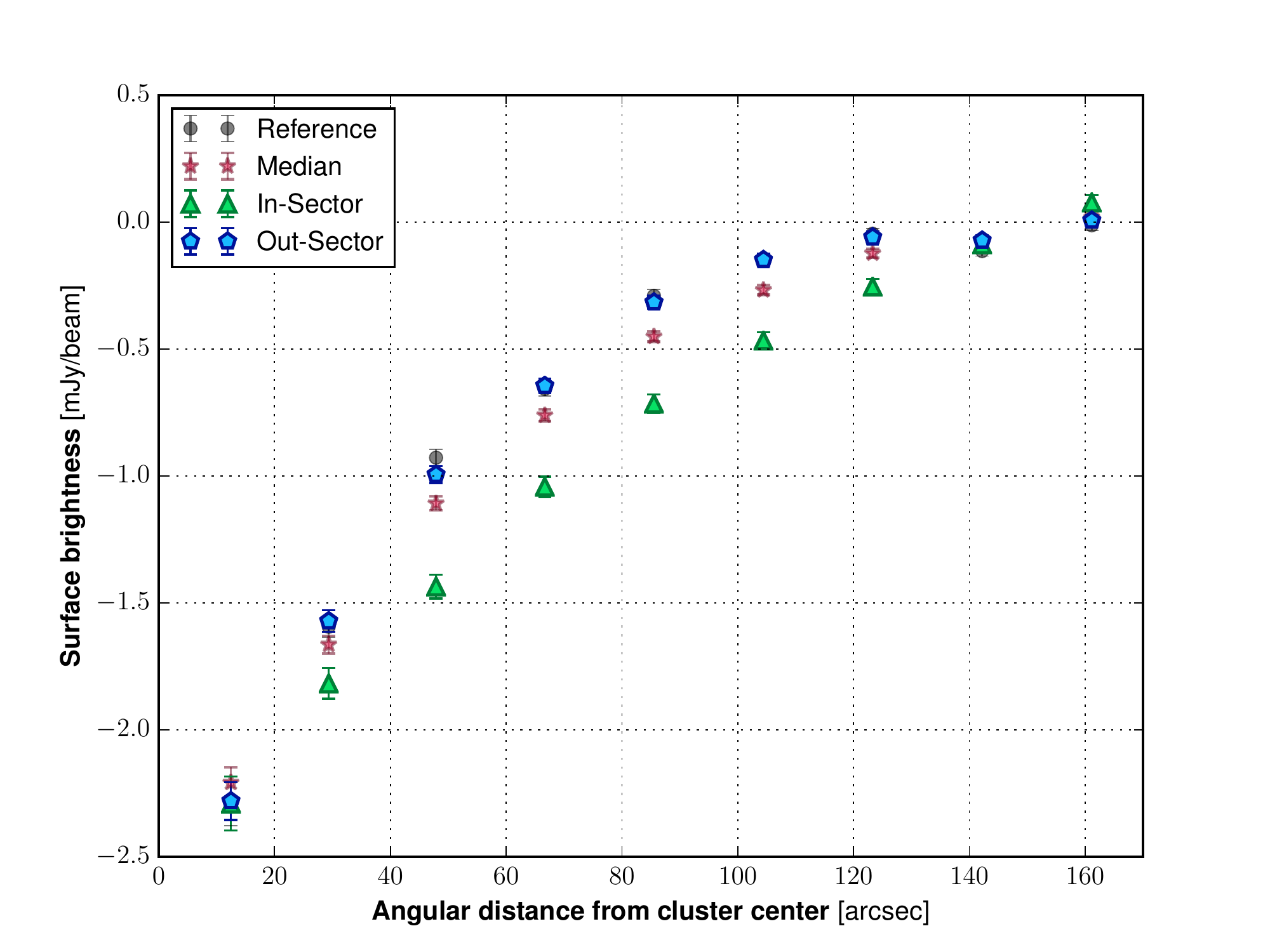}
\caption{{\footnotesize \textbf{Left:} NIKA2 tSZ surface brightness map at 150~GHz after the subtraction of the submillimeter point source contaminant. The best circular sector, which encloses the over-pressure region within \psz\ , is darkened and delimited by the white lines. \textbf{Right:} tSZ surface brightness profiles computed inside (green) and outside (blue) the dark sector shown in the left panel. The inner and outer radii of the mask are defined by the region where the two profiles are significantly different. The reference profile computed in the north-east quarter of the NIKA2 150 GHz map is shown in black and the median profile computed on the whole map is represented in red.}}
\label{fig:mask_defintion}
\end{figure*}

\subsection{Submillimeter sources}\label{sec:submm}
We considered the Herschel Photoconductor Array Camera and Spectrometer (PACS) \citep{pog10}, Spectral and Photometric Imaging Receiver (SPIRE) \citep{gri10}, and NIKA2 260~GHz data to identify submillimeter point sources in the observed field and compute their expected fluxes at 150~GHz. We found 27 point sources in the region observed by NIKA2 in the SPIRE 250~$\mu$m catalog. The source identified in the NIKA2 260~GHz channel (see Sect. \ref{sec:Raw_NIKA_observations}) is not listed in the SPIRE catalog. Therefore we added this source to the sample and considered the surface brightness peak location as a first estimate of the source position.\\
\indent The procedure used to fit the flux of each source in the Herschel and NIKA2 channels is based on the one described in \cite{ada16a}. We perform a Markov chain Monte Carlo (MCMC) analysis to jointly fit the flux of the 28 sources at each frequency independently. We consider the five Herschel bands as well as the NIKA2 260~GHz channel to constrain the fluxes. The NIKA2 map at 150~GHz is not used as most of the source fluxes are contaminated by negative tSZ signal. The uncertainty on the position of each source is taken into account by using Gaussian priors with both mean and standard deviation given by the positions and uncertainties of the SPIRE 250~$\mu$m catalog. A local background is also fitted. At each step of the MCMC, a surface brightness map is computed by adding the local background and Gaussian functions at the source positions with a fixed FWHM given by the Herschel resolution in the considered channel (35.2, 23.9, 17.6, 9.9, and 6.1 arcsec at 500, 350, 250, 160, and 100 $\mu$m respectively). This procedure enables us to take into account the correlations between the fluxes of close unresolved sources in the error bar estimation for each source flux especially in the low frequency channels of Herschel. The amplitude of each Gaussian in the best-fit surface brightness maps corresponds to the measured fluxes that are reported for the 28 sources in Table \ref{tab:Submm_ps_flux}. The error bars are computed using the standard deviation of the Markov chains after their convergence.\\
We use a modified blackbody spectrum
\begin{equation}
F_{\nu} = A_0\left(\frac{\nu}{\nu_0}\right)^{\beta_{\mathrm{dust}}}B_{\nu}(T_{\mathrm{dust}})
\end{equation}
to model the SED of the identified submillimeter point sources from their estimated fluxes. A second MCMC analysis is performed to estimate the normalization $A_0$, dust spectral index $\beta_{\mathrm{dust}}$ , and dust temperature $T_{\mathrm{dust}}$ for each source. The reference frequency $\nu_0$ is fixed to 500~GHz. At each step of the MCMC, the measured fluxes are color corrected using both PACS \citep{pog10} and SPIRE\footnote{http://herschel.esac.esa.int/Docs/SPIRE/html/spire\_om.html.} calibration files. The best-fit SEDs are then integrated within the NIKA2 bandpass at 150~GHz to estimate the expected flux of each source at this frequency. The corresponding fluxes are reported in the last column of Table \ref{tab:Submm_ps_flux}. The surface brightness map of the point source contamination at 150~GHz is shown in Fig. \ref{fig:Expected_pts_NIKA2}, left panel. This map is subtracted from the NIKA2 map at 150~GHz to minimize the bias induced by the submillimeter point source contaminant on the measurement of the tSZ signal of \psz.\\
\indent The best-fit SED of the submillimeter point source identified in the NIKA2 map at 260~GHz (SMG28 in Table \ref{tab:Submm_ps_flux}) is shown in Fig. \ref{fig:Expected_pts_NIKA2}, right panel. The dust temperature is found to be $T_{\mathrm{28}}^{\mathrm{fit}} = (8.1 \pm 0.6)~\mathrm{K}$ and the spectral index is $\beta_{\mathrm{28}}^{\mathrm{fit}} = 1.35 \pm 0.23$ at the maximum likelihood. Assuming a typical SED with the same spectral index and a physical dust temperature $T_{\mathrm{28}}^{\mathrm{dust}} = (35 \pm 5)~\mathrm{K}$ \citep[see, e.g.][]{pla16e}, the redshift $z$ of this source is given by
\begin{equation}
        1+z = \frac{T_{\mathrm{28}}^{\mathrm{dust}}}{T_{\mathrm{28}}^{\mathrm{fit}}}.
\label{eq:redshift_sou}
\end{equation}
Therefore, the redshift of the point source identified in the NIKA2 260~GHz map is found to be $z = 3.32 \pm 0.72$. We emphasize that this estimate does not take into account the degeneracy between the source redshift and its physical temperature. Disentangling these two parameters requires taking into account other source properties, such as the neutral hydrogen column density, which have not been measured yet. Nevertheless, this result highlights the potential of the NIKA2 camera to constrain the SED of high-redshift dusty galaxies by providing additional constraints to existing data sets obtained at higher frequencies.

\section{Over-pressure region in \psz}\label{sec:overp}
Over-pressure regions caused by merger events, substructure accretion, or turbulence are expected to occur more frequently at high redshift in the standard scenario of the formation of the large-scale structures \citep[see, e.g.][]{poo07}. Although \cite{mac17} do not find more morphologically disturbed clusters at high redshift from X-ray observations, the tSZ properties of high-redshift clusters could still be significantly different. For example, thermal pressure excesses caused by clumps of low density and high temperature would be challenging to identify from X-ray observations at high redshift whereas tSZ observations could identify such over-pressure regions. Such an increase in thermal pressure is challenging to link to a modification of the main halo mass because it is due to local perturbations of the gas thermodynamic properties and the related thermal energy is not redistributed evenly within the cluster. It is therefore essential to identify and mask over-pressure regions in galaxy clusters to study their impact on the characterization of the thermal pressure distribution that can be related to the underlying gravitational potential of the main halo. This section describes the analysis that has been done to identify an over-pressure region within the ICM of \psz\ taking advantage of the high angular resolution and sensitivity of NIKA2.\\
\begin{figure*}[h!]
\centering
\includegraphics[height=6.8cm]{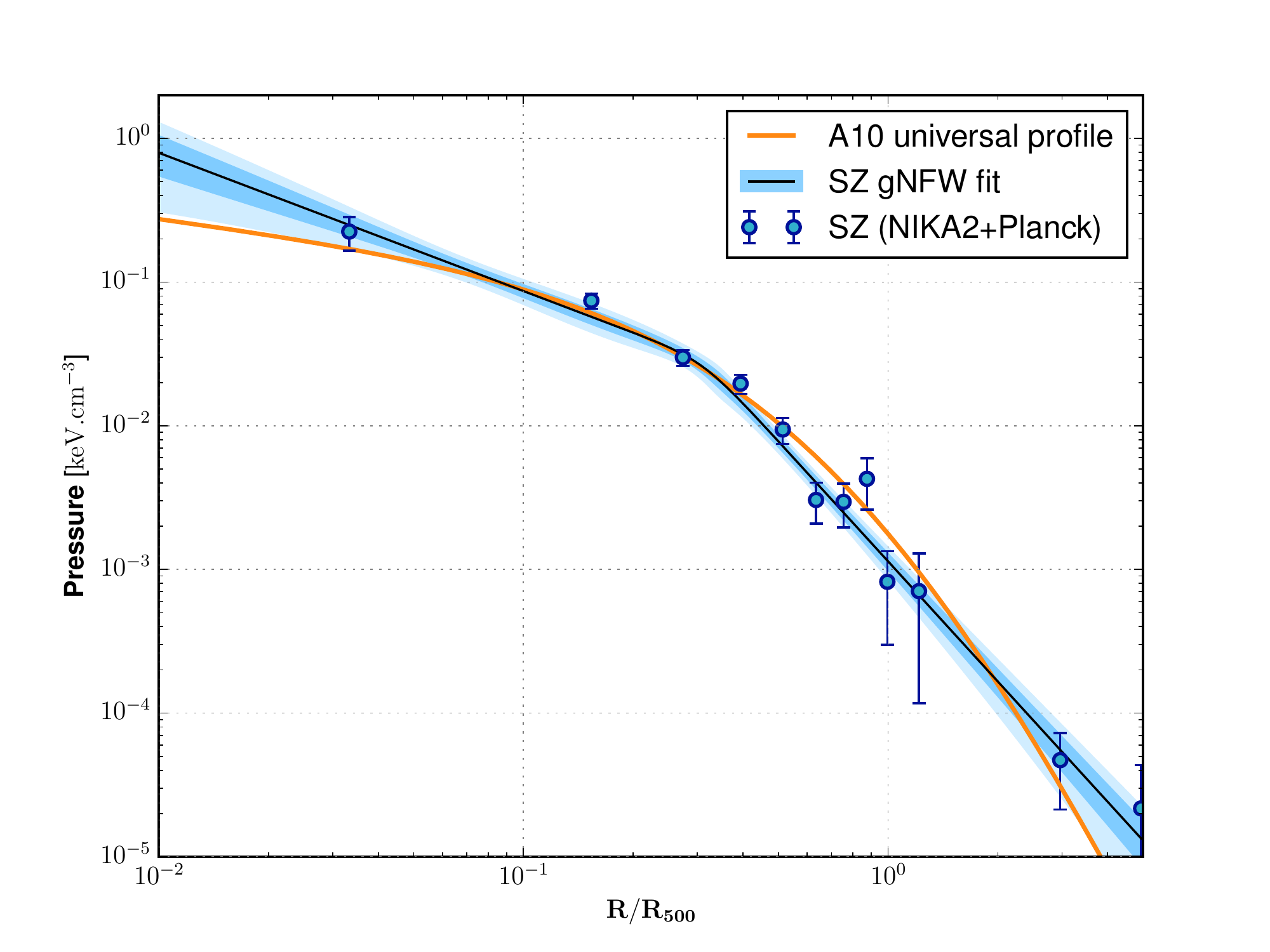}
\includegraphics[height=6.8cm]{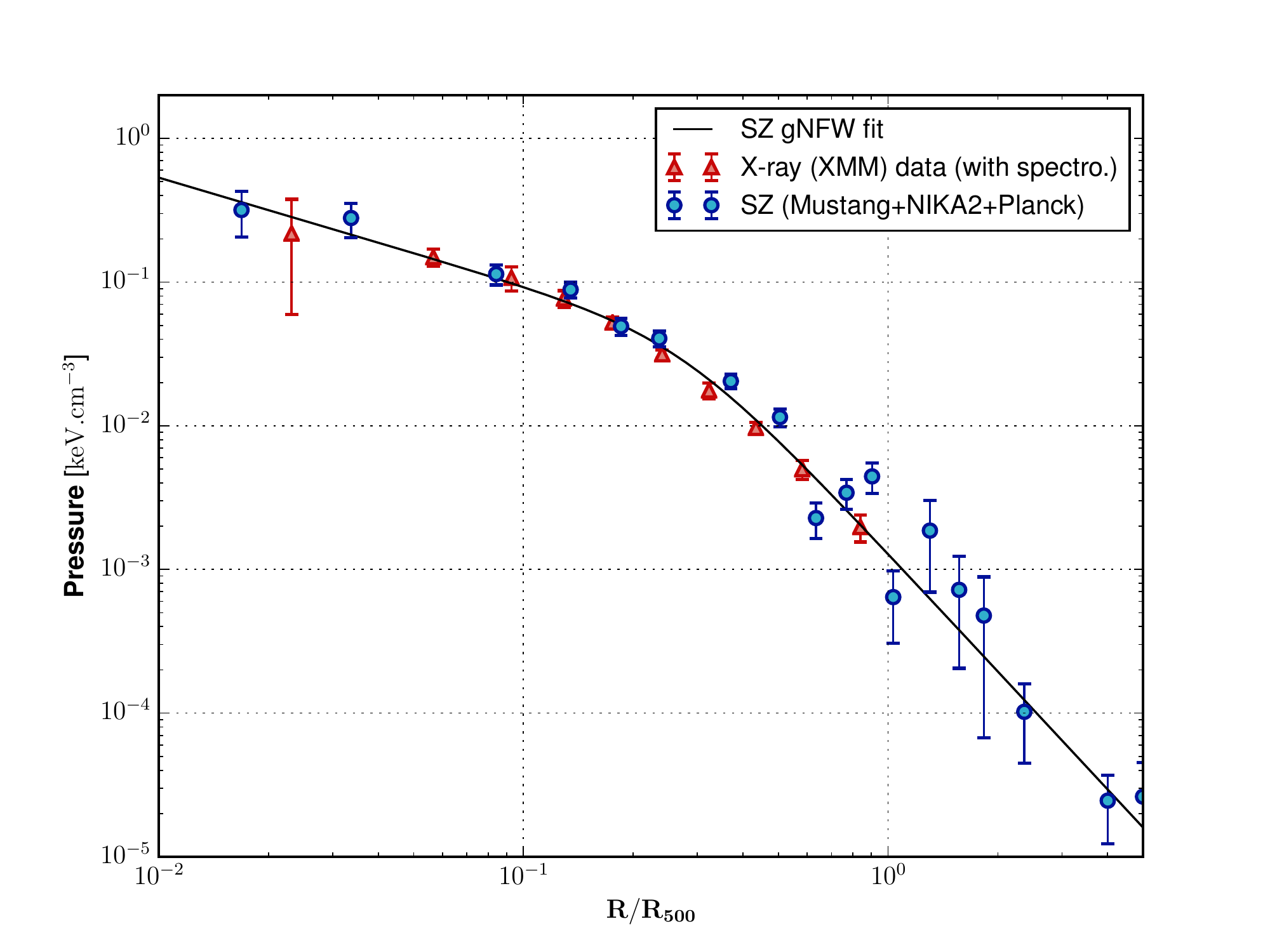}
\caption{{\footnotesize \textbf{Left:} Non-parametric pressure profile (blue points) constrained from the combination of the NIKA2 tSZ surface brightness map with the over-pressure region masked and the \planck\ Compton parameter map only. The deprojected profile has been fitted using a generalized Navarro-Frenk-White \citep[gNFW, ][]{nag07} model (black line), which is compared to the universal pressure profile obtained by \cite{arn10} (orange line). \textbf{Right:} Non-parametric pressure profile (blue) deprojected from the MUSTANG and NIKA2 tSZ surface brightness map with the over-pressure region masked and the \planck\ Compton parameter map. A gNFW pressure profile model has been fitted on the MUSTANG+NIKA2+\planck\ deprojected pressure points (black solid line). We also show the \xmm\ estimated pressure profile (red) based on the deprojected density profile and the temperature estimation from X-ray spectroscopic observations. The MUSTANG/NIKA2/\planck\ and XMM-{\it Newton} estimates are compatible within error bars.}}
\label{fig:Nonparam_profiles}
\end{figure*}
\indent First, from a visual inspection of the maps shown in Figs. \ref{fig:NIKA2_maps} and \ref{fig:NIKA2_maps_iterative} we identify a deviation from the tSZ signal circular symmetry in the south-west region of \psz. The fact that the \xmm\ X-ray surface brightness map of \psz\ (see middle left panel in Fig. \ref{fig:NIKA2_maps}) does not show any significant overdensity substructure in this cluster region could be a hint that this thermal pressure excess is caused by a low density infalling sub clump with a high temperature. Another plausible explanation would be that this tSZ surface brightness excess is caused by a low density and very extended structure along the line of sight. A precise measurement of the gas temperature in this cluster region will help to discriminate between these two descriptions.\\
\indent We consider the north-east quarter of the NIKA2 150 GHz map as a reference region where the tSZ signal is assumed to trace the thermal pressure distribution of the main halo without a merging event. The location and size of the over-pressure region are characterized by computing the tSZ surface brightness profiles inside ($y_{\mathrm{in}}$) and outside ($y_{\mathrm{out}}$) circular sectors defined by a central position and a central angle. Each profile is compared to the one computed within the reference region ($y_{\mathrm{ref}}$) for different values of the central position and central angle. The best sector parameters are defined as the ones that maximize the ratio
\begin{equation}
R = \frac{\int |y_{\mathrm{ref}} - y_{\mathrm{in}}| \, d\theta}{\int |y_{\mathrm{ref}} - y_{\mathrm{out}}| \, d\theta}.
\label{eq:ratio_mask}
\end{equation}
This ensures the maximization of the difference between the tSZ profile computed within the masked sector (dark region in Fig. \ref{fig:mask_defintion} left panel) and the reference profile, while minimizing the difference between the tSZ profile computed outside the mask (bright region in Fig. \ref{fig:mask_defintion} left panel) and the reference profile for all angular distances.\\
\indent We find that the smallest sector that encloses the whole over-pressure region is centered at a position of $-30^{\circ}$ with respect to the R.A axis and has an aperture angle of $120^{\circ}$ (dark region in Fig. \ref{fig:mask_defintion} left panel). The tSZ surface brightness profiles computed inside and outside this sector are shown in Fig. \ref{fig:mask_defintion} right panel. The difference between the profiles is significant for angular distances between 30~arcsec and 130~arcsec. The profile computed outside the mask is consistent with the reference profile in black. We also investigate the possibility of considering the median tSZ profile in red computed over the whole map to minimize the impact of the disturbed region. As expected, the latter is much more consistent with the reference profile than the mean profile computed within the mask. However, as the over-pressure region represents a third of the cluster tSZ signal constrained by NIKA2 and the variation of the signal is continuous between the masked and the unmasked regions, the median value computed in each bin is slightly biased by the over-pressure. The median profile computed over the whole map is therefore lower than the mean profile computed outside the masked sector. We note, however, that in the presence of small spatial scale fluctuations of the ICM, computing the median profile represents a good alternative to the definition of a masked sector that can be challenging to undertake for a large sample of galaxy clusters.\\

Considering the whole NIKA2 map at 150~GHz for the ICM analysis under the hydrostatic equilibrium hypothesis would not enable the characterization of the impact of the south-west disturbed region on the estimation of the whole ICM pressure distribution. Therefore, a mask of the over-pressure region is defined in the following by using the best sector parameters, an inner radius of 30~arcsec and an outer radius of 130~arcsec.\\
\indent The ICM analysis described in Sect. \ref{sec:MCMC} considers both the whole NIKA2 map at 150~GHz and the map with a mask on the over-pressure region. This enables us to constrain the impact of the thermal pressure excess in the south-west of \psz\ on the pressure profile characterization.


\section{ICM characterization from a multi-instrument MCMC analysis}\label{sec:MCMC}

This section is dedicated to the characterization of the ICM thermodynamic properties of \psz. We first deprojected the tSZ signal mapped by NIKA2 and \planck\ to show the typical constraints that will be obtained on the pressure profile characterization for each cluster of the NIKA2 SZ large program. We then combined the MUSTANG, NIKA2, Bolocam, and \planck\ data in a multi-instrument analysis in order to get the best constraints on the cluster pressure profile from its core to its outskirts.

\subsection{Pressure profile reconstruction}\label{sec:MCMC_pressure}

\begin{figure*}[h!]
\centering
\includegraphics[height=4.4cm]{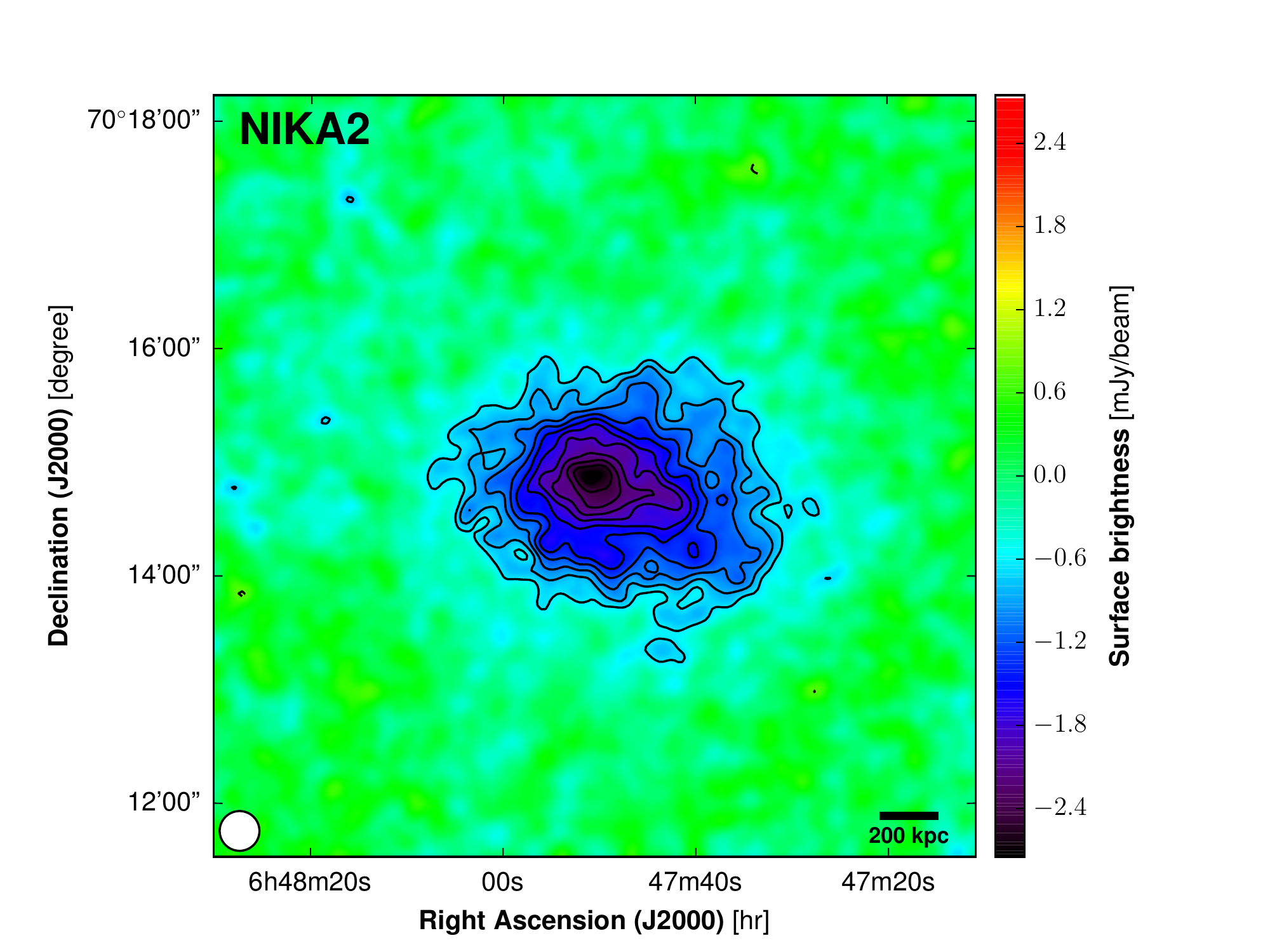}
\includegraphics[height=4.4cm]{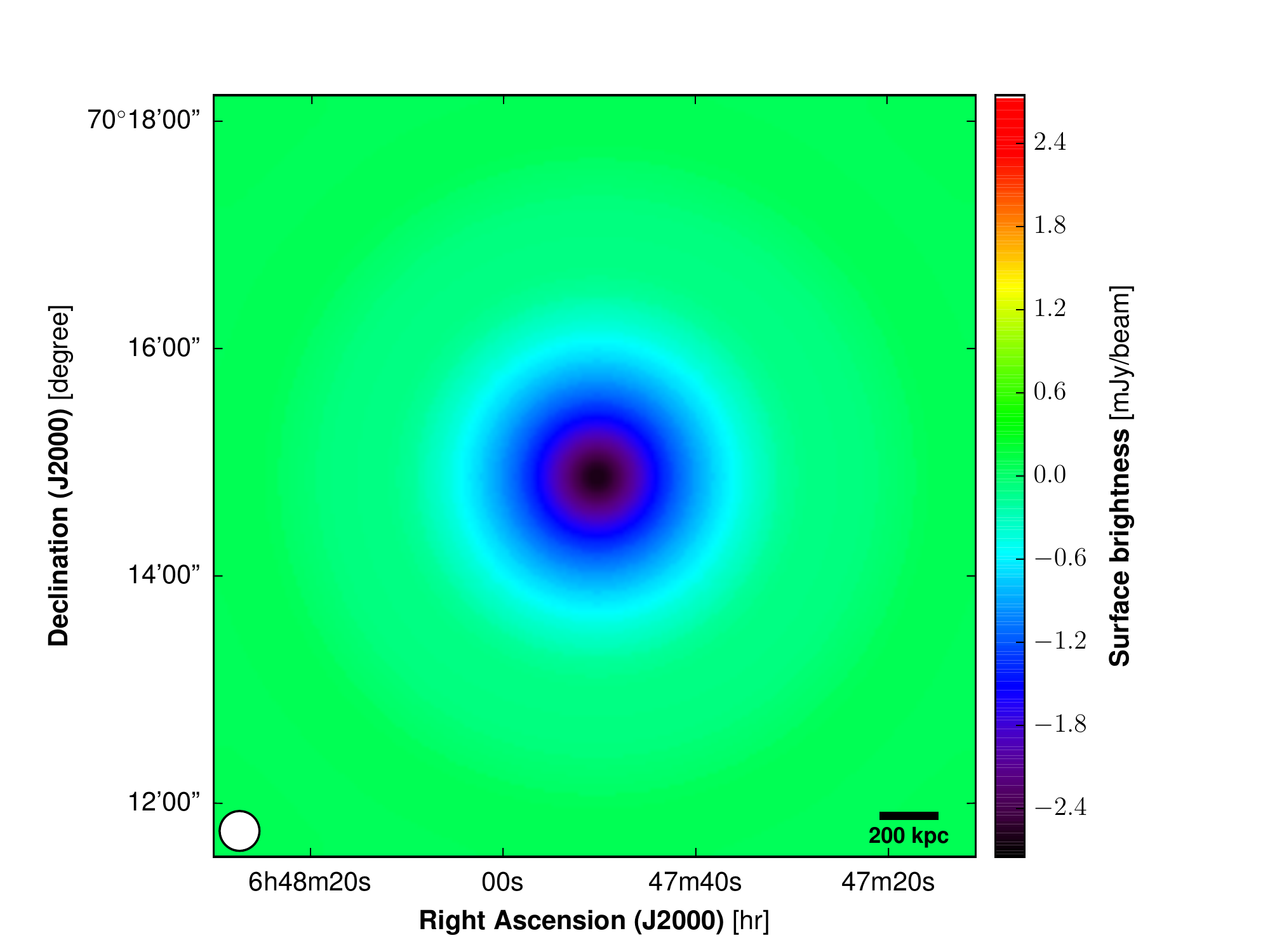}
\includegraphics[height=4.4cm]{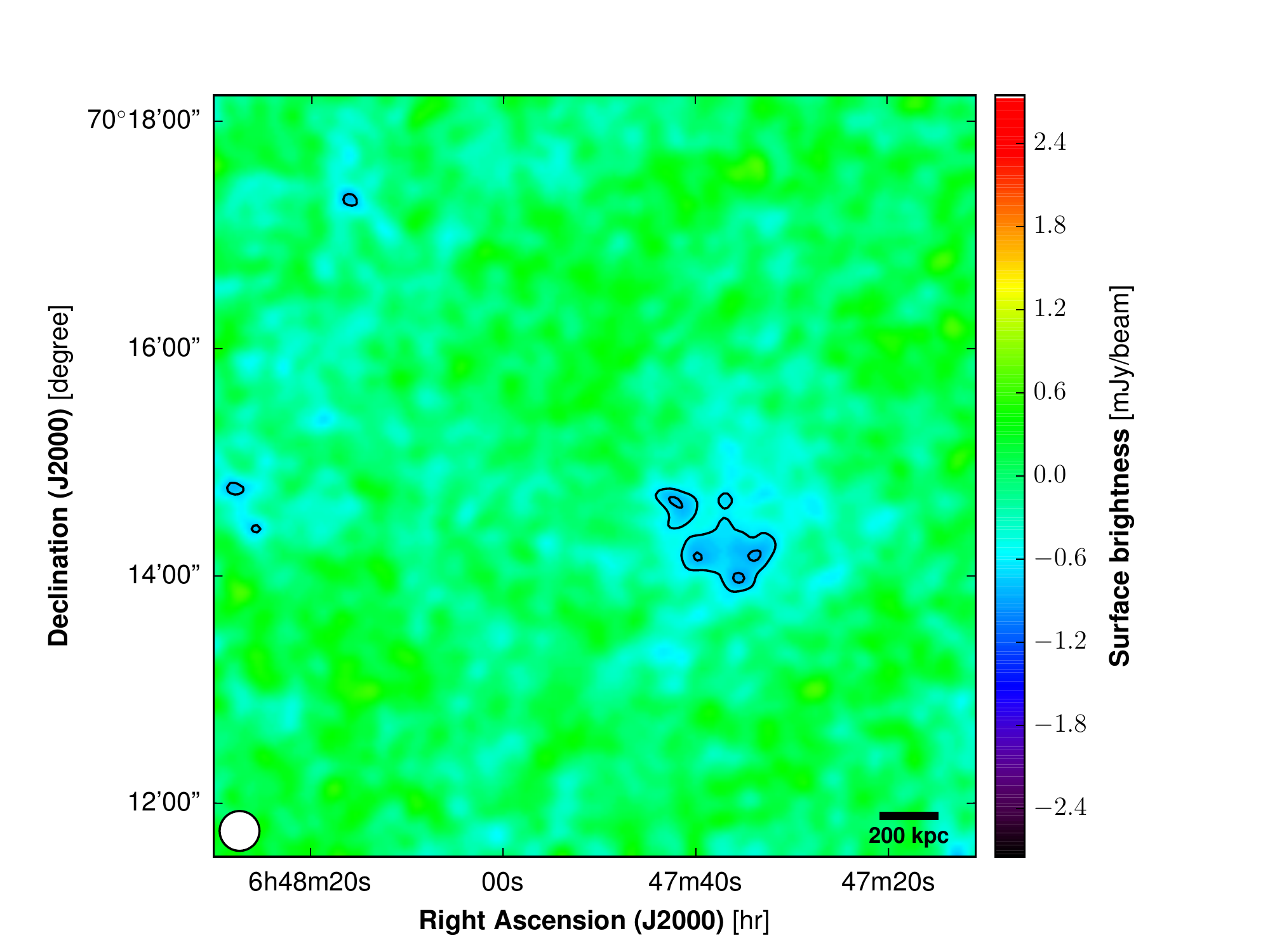}
\includegraphics[height=4.4cm]{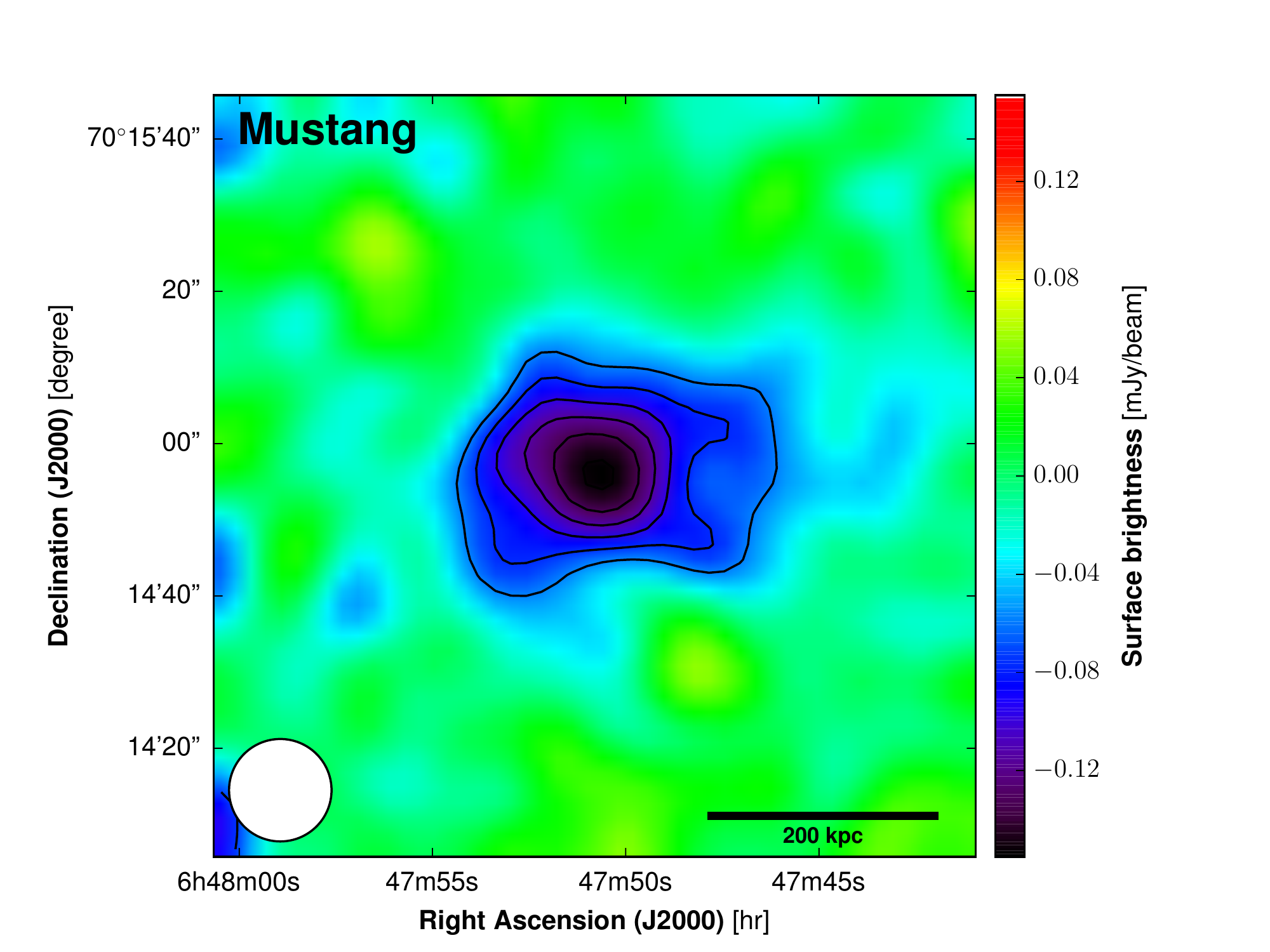}
\includegraphics[height=4.4cm]{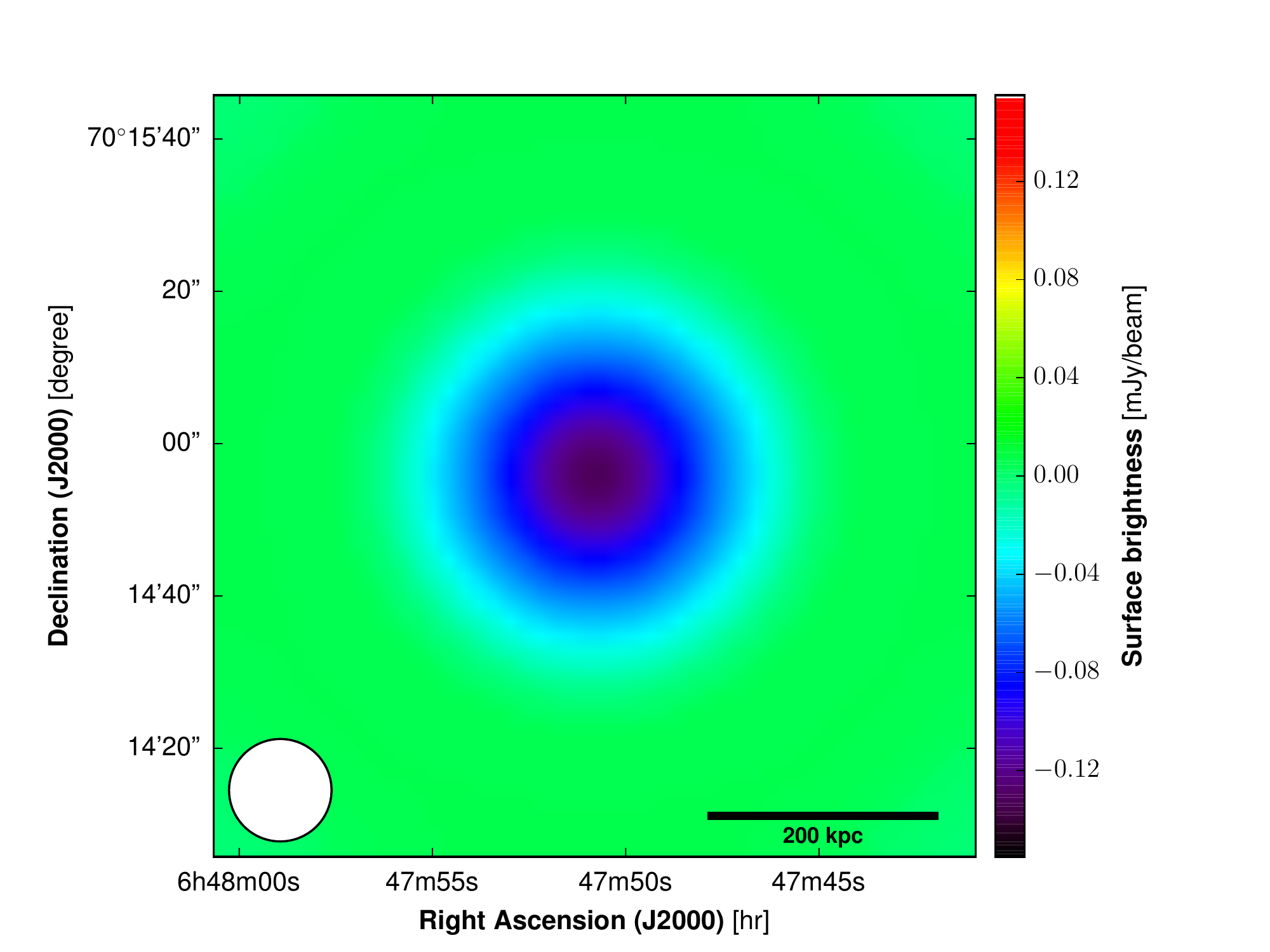}
\includegraphics[height=4.4cm]{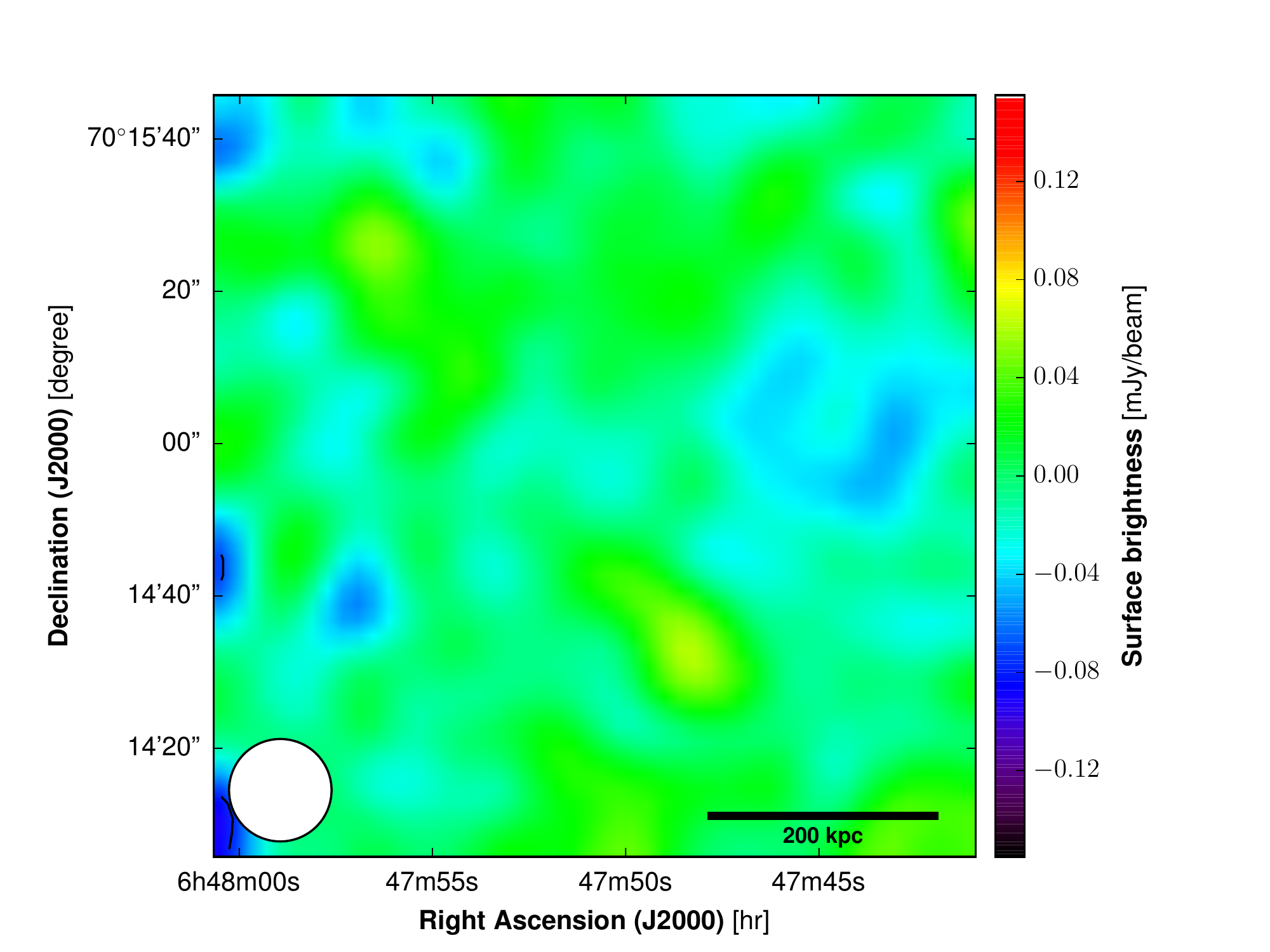}
\includegraphics[height=4.4cm]{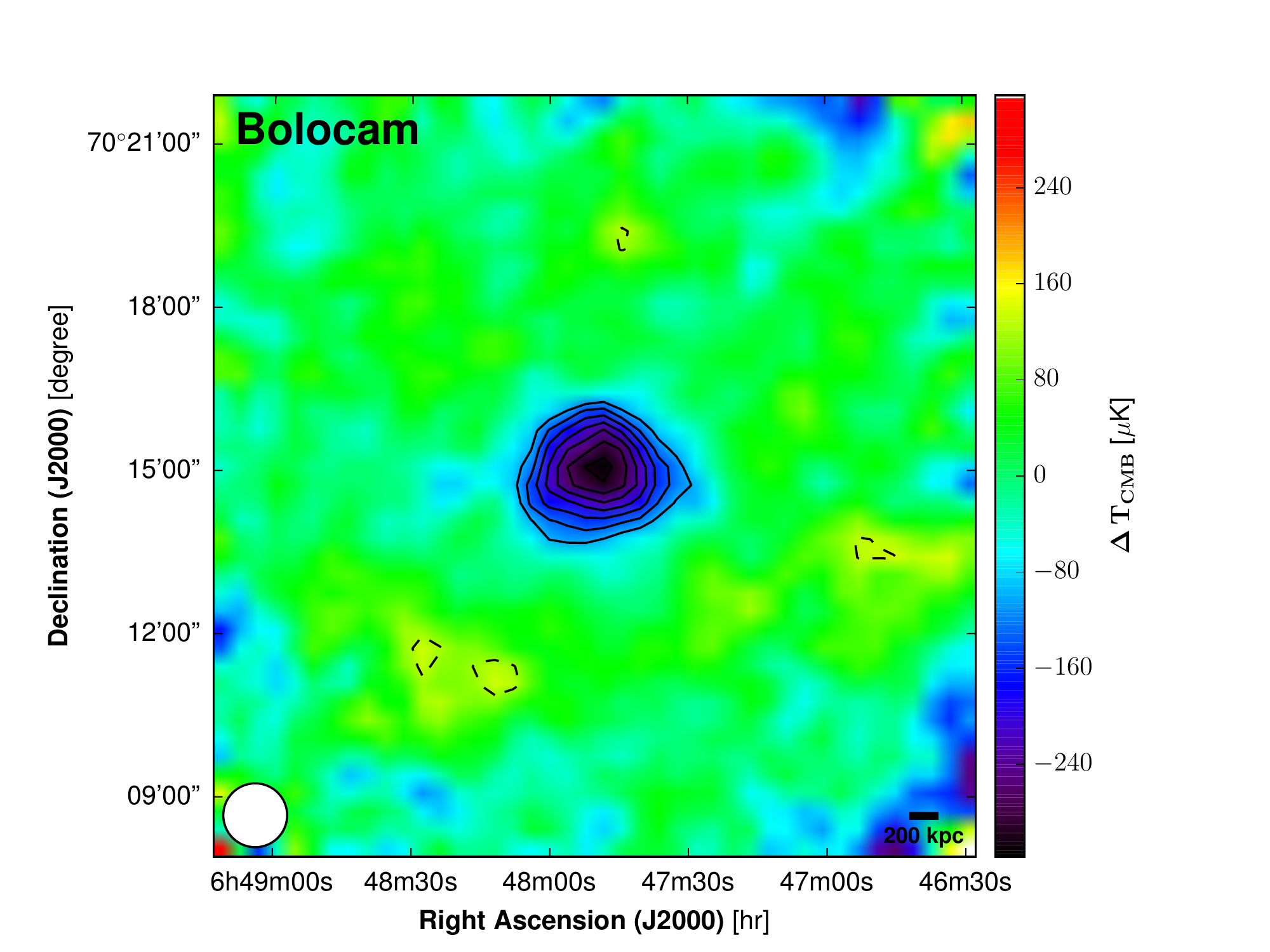}
\includegraphics[height=4.4cm]{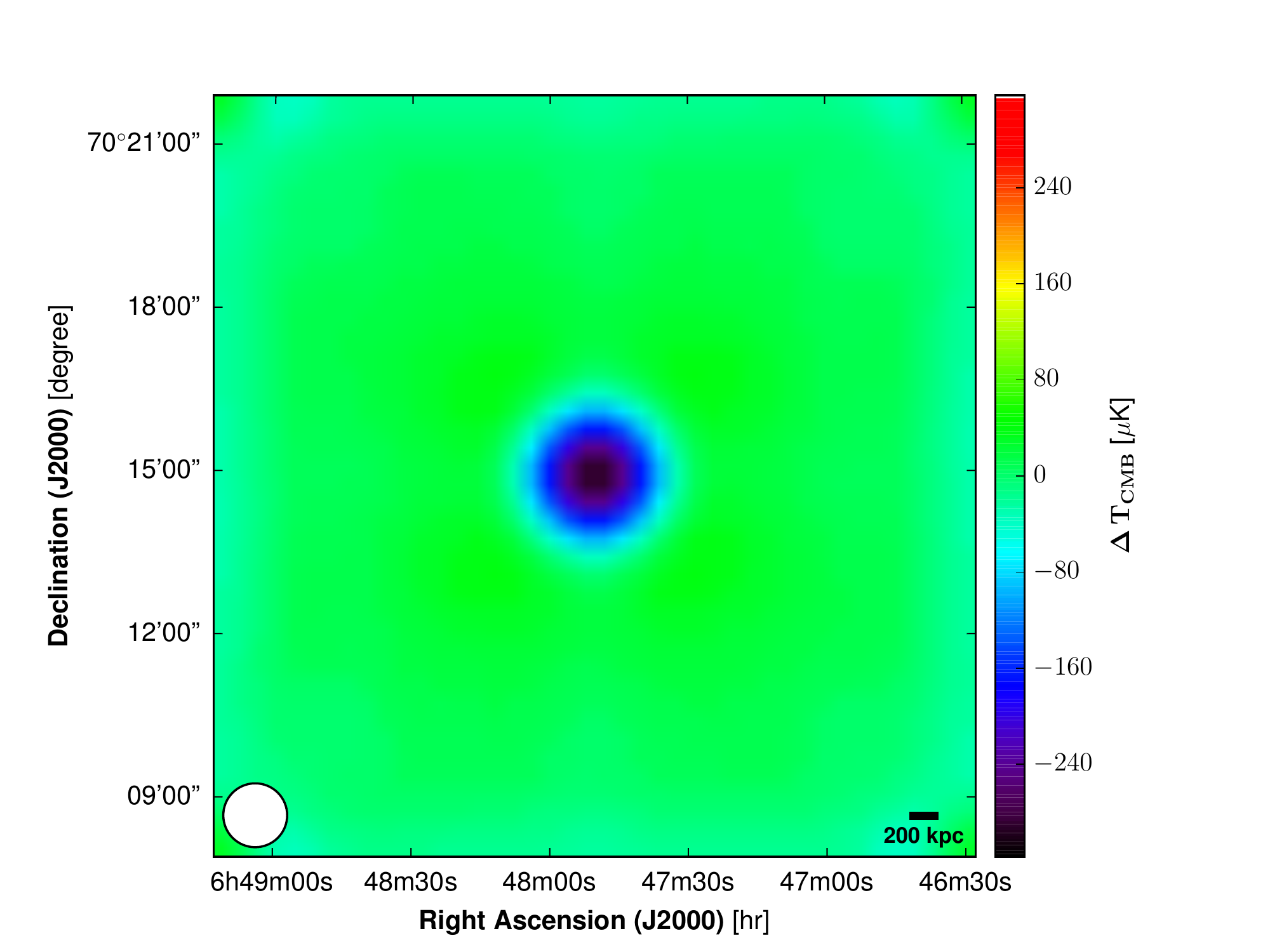}
\includegraphics[height=4.4cm]{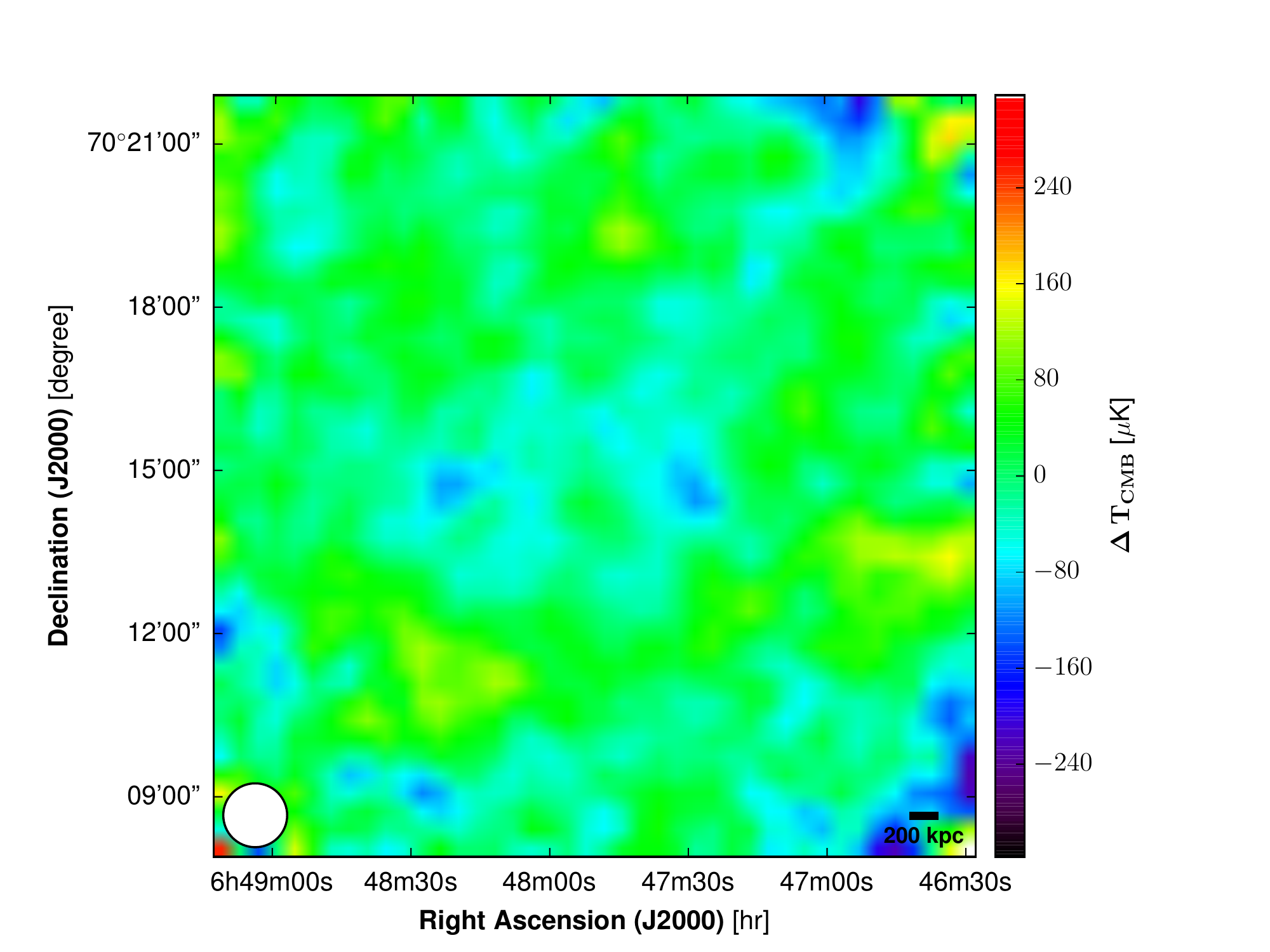}
\includegraphics[height=4.4cm]{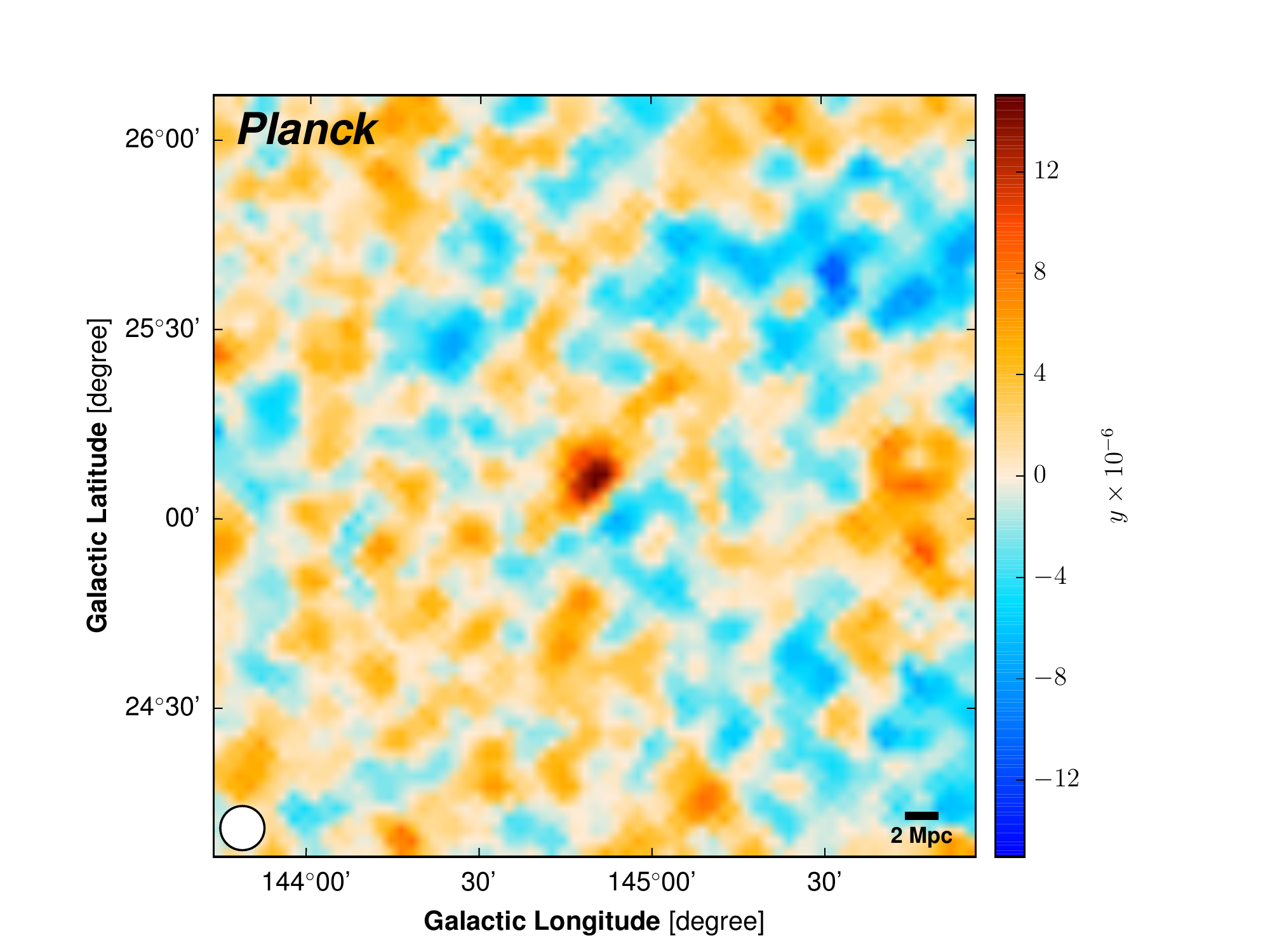}
\includegraphics[height=4.4cm]{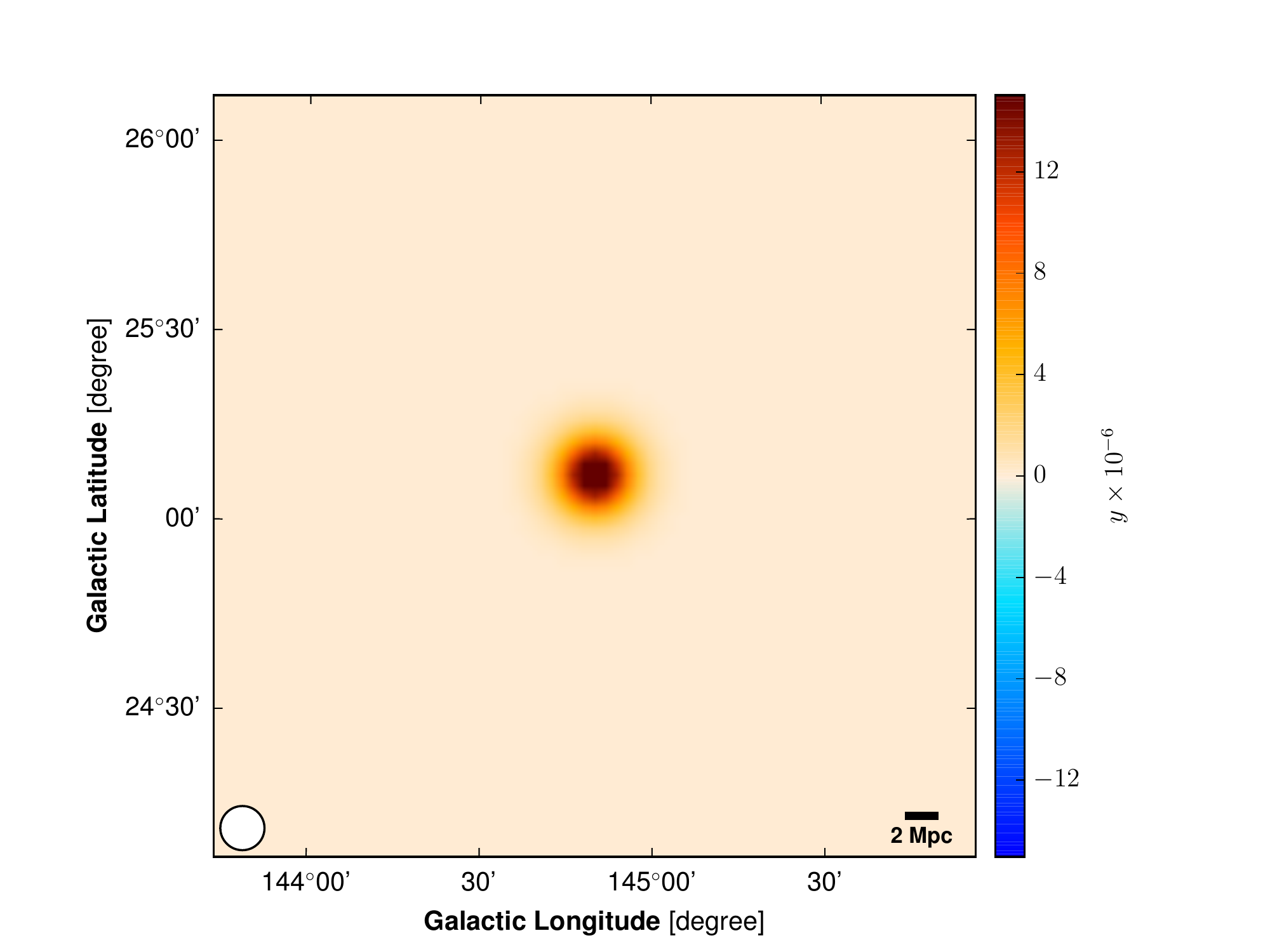}
\includegraphics[height=4.4cm]{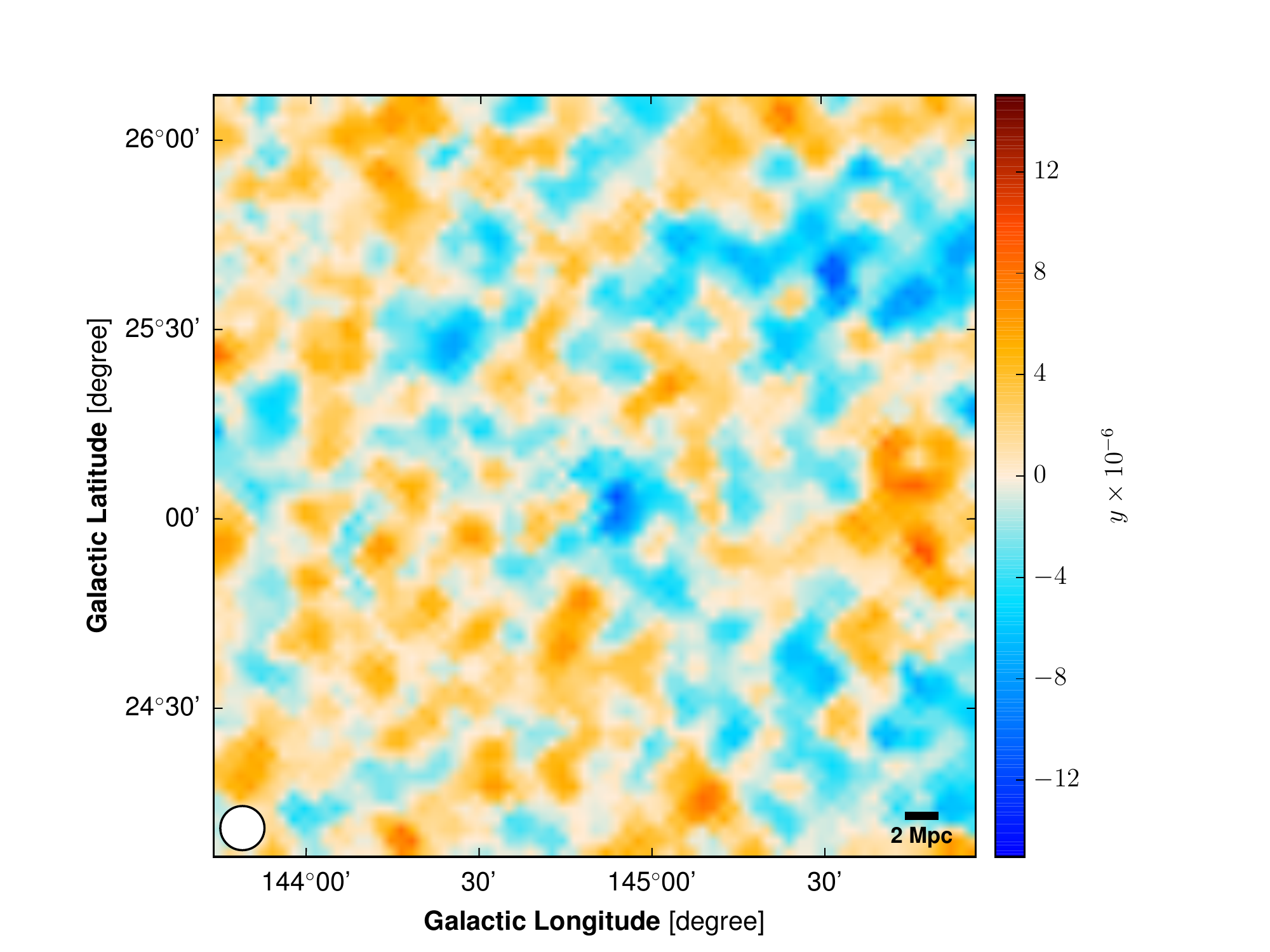}
\caption{{\footnotesize \textbf{From top to bottom:} NIKA2, MUSTANG, Bolocam, and \planck\ tSZ data (left), maximum likelihood tSZ map (middle), and residual (right) computed from a MCMC analysis based on a non-parametric model and masking the over-pressure region in the south-west of \psz. The Bolocam row is shown as a cross-check even if this data is not used for this analysis. The significance contours start at $3\sigma$ with $1\sigma$ increments. A $4\sigma$ tSZ signal is identified in the NIKA2 residual map within the region that is masked for this analysis.}}
\label{fig:NIKA2_Data_fit_residual}
\end{figure*}

\subsubsection{Method}\label{sec:MCMC_method}

We use a dedicated tSZ pipeline to constrain the pressure profile of \psz\ from a multi-instrument analysis combining the tSZ data-sets presented in Sect. \ref{sec:ancillary}. The methodology used in the pipeline is based on a forward fitting Bayesian framework to account for the instrumental beam and analysis transfer function convolution of the true tSZ signal of the cluster. The details of the deprojection procedure can be found in \cite{ada15,ada16a} and are briefly recalled in this section. To present the method, we consider the joint analysis combining the largest data set (MUSTANG, NIKA2, Bolocam, and \planck\ data).\\
The pressure distribution within the ICM of \psz\ is modeled by a non-parametric pressure profile \citep[see][]{rup17} defined by constrained pressure values $P(r_i)$ at given distances from the cluster center, $r_i$, and by a power law interpolation for $r \in [r_i, \, r_{i+1}]$:
\begin{equation}
        P(r) = P(r_i)\times 10^{\alpha_i}~~~\mathrm{with}~~~ \alpha_i = \frac{\mathrm{log_{10}}\left(\frac{P(r_{i+1})}{P(r_i)}\right) \times \mathrm{log_{10}}\left(\frac{r}{r_i}\right)}{\mathrm{log_{10}}\left(\frac{r_{i+1}}{r_i}\right)}
\label{eq:P_profile}
.\end{equation} 
Following the work of \cite{rom17}, the radial positions where the pressure is constrained are based on the instrumental performance in terms of angular resolution and FOV of the considered experiments. The separation between each radial bin is set to the physical distance corresponding to the beam FWHM of the MUSTANG, NIKA2, and Bolocam instruments for the inner, intermediate, and outer part of the profile respectively. We also set one pressure radial bin outside each experiment FOV and the outermost bins are set to 4 and $5R_{500}$ to constrain the total integrated Compton parameter.\\
\indent A MCMC analysis is performed to estimate the best-fit pressure profile of \psz\ using the considered data sets. The sampling of the pressure within each bin is made by using the python package emcee \citep{for13}. At each iteration of the MCMC algorithm, a pressure profile is generated using Eq. \ref{eq:P_profile} and integrated along the line of sight to get a model of the Compton parameter profile of the cluster (see Eq. \ref{eq:y_compton}). The Compton parameter profile is then convolved with the respective beam profile of each instrument to account for the beam smoothing of the signal following the methodology presented in \cite{bir94}. We use the parameters of the double-Gaussian beam defined in \cite{rom15} to smooth the Compton parameter profile to the MUSTANG angular resolution. The NIKA2, Bolocam, and \planck\ Compton parameter profiles are obtained by convolving the initial profile with a 17.7~arcsec, 58~arcsec, and 7~arcmin FWHM Gaussian beam respectively. Each convolved profile is then projected on a pixelized grid identical to the one used for the projection of the data measured by each instrument. We apply the transfer function of MUSTANG, NIKA2, and Bolocam to the computed Compton parameter maps to account for the processing filtering specific to each experiment. We assume the filtering of the tSZ signal to be negligible for the \planck\ data. The filtered Compton parameter maps are converted into tSZ surface brightness maps for MUSTANG and Bolocam by using the conversion coefficients provided along with the MUSTANG and Bolocam data. These coefficients take into account the relativistic corrections by assuming a mean ICM temperature of 11~keV (see Sect. \ref{sec:comparison_X_tSZ}). The coefficient that converts the NIKA2 Compton parameter map into a tSZ surface brightness map is computed at each iteration of the MCMC sampling. It is obtained by integrating the tSZ spectrum within the 150~GHz NIKA2 bandpass and by considering the temperature profile computed from the combination of the current pressure profile and the \xmm\ density profile for the relativistic corrections (see Sect. \ref{sec:X_ray_profiles}).\\
\indent The tSZ map model of each experiment is then compared with the data using the following likelihood function:
\begin{equation}
\begin{tabular}{r@{\hskip 1mm}l}
       $-2 \mathrm{ln} \, \mathscr{L}$  & $ =\chi^2_{\mathrm{MUSTANG}} + \chi^2_{\mathrm{NIKA2}} + \chi^2_{\mathrm{Bolocam}} + \chi^2_{\mathrm{Planck}}$\\[0.2cm]
       & $=\sum\limits_{\substack{i=1}}^{N_{\mathrm{pixels}}^{\mathrm{MUSTANG}}} [(M_{\mathrm{MUSTANG}}^{\mathrm{data}} - M_{\mathrm{MUSTANG}}^{\mathrm{model}})/(M_{\mathrm{MUSTANG}}^{\mathrm{rms}})]_i^2$ \\[0.2cm]
       & $+\sum\limits_{\substack{j=1}}^{N_{\mathrm{pixels}}^{\mathrm{NIKA2}}} [(M_{\mathrm{NIKA2}}^{\mathrm{data}} - M_{\mathrm{NIKA2}}^{\mathrm{model}})^T C_{\mathrm{NIKA2}}^{-1} (M_{\mathrm{NIKA2}}^{\mathrm{data}} - M_{\mathrm{NIKA2}}^{\mathrm{model}})]_j $ \\[0.2cm]
       & $+\sum\limits_{\substack{k=1}}^{N_{\mathrm{pixels}}^{\mathrm{Bolocam}}} [(M_{\mathrm{Bolocam}}^{\mathrm{data}} - M_{\mathrm{Bolocam}}^{\mathrm{model}})/(M_{\mathrm{Bolocam}}^{\mathrm{rms}})]_k^2$ \\[0.2cm]
       & $+ \sum\limits_{\substack{l=1}}^{N_{\mathrm{pixels}}^{\mathrm{Planck}}} [(y_{\mathrm{Planck}} - y_{\mathrm{model}})^T C_{\mathrm{Planck}}^{-1} (y_{\mathrm{Planck}} - y_{\mathrm{model}})]_l,$
\label{eq:chi2_SZ_combined}
\end{tabular}
\end{equation}
where $M^{\mathrm{data}}$ and $M^{\mathrm{model}}$ are the measured and modelized tSZ surface brightness maps for MUSTANG, NIKA2, and Bolocam. The \planck\ Compton parameter map of \psz\ and its model is given by $y_{\mathrm{Planck}}$ and $y_{\mathrm{model}}$ respectively. The map $M^{\mathrm{rms}}$ gives the RMS noise of each pixel in the MUSTANG and Bolocam maps assuming white noise. The pixel-to-pixel correlation induced by residual correlated noise in both the NIKA2 and \planck\ maps is taken into account by using the inverse covariance matrices $C_{\mathrm{NIKA2}}^{-1}$ and $C_{\mathrm{Planck}}^{-1}$ in the likelihood estimation. In the case of a standard analysis of a NIKA2 SZ large program cluster, the contributions of MUSTANG and Bolocam are simply removed from the likelihood function and the pressure profile is deprojected from the NIKA2 and \planck\ data only.\\
\indent The universal pressure profile \citep{arn10} and the AMI total integrated Compton parameter (see Sect. \ref{sec:ancillary}) are used to define the initial pressure value of each chain. A Gaussian random drawing is made around the mean pressure values with a relative standard deviation of 50\%. This enables us to explore the model parameter space to find potential deviations from the universal pressure profile without starting the chains with extreme pressure values. We consider uniform priors to impose the pressure to remain positive for each bin. The MCMC procedure also marginalizes over the zero level of the MUSTANG, NIKA2, and Bolocam maps and the NIKA2 calibration coefficient uncertainty. The convergence of the chains is assessed by using the convergence test proposed by \cite{gel92} for all fitted parameters. Once the convergence is ensured, the final posterior marginalized distributions are given by the remaining chain samples after the burn-in cutoff, which discards the first 50\% of each chain. These distributions are then used to compute the best-fit pressure profile of \psz\ and the error bars of each constrained point.

\subsubsection{Results}\label{sec:MCMC_results}

\noindent\textbf{NIKA2+\textit{Planck}:}\\
We first perform the MCMC analysis by considering only the NIKA2 150~GHz map and the \planck\ data to show the typical constraints that will be obtained on the pressure profile characterization for all the clusters of the NIKA2 SZ large program. We follow the methodology presented in Sect. \ref{sec:MCMC_method} and simply remove the MUSTANG and Bolocam contributions in the likelihood function (Eq. \ref{eq:chi2_SZ_combined}). The position of the first radial bin is set at $40~\rm{kpc}$ from the cluster center, which corresponds to an angular distance of 6~arcsec. This enables us to constrain the pressure within the NIKA2 17.7~arcsec FWHM beam. The angular distance between each bin is set to the NIKA2 FWHM to efficiently sample the pressure profile while minimizing the correlation between each bin. Two radial bins are also set at 3 and $5R_{500}$ to constrain the pressure profile outer slope. We take into account the thermal pressure excess identified in the south-west of \psz\ by masking this region in the NIKA2 150~GHz map (see Sect. \ref{sec:overp}). The \planck\ map also contains all the tSZ information but the uncertainty on $Y_{\mathrm{5R500}}$ (see Sect. \ref{sec:ancillary}) is much larger than the contribution induced by the over-pressure region on the integrated Compton parameter (see Sect. \ref{sec:Impact_mass}). Thus, we neglect the over-pressure contribution in the \planck\ $y$-map.\\
\indent The best-fit non-parametric pressure profile obtained from this analysis is shown in the left panel of Fig. \ref{fig:Nonparam_profiles}. The tight constraints on the ICM pressure distribution from $\rm{R} \sim 0.03 \rm{R_{500}}$ to $\rm{R} \sim \rm{R_{500}}$ are mostly due to the NIKA2 150~GHz map. The \planck\ data enable us to constrain the pressure profile outer slope. This shows the potential of the NIKA2 SZ large program to estimate the ICM pressure distribution of high-redshift clusters from their core to their outskirts ($\rm{R} \sim 3\rm{R_{500}}$). The best-fit non-parametric pressure profile is fitted by a parametric generalized Navarro-Frenk-White model \citep[gNFW, ][]{nag07},
\begin{figure*}[h!]
\centering
\includegraphics[height=6.8cm]{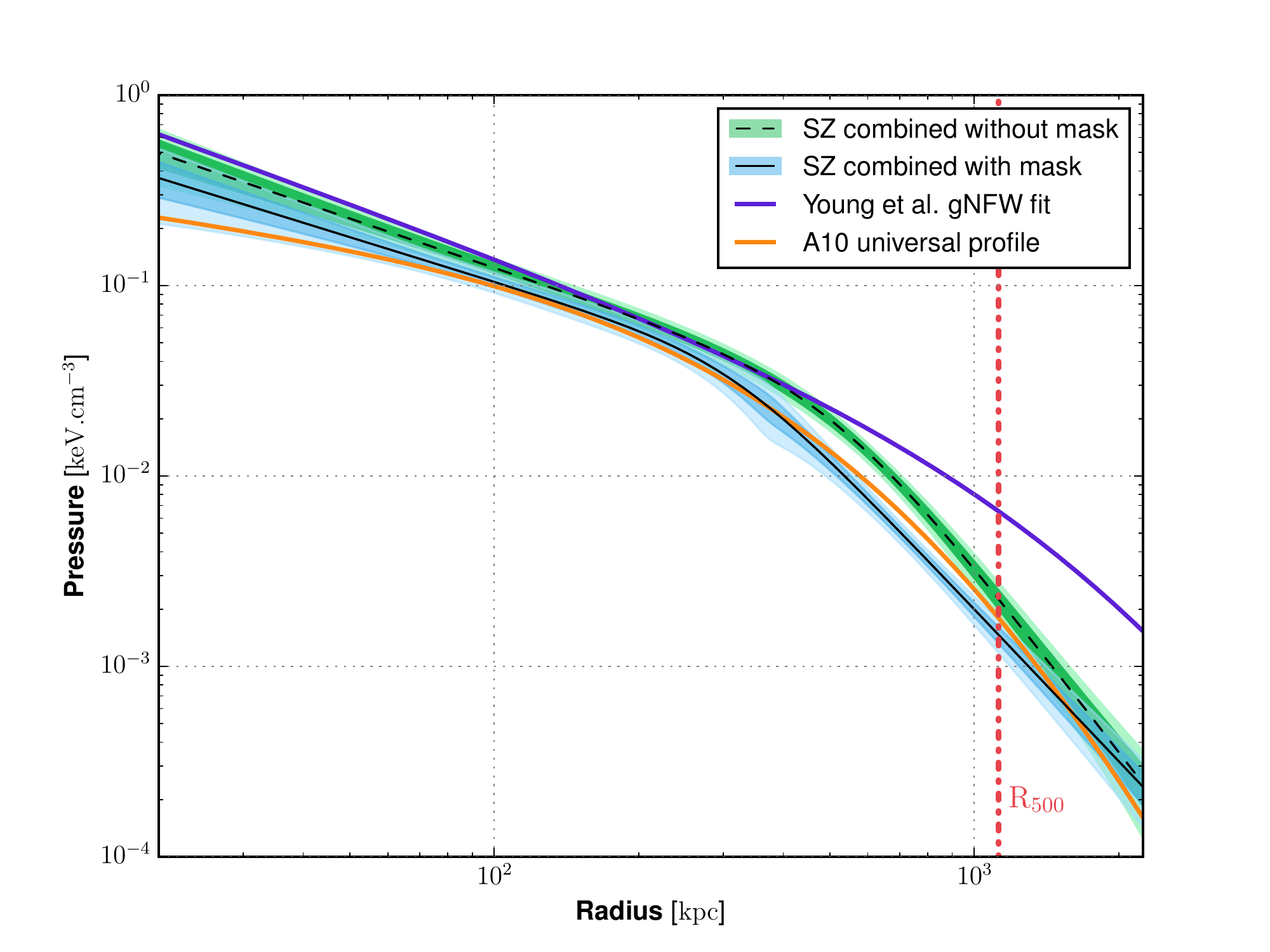}
\includegraphics[height=6.8cm]{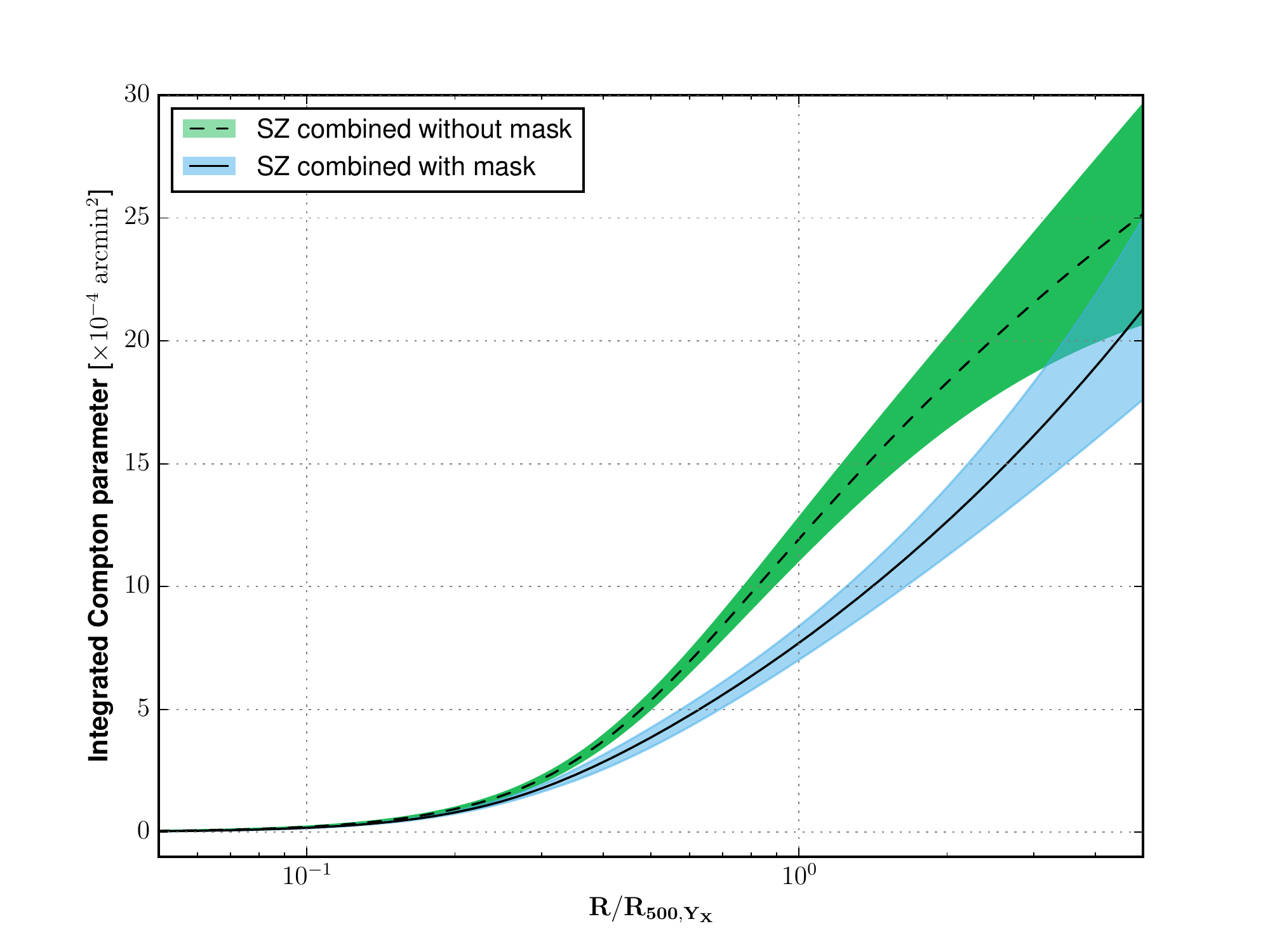}
\caption{{\footnotesize \textbf{Left:} Comparison of the gNFW pressure profiles fitted on the pressure points constrained from a non-parametric deprojection of the tSZ data with (blue) and without (green) masking the over-pressure region. The Bolocam data are used along with the MUSTANG, NIKA2, and \planck\ data for the estimation of the green profile. In this case, the pressure is significantly higher than the constraints obtained masking the over-pressure between 500 and 900~kpc because of the excess of thermal pressure within the south-west region of \psz. The pressure profile from \cite{you15} as well as the universal pressure profile from \cite{arn10} (A10) are also shown in magenta and orange respectively. The red dashed line shows an estimate of $\rm{R_{500}}$ computed using the A10 profile, the AMI integrated Compton parameter, and the scaling relation used in \cite{pla14}. \textbf{Right:} Profiles of the spherically integrated Compton parameter computed from the pressure profiles obtained with (blue) and without (green) masking the over-pressure region. The radius has been normalized by $\rm{R_{500,Y_X}}$ constrained by \xmm.}}
\label{fig:compare_prof}
\end{figure*}
\begin{equation}
        P_e(r) = \frac{P_0}{\left(\frac{r}{r_p}\right)^c \left(1+\left(\frac{r}{r_p}\right)^a\right)^{\frac{b-c}{a}}},
\label{eq:gNFW}
\end{equation}
where $P_0$ is a normalization constant, $r_p$ is a characteristic radius, $b$ and $c$ give the slope of the profile at large and small radii respectively, and $a$ characterizes the transition between the two slopes. Following \cite{rom17}, we take into account the correlation between the pressure bins by computing the covariance matrix of the pressure bins using the final Markov chains of the non-parametric fit. The best-fit gNFW profile (black line in the left panel of Fig. \ref{fig:Nonparam_profiles}) is compared with the universal pressure profile of \cite{arn10} (A10) computed by considering the \xmm\ estimates of both $\rm{R_{500}}$ and $\rm{M_{500}}$ (see Sect. \ref{sec:comparison_X_tSZ}). While the overall normalization is consistent between the two gNFW models, we identify significant differences between their slope parameters. The difference between the outer slopes of the deprojected profile and the A10 profile is entirely consistent with the large range of outer slopes found in \cite{pla13b,say13}. While no conclusion can be drawn for a single cluster analysis, this highlights that the NIKA2 SZ large program will be able to identify a potential redshift evolution of the universal pressure profile parameters and constrain the scatter in the outer profile slopes at high redshift, for which no current constraints exist. Furthermore, we estimate the cluster integrated Compton parameter from the integration of the best-fit gNFW profile and find that $\rm{Y_{500}} = (8.12 \pm 0.60) \mathrm{\times 10^{-4}~arcmin^2}$. Such a relative uncertainty of $\sim$7\% on the estimate of $\rm{Y_{500}}$ is the typical constraint that will be obtained for each cluster of the NIKA2 SZ large program.\\

\indent For this cluster, as both MUSTANG and Bolocam tSZ data are available, we choose to add this additional information to increase our constraining power on the characterization of the ICM pressure profile. We therefore optimize the pressure point positions for this combination of instruments using the method described in Sect. \ref{sec:MCMC_pressure}.\\

\noindent\textbf{MUSTANG+NIKA2+\textit{Planck}:}\\
\noindent We mask the region of the NIKA2 map defined in Sect. \ref{sec:overp} when constraining the thermal pressure distribution of the cluster to exclude the over-pressure region. As the Bolocam angular resolution is too low to spatially resolve the over-pressure region, we decided to perform the MCMC analysis by combining only the MUSTANG, NIKA2, and \planck\ data. 

The best-fit tSZ map models and residuals obtained at the end of the MCMC analysis are shown in Fig. \ref{fig:NIKA2_Data_fit_residual}. We observe a $4\sigma$ tSZ residual in the south-west region of \psz\ in the NIKA2 map after subtracting the best-fit tSZ surface brightness map model. The corresponding Bolocam tSZ map model and residual are shown as a cross check although Bolocam data are not used for this analysis. A negative surface brightness is also observed in the Bolocam residual at the location of the tSZ excess in the NIKA2 residual map although it is not significant. We compute the integrated Compton parameter within the mask defined in Sect. \ref{sec:overp} in the NIKA2 map residual and find that $\rm{Y_{residual}} = (1.2 \pm 0.4) \times 10^{-4}~\mathrm{arcmin}^2$. The uncertainty on this estimate is computed by performing the integration in the same region of noise map realizations and by computing the standard deviation of the estimated values. The residual integrated Compton parameter is therefore four times smaller than the uncertainty on the cluster integrated Compton parameter given by {\planck}/AMI (see Sect. \ref{sec:ancillary}). 

The best-fit non-parametric pressure profile constrained within this framework is shown in the right panel of Fig. \ref{fig:Nonparam_profiles}. The pressure distribution is constrained from the cluster center ($\rm{R} \sim 0.02 \rm{R_{500}}$) to its outskirts ($\rm{R} \sim 3 \rm{R_{500}}$). The tightest constraints are obtained in the intermediate region of the profile and are mostly driven by the NIKA2 data. The inner points are mostly constrained by the MUSTANG data. The tight constraints on the pressure distribution given by the non-parametric analysis of the tSZ data enable us to characterize precisely the parameters of the best-fit gNFW profile shown in black. The information brought by the MUSTANG data improves on the constraints of the inner slope of the profile, which is less steep than the one obtained by the analysis of the NIKA2 and \planck\ data only. The best-fit gNFW profiles obtained by both analyses are still fully compatible within their $1\sigma$ uncertainties.\\

\noindent\textbf{MUSTANG+NIKA2+Bolocam+\textit{Planck}:}\\
\noindent We then perform the MCMC analysis by considering all the tSZ experiments without masking the over-pressure region in the NIKA2 map. The best-fit gNFW profile of the pressure points contrained by MUSTANG, NIKA2, Bolocam, and \planck\ is shown in Fig. \ref{fig:compare_prof} left panel along with its 1 and $2\sigma$ uncertainties. It is compared with the best-fit gNFW profile obtained from the MCMC analysis performed by combining the MUSTANG, NIKA2, and \planck\ data and by masking the over-pressure region (blue profile). We also perform the MCMC analysis without masking the over-pressure region by using only the MUSTANG, NIKA2, and \planck\ data and check that the best-fit pressure profile is not modified by the addition of the Bolocam data in this analysis. However, the additional information brought by the Bolocam tSZ map enables us to lower the error bars of the pressure points in the cluster outskirt by about 30\%.
\begin{figure*}[h!]
\centering
\includegraphics[height=6.8cm]{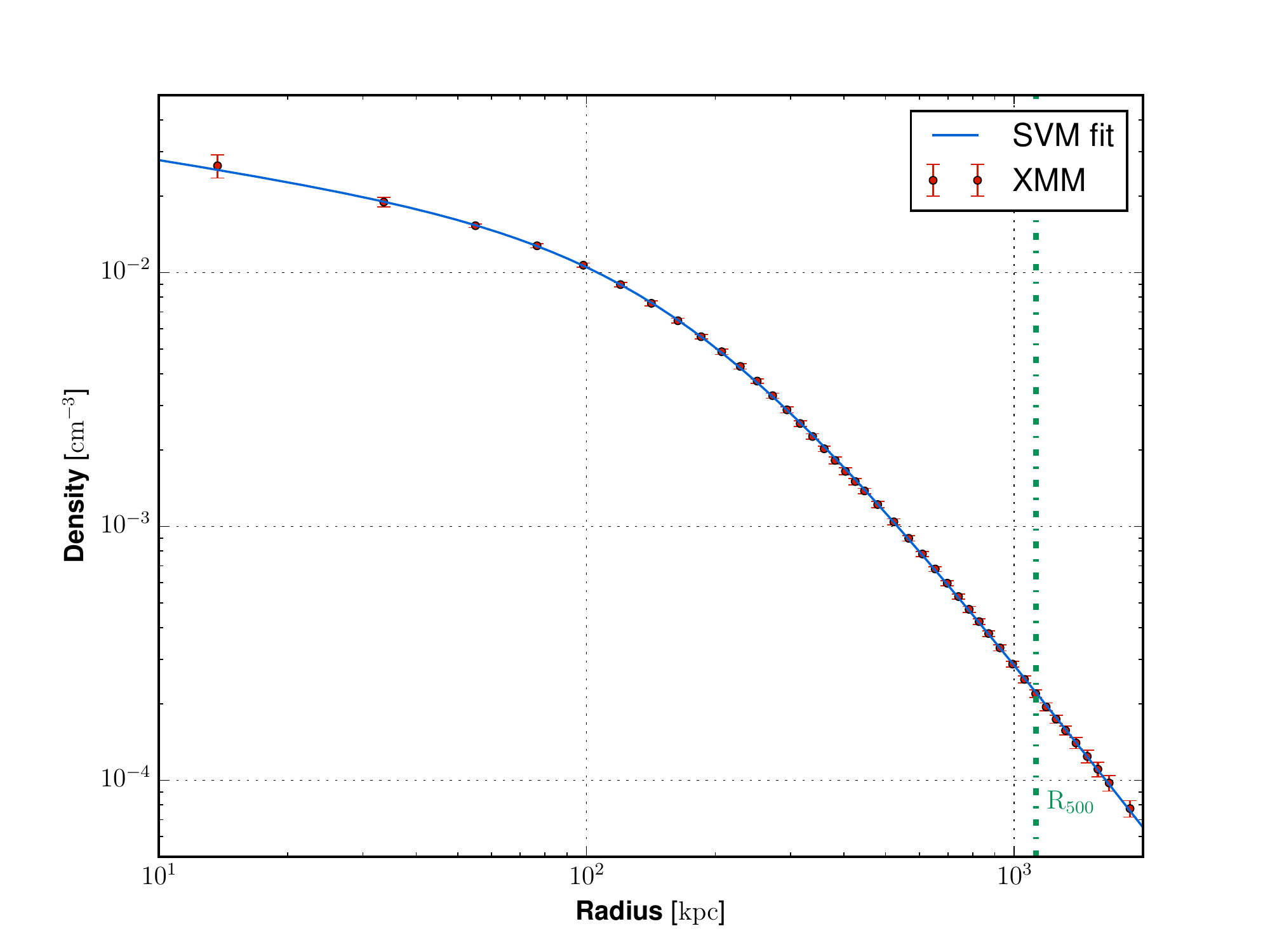}
\includegraphics[height=6.8cm]{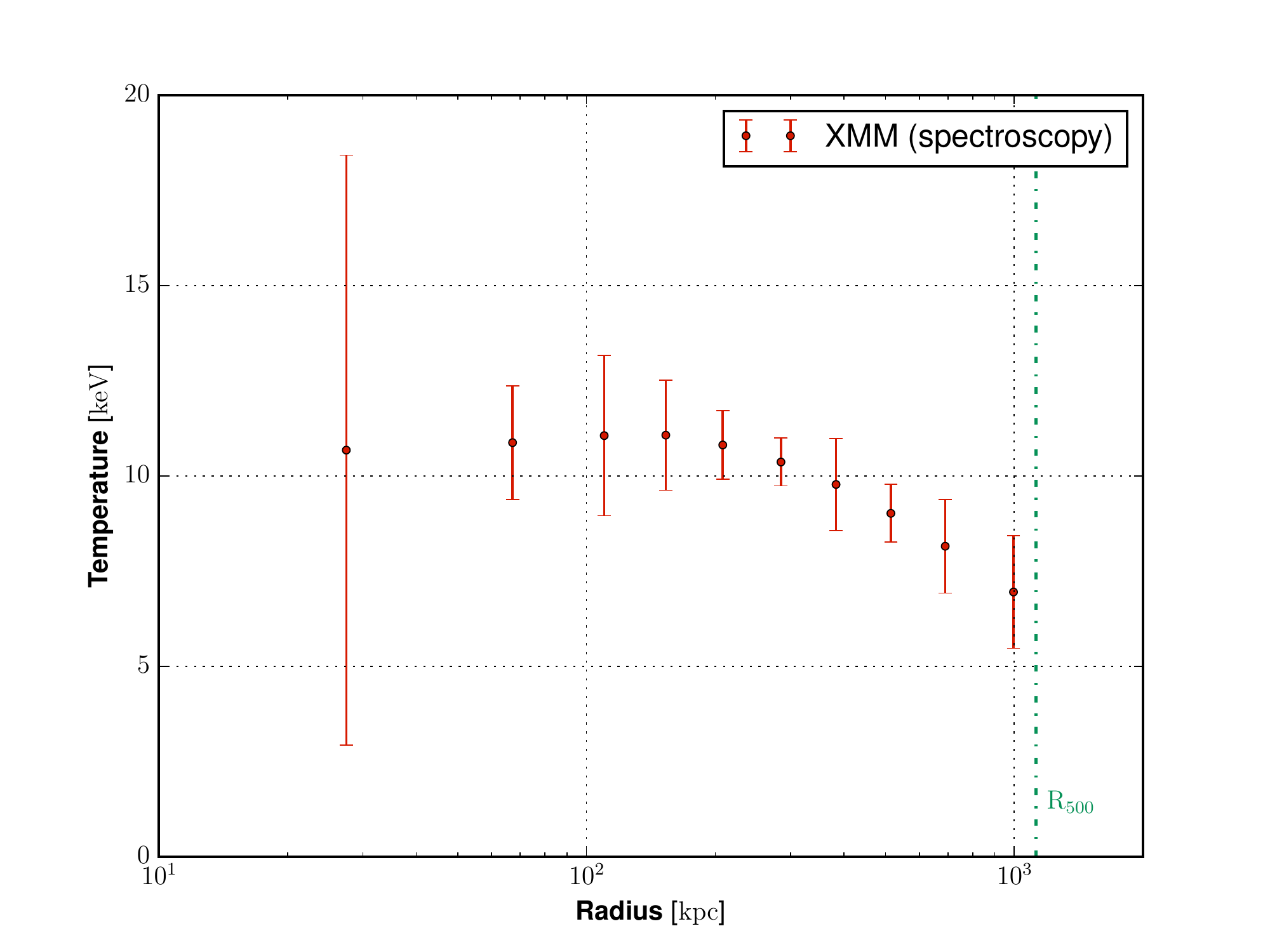}
\caption{{\footnotesize \textbf{Left:} Electronic density profile constrained from the \xmm\ data (red dots). The best-fit SVM model \citep{vik06} is shown as a blue line. \textbf{Right:} \xmm\ deprojected temperature profile estimated from the X-ray spectroscopy data. The characteristic radius obtained by \xmm, $\rm{R_{500}} = 1124 \pm 12~\mathrm{kpc}$ is measured from the $\rm{Y_X} - \rm{M_{500}}$ relation of \cite{arn10} and is represented by a dashed vertical green line.}}
\label{fig:XMM_profiles}
\end{figure*}

\subsection{X-ray radial thermodynamic profiles}\label{sec:X_ray_profiles}
The ICM thermodynamic radial profiles are estimated with and wihtout masking the south-west region of the cluster in the \xmm\ event files to be able to combine the X-ray density profile with the pressure profile obtained by masking the over-pressure region in the tSZ data. Gas density and temperature profiles were extracted from the \xmm\ observations following the procedures described in detail in \cite{pra10} and \cite{plaint13} and are shown in Fig. \ref{fig:XMM_profiles}. The gas pressure and entropy radial profiles were computed directly from the deprojected density and temperature profiles and are shown in Fig. \ref{fig:ICM_profiles}. The hydrostatic mass profile, derived assuming a spherical gas distribution in hydrostatic equilibrium, was calculated using the Monte Carlo procedure described in \cite{dem10}, and references therein.\\
\indent The density profile, obtained using the regularized deprojection and PSF-correction procedure described in \cite{cro06}, exhibits structure that is amplified in the radial derivative. We thus use the simplified Vikhlinin parametric model (SVM) \citep{vik06} to fit the \xmm\ deprojected density profile:
\begin{equation}
        n_e(r) = n_{e0} \left[1+\left(\frac{r}{r_c}\right)^2 \right]^{-3 \beta /2} \left[ 1+\left(\frac{r}{r_s}\right)^{\gamma} \right]^{-\epsilon/2 \gamma},
\label{eq:SVM}
\end{equation}
where $n_{e0}$ is the cluster central density, $r_s$ is the transition radius at which an additional steepening in the profile occurs, and $r_c$ is the core radius. The $\beta$ parameter gives the inner profile slope and $\epsilon$ the outer one. The $\gamma$ parameter gives the width of the transition in the profile. We fix the value of the $\gamma$ parameter at three since it provides a good description of all clusters considered in the analysis of \cite{vik06b}. The best-fit SVM profile of \psz\ is shown along with the \xmm\ density profile points in the left panel of Fig. \ref{fig:XMM_profiles}. The agreement between the parametric model and the deprojected profile is excellent. We therefore use the best-fit SVM profile in combination with the pressure profile constrained by the tSZ analysis to estimate the radial thermodynamic profiles of \psz\ (see Sect. \ref{sec:comparison_X_tSZ}).

\subsection{Thermodynamics of the cluster}\label{sec:comparison_X_tSZ}

In this section, we undertake a hydrostatic mass analysis to recover the thermodynamic properties of \psz\ obtained by the combination of the \xmm\ density profile and the pressure profile constrained by the tSZ analysis described in Sect. \ref{sec:MCMC_pressure}.\\
\begin{figure*}[h!]
\centering
\includegraphics[height=6.8cm]{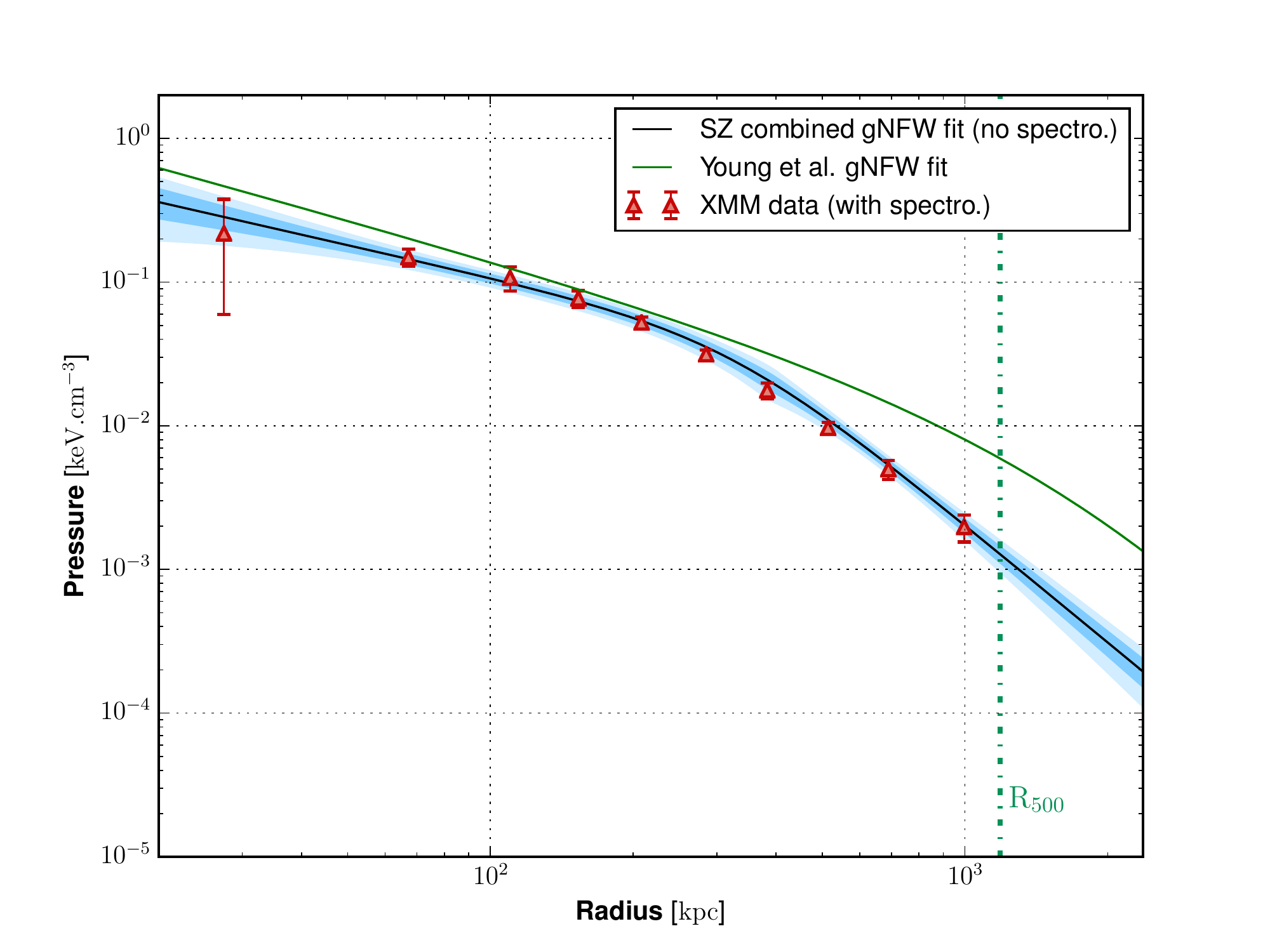}
\includegraphics[height=6.8cm]{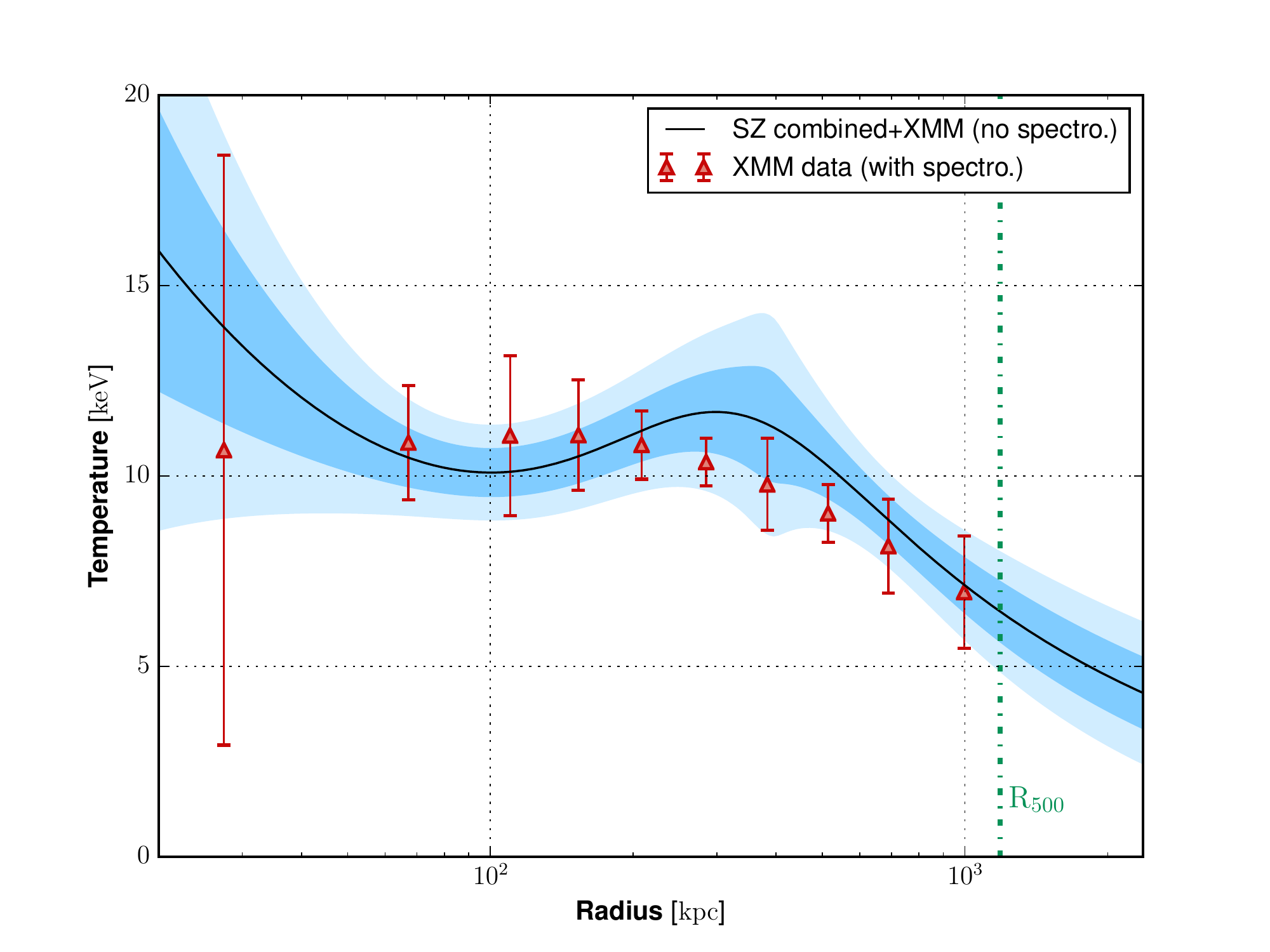}
\includegraphics[height=6.8cm]{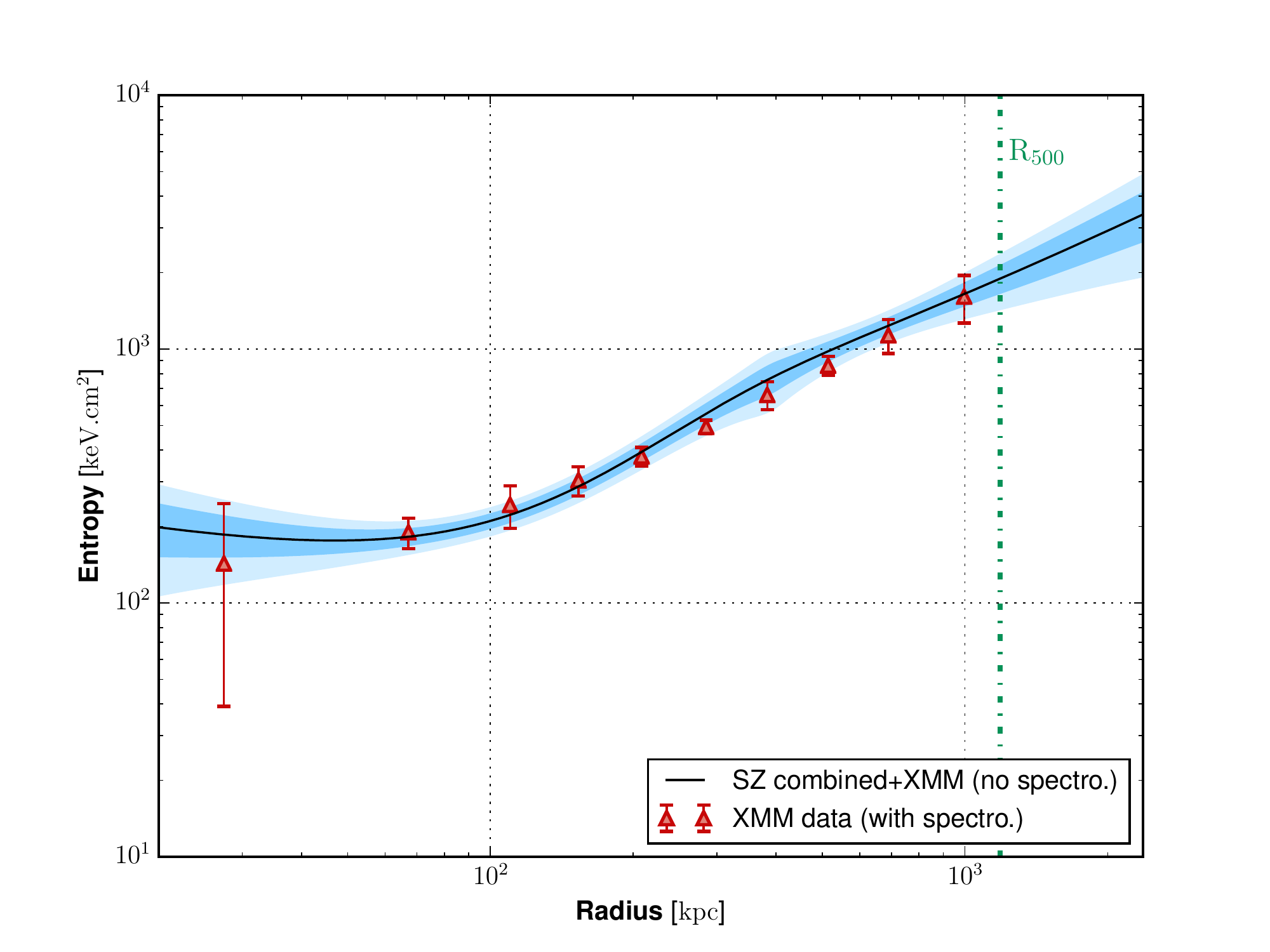}
\includegraphics[height=6.8cm]{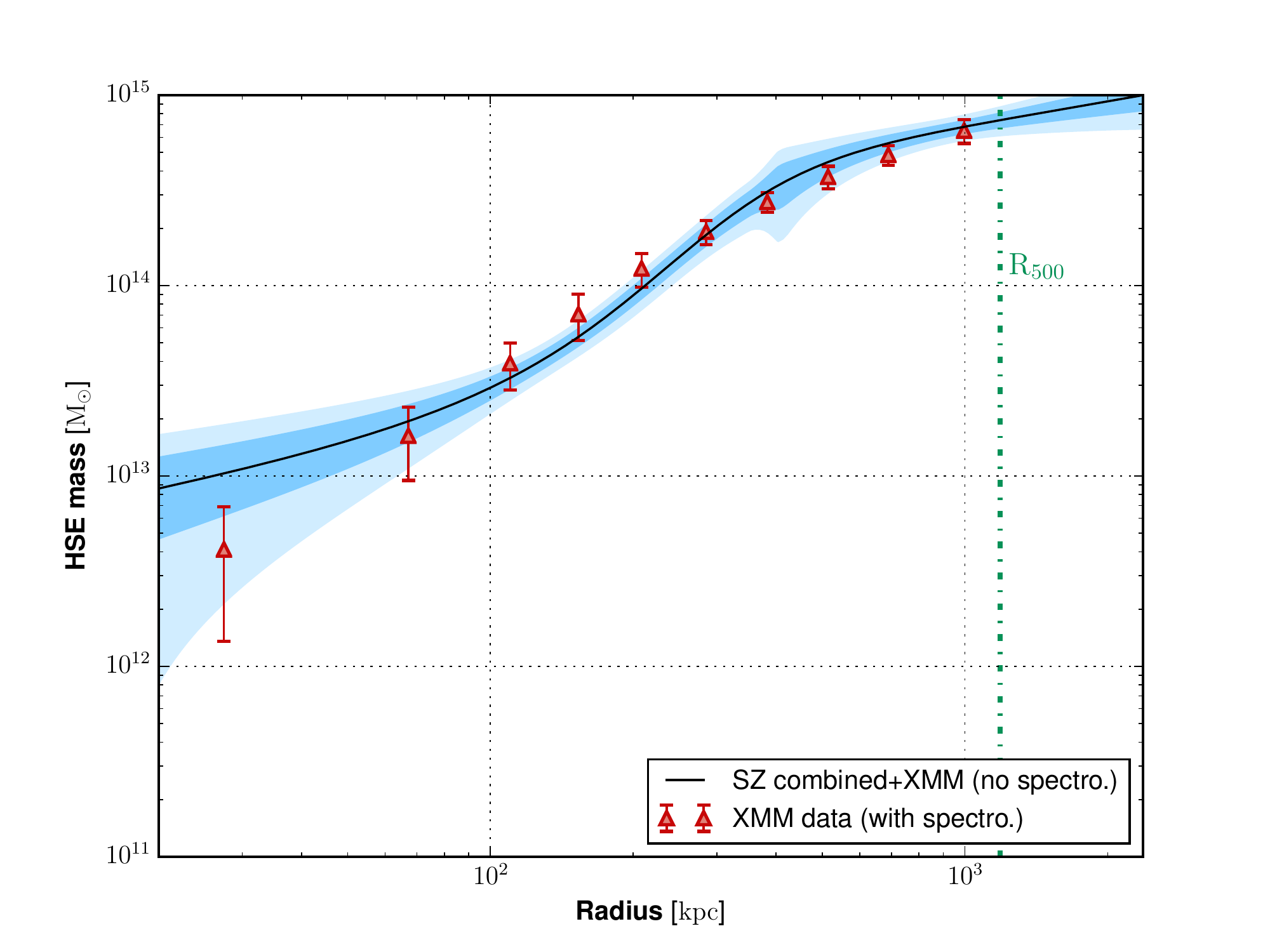}
\caption{{\footnotesize Constraints on the deprojected radial profiles of the pressure (top left), temperature (top right), entropy (bottom left), and hydrostatic mass (bottom right) obtained from the MCMC analysis masking the over-pressure region. The tSZ pressure profile corresponds to the best-fit gNFW model of the non-parametric pressure profile constrained from the analysis masking the over-pressure region. The pressure profile from \cite{you15} is also shown in green. The XMM-{\it Newton} measurements are indicated with red triangles. The dark and light blue regions show the 68\% and 95\% confidence region on the ICM thermodynamic radial profiles estimated with this multi-probe analysis. The best fit is indicated by the black line.}}
\label{fig:ICM_profiles}
\end{figure*}

\indent As the whole ICM cannot be considered in hydrostatic equilibrium (see Sect. \ref{sec:overp}), we use the non-parametric pressure profile constrained by the MCMC analysis that combines the MUSTANG, NIKA2, and \planck\ data masking the over-pressure region along with its best-fit gNFW model to describe the pressure distribution within the main halo. The pressure points obtained from the tSZ data (see the right panel in Fig. \ref{fig:Nonparam_profiles}) are compared with the \xmm\ deprojected pressure profile (red points). The points are compatible in the cluster center and around $\rm{R_{500}}$ but we notice a slight difference between the pressure estimates between $0.3 \rm{R_{500}}$ and $0.6 \rm{R_{500}}$. Such a discrepancy could be the result of projection effects induced by a deviation of the gas distribution from the spherical symmetry along the line of sight. The two-dimensional analysis of the ICM combining the NIKA2 and \xmm\ maps of \psz\ assuming a triaxial morphology is beyond the scope of this study and will be developed in a forthcoming paper.\\
\begin{table*}[h!]
\begin{center}
\begin{tabular}{ccc}
\hline
\hline
 & \textbf{XMM} & \textbf{tSZ+XMM} \\
 & (spectro.) & (no spectro.) \\
\hline
$\rm{R_{500}}$ [kpc] & $1124 \pm 11$ & $1107 \pm 30$\\
$\rm{M_{500}}~[\mathrm{\times 10^{14}~M_{\odot}}]$ & $7.64 \pm 0.24$ & $6.95 \pm 0.56$\\
\hline
\hline
\end{tabular}
\end{center}
\caption{{\footnotesize Comparison between \psz\ $\rm{R_{500}}$ radius and total mass ($\mathrm{M_{500}}$) obtained from the \xmm\ analysis based on X-ray spectroscopy data and the constraints from the tSZ/X-ray multi-probe analysis masking the over-pressure region.}}
\label{tab:compare_xmm_nika2}
\end{table*}
\indent The gNFW profile fitted on the non-parametric tSZ pressure profile is shown in black with its 1 and $2\sigma$ uncertainties in blue in the top left panel of Fig. \ref{fig:ICM_profiles}. We compare this estimate of the pressure profile with the one obtained by \cite{you15} (Y15 in the following) using a combination of MUSTANG and Bolocam data (see green profile in the top left panel of Fig. \ref{fig:ICM_profiles}). As the thermal pressure excess in the south-west region of \psz\ could not be identified either with the Bolocam or the MUSTANG data, the pressure distribution given by the Y15 profile is over-estimated in the inner and intermediate regions of the cluster. It is therefore compatible with the estimate that we obtain from the MCMC analysis without masking the over-pressure region from 0 to about 700~kpc (see green profile in Fig. \ref{fig:compare_prof}), but it is significantly different from the pressure profile constrained by masking the south-west region of the cluster. Furthermore, as the \planck\ data are not used by \cite{you15} to constrain the outer slope of the profile, the latter is much shallower than our estimate. This effect has already been identified by \cite{say16} who showed that the pressure profiles obtained by the analysis of the Bolocam data in \cite{say13} are significantly shallower than the one constrained by adding the \planck\ information to the analysis although the considered cluster samples are nearly identical. The comparison between the Y15 pressure profile and our estimate shown in the top left panel of Fig. \ref{fig:ICM_profiles} emphasizes the importance of high angular resolution tSZ observations carried out with a large FOV instrument to understand the impact of ICM perturbations on the estimate of the pressure profile of galaxy clusters.\\
\indent The best-fit gNFW pressure profile shown in Fig. \ref{fig:ICM_profiles} is combined with the best-fit SVM density profile shown in the left panel of Fig. \ref{fig:XMM_profiles} to estimate the cluster thermodynamic properties without relying on X-ray spectroscopy. Considering the ICM as an ideal gas, we compute both the temperature and the entropy radial profiles of \psz\ as 
\begin{equation}
k_B T_e(r) = \frac{P_e(r)}{n_e(r)}~~~\rm{and}~~~K(r) =  \frac{P_e(r)}{n_e(r)^{5/3}},
\label{eq:temp_entro}
\end{equation}
where $k_B$ is the Boltzmann constant. Assuming the hydrostatic equilibrium, we also compute the total mass enclosed within the radius $r$ given by
\begin{equation}
M_{\rm HSE}(r) = -\frac{r^2}{\mu_{gas} m_p n_e(r) G} \frac{dP_e(r)}{dr},
\label{eq:hse_mass}
\end{equation}
where $\mu_{\rm{gas}} = 0.61$ is the mean molecular weight of the gas, $G$ the Newton constant, and $m_p$ the proton mass.\\
\indent The computed temperature profile is shown in black in the top right panel of Fig. \ref{fig:ICM_profiles}. It is compatible with the deprojected profile given by \xmm\ using X-ray spectroscopy data (red points). Both estimates describe a rather constant ICM temperature of about 11~keV from the cluster center to a radial distance of 400~kpc. The error bars of both profiles are too large in the cluster core to constrain the temperature evolution below 40~kpc although a significant decrease of the temperature toward the center seems to be excluded by our estimate. This seems to indicate that \psz\ has a disturbed core. The mass-weighted temperature provided by the combination of the pressure profile constrained with the tSZ data set and the density profile of \xmm\ is slightly higher than the X-ray spectroscopic temperature between 200 and 600~kpc although both estimates are consistent within the $1\sigma$ uncertainties. This could be a hint for the presence of gas clumps or a deviation from spherical symmetry of the ICM along the line of sight. The temperature decreases for radii larger than 400~kpc with a consistent trend between the X-ray spectroscopic and the mass-weighted profile. The relative uncertainties on our estimate of the temperature profile are comparable with the ones that affect the \xmm\ estimate. Constraining the ICM temperature profile without relying on X-ray spectroscopy is of key importance to study the ICM thermodynamics of high-redshift galaxy clusters. This result shows the potential of joint tSZ/X-ray analyses to estimate the ICM temperature of distant clusters with a reasonable amount of telescope time.\\
\indent The entropy profile computed from the combination of the pressure profile constrained by the tSZ analysis and the \xmm\ density profile is shown in black in the bottom left panel of Fig. \ref{fig:ICM_profiles}. Both our estimate and the \xmm\ one are consistent within error bars. The radial evolution of the entropy is well-described by a power law for radii larger than 100~kpc. However, we observe a flattening of the entropy profile in the cluster core with a constant central entropy value of $(200 \pm 30)~\mathrm{keV.cm^2}$ for radii lower than 100~kpc. This characteristic of the entropy profile is known to be a clear indicator of a morphologically disturbed cluster core \citep[see, e.g.][]{pra10}. This strengthens the previous conclusion that \psz\ has a disturbed core.\\
\indent The hydrostatic mass profile computed by using Equation \ref{eq:hse_mass} is shown in black in the bottom right panel of Fig. \ref{fig:ICM_profiles}. As the pressure profile constrained by the tSZ analysis is steeper than the one estimated by \xmm\ in the core region, our estimate of the mass profile shows a higher mass enclosed within the central part of \psz. Both estimates are, however, compatible within their $1\sigma$ error bars. We use the hydrostatic mass profile of \psz\ to compute its density contrast profile $\langle \, \rho(r) \, \rangle/\rho_c$, where $\rho_c$ is the critical density of the universe at the cluster redshift, and we estimate the value of $\rm{R_{500}}$. The constraints on $\rm{R_{500}}$ and $\rm{M_{500}}$ computed from the tSZ MCMC analysis masking the over-pressure region of \psz\ are compatible with the ones obtained using the \xmm\ data and excluding the same region of the cluster (see Table \ref{tab:compare_xmm_nika2}).
\begin{table*}[h]
\begin{center}
\begin{tabular}{ccc}
\hline
\hline
 & \textbf{With mask} & \textbf{Without mask} \\
\hline
$\rm{R_{500}}$ [kpc] & $1107 \pm 30$ & $1342 \pm 61$\\
$\rm{Y_{500}}~[\mathrm{\times 10^{-4}~arcmin^2}]$ & $8.06 \pm 0.46$ & $13.31 \pm 0.85$ \\
$\rm{M_{500}}~[\mathrm{\times 10^{14}~M_{\odot}}]$ & $6.95 \pm 0.56$ & $12.42 \pm 1.43$ \\
\hline
\hline
\end{tabular}
\end{center}
\caption{{\footnotesize Estimations of \psz\ $\rm{R_{500}}$ radius, integrated Compton parameter ($\mathrm{Y_{500}}$), and total mass ($\mathrm{M_{500}}$) from the multi-instrument MCMC analysis with and without masking the over-pressure region. The estimates of $\mathrm{M_{500}}$ are computed assuming hydrostatic equilibrium and using the combination of the \xmm\ density profile and the tSZ pressure profiles shown in the right panel of in Fig. \ref{fig:Nonparam_profiles}.}}
\label{tab:Cluster_global}
\end{table*}\\
We emphasize the fact that the constraints obtained from the joint tSZ/X-ray analysis on the radial thermodynamic profiles of \psz\ are almost as stringent as the ones computed from the X-ray only analysis that uses X-ray spectroscopic data.

\subsection{Impact of the over-pressure region on the pressure profile characterization}\label{sec:Impact_mass}

As shown in the left panel of Fig. \ref{fig:compare_prof}, the pressure profiles obtained with and without masking the over-pressure region are significantly different. The thermal pressure constrained without masking the over-pressure region is higher than the one obtained using the mask defined in Sect. \ref{sec:overp} from the cluster core to about $2\rm{R_{500}}$. While the profile uncertainties overlap in the cluster core, the significance of the difference between the pressure profile estimates is higher than $3\sigma$ between 500 and 900~kpc because of the excess of thermal pressure within the south-west region of \psz.\\
\indent This difference is also identified in the radial profiles of the integrated Compton parameter computed from the spherical integration of the pressure profiles obtained with and without masking the over-pressure region (see Fig. \ref{fig:compare_prof} right panel). The radius is normalized by the estimate of $\rm{R_{500,Y_X}}$ constrained from the \xmm\ data to give a common reference to both profiles. While the value of the integrated Compton parameter $\rm{Y}$ is consistent between the two profiles at $5\rm{R_{500,Y_X}}$ in order to be compatible with the measurement made by \planck, we observe a significant difference of 60\% between the values of $\rm{Y}$ at $\rm{R_{500,Y_X}}$.\\
\indent We use Equation \ref{eq:hse_mass} and the \xmm\ deprojected density profiles estimated with and without masking the disturbed region (see Sect. \ref{sec:X_ray_profiles}) to compute the characteristic radius $\rm{R_{500}}$ for both the MCMC analysis. The transition between the inner and the outer slope of the green pressure profile (without mask) in Fig. \ref{fig:compare_prof} is more distant from the cluster center than that of the blue profile (with mask). Using the whole NIKA2 map to constrain the cluster pressure profile leads therefore to a concentration parameter lower than the one that we obtain if the over-pressure region is masked. This explains why the $\rm{R_{500}}$ radius associated to the green profile is larger than the one computed from the blue one (see Table \ref{tab:Cluster_global}). Such a difference in the estimation of $\rm{R_{500}}$ further increases the discrepancy between both the integrated Compton parameter and the hydrostatic mass estimated from the two different analyses. The estimates of $\rm{Y_{500}}$ and $\rm{M_{500}}$ obtained from the analysis without masking the over-pressure region deviate from the values that are obtained by masking the south-west region of \psz\ by 65 and 79\% respectively. This emphasizes that the high angular resolution of NIKA2 enables the characterization of the systematic uncertainty induced by disturbed ICM regions on the estimation of $\rm{Y_{500}}$ and $\rm{M_{500}}$.\\
\indent While the impact of the thermal pressure excess in the south-west region of \psz\ is negligible with respect to the total integrated Compton parameter (see Sect. \ref{sec:MCMC_results}), its effect on both the cluster pressure profile characterization and the $\rm{Y_{500}}$ estimate happens to be very significant. The consequence of such a result for the NIKA2 tSZ large program and its repercussions for future cosmology studies are discussed in Sect. \ref{sec:discussion}. 

\begin{figure}[h!]
\centering
\includegraphics[height=6.8cm]{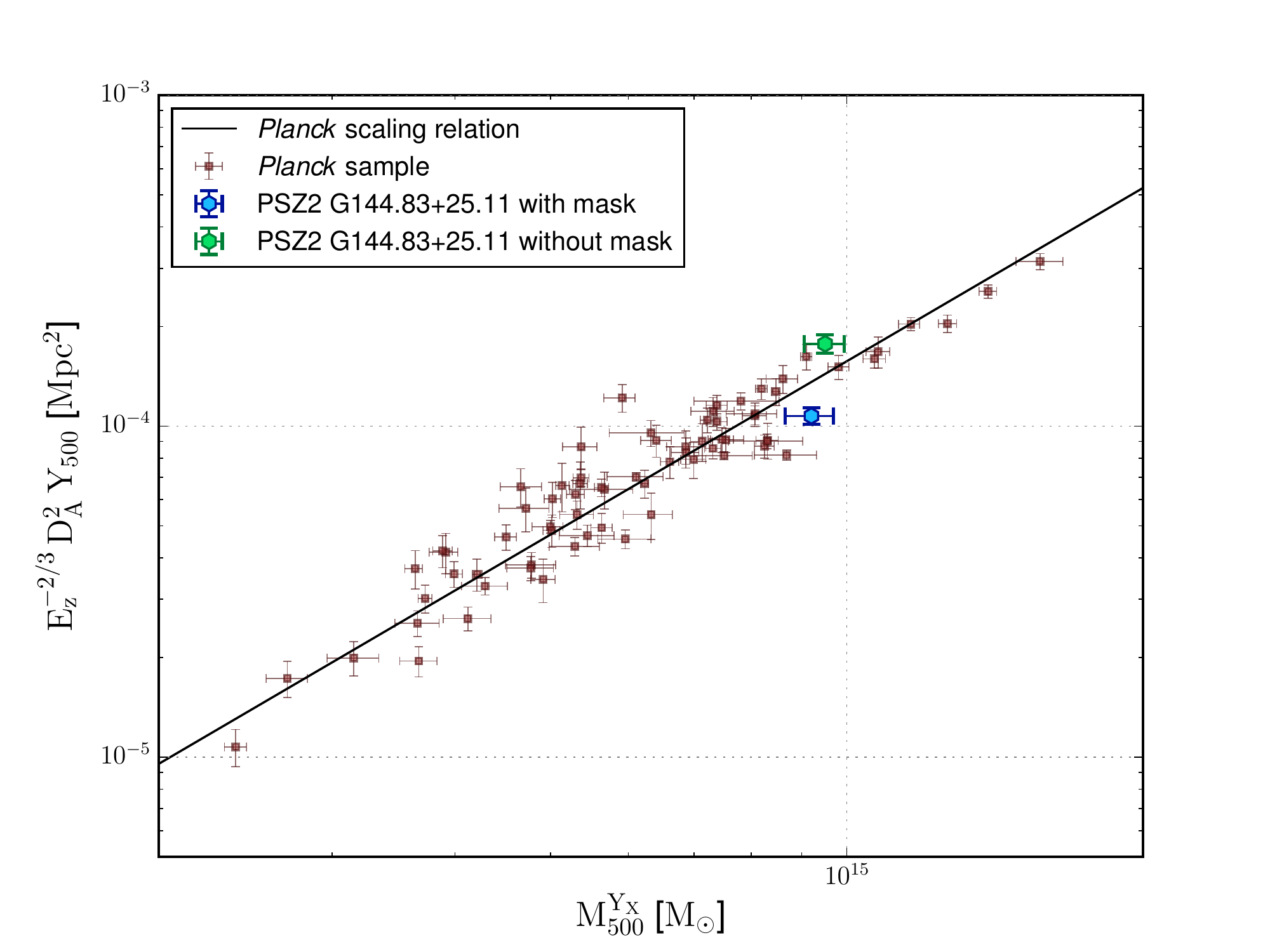}
\caption{{\footnotesize \planck\  $\rm{Y_{500}}-\rm{M_{500}}$ scaling relation (black line) \citep{pla14} together with the 71 points (red squares) used for its calibration from the study of low-redshift clusters ($z < 0.45$). The two estimates of $\rm{Y_{500}}$ based on the joint tSZ analysis with and without masking the over-pressure region in \psz\ are shown with the blue and the green points respectively. All the masses are estimated using the $\rm{Y_X}$ mass proxy.}}
\label{fig:scaling_law}
\end{figure}

\section{Discussion}\label{sec:discussion}

In this section, we review the main results of this paper in light of future cosmology analyses based on tSZ cluster statistics. As large tSZ-cluster catalogs are currently established only by low angular resolution instruments (FWHM $>1~\rm{arcmin}$), both a universal pressure profile and a tSZ-mass scaling relation are needed to constrain the cluster mass function and to estimate cosmological parameters from these observations.\\
\indent The tSZ flux is not to be taken as an accurate indicator of the mass of a galaxy cluster if thermal pressure excesses, caused by infalling substructures or shocks, are identified within the ICM. Indeed, this observable is directly related to the total energy content of a galaxy cluster only if the ICM is in hydrostatic equilibrium within the cluster gravitational potential. Therefore, the calibration of the $\rm{Y_{500}^{\rm{HSE}}}-\rm{M_{500}^{\rm{HSE}}}$ scaling relation has to be done by high angular resolution tSZ instruments, such as NIKA2, in order to precisely characterize deviations from hydrostatic equilibrium within the ICM.\\
\indent The results presented in Sect. \ref{sec:Impact_mass} show that over-pressure regions within the ICM of a given cluster are expected to induce a substantial variation of the estimate of $\rm{Y_{500}^{\rm{HSE}}}$ related to the value of the hydrostatic mass $\rm{M_{500}^{\rm{HSE}}}$ with respect to the constraints obtained by masking the disturbed ICM region. We show in Fig. \ref{fig:scaling_law} the 71 points  \cite{pla14} used to calibrate the $\rm{Y_{500}^{\rm{HSE}}}-\rm{M_{500}^{\rm{HSE}}}$ scaling relation at low redshift (black line). The cluster total masses are given by the $\rm{Y_X}$ mass proxy \citep[see, e.g.][]{arn07} for all the displayed points. The two points that we obtain from the tSZ analysis with (blue) and without (green) masking the thermal pressure excess in the south-west region of \psz\ are also shown and compared to the best-fit scaling relation.\\ 
\indent While no conclusion can be drawn from a single cluster, we note that our estimates of $\rm{Y_{500}^{\rm{HSE}}}$ are consistent with the scaling relation of \cite{pla14} for the corresponding hydrostatic masses given by the $\rm{Y_X}$ mass proxy. However, the (blue) point obtained by masking the over-pressure region has a much lower value of $\rm{Y_{500}^{\rm{HSE}}}$ than the one obtained by considering the green pressure profile in Fig. \ref{fig:compare_prof}, which gives a better description of the overall ICM pressure distribution. While the NIKA2 performance in terms of angular resolution, FOV, and sensitivity enables us to reach relative uncertainties associated to both estimates of $\rm{Y_{500}^{\rm{HSE}}}$ of the order of 7\%, the difference between the two values of $\rm{Y_{500}^{\rm{HSE}}}$ estimated in this study is of the order of magnitude of the current scatter of the scaling relation. This result shows that the NIKA2 tSZ large program will be able to take into account the impact of the systematic uncertainties induced by disturbed ICM morphologies on the calibration of the $\rm{Y_{500}^{\rm{HSE}}}-\rm{M_{500}^{\rm{HSE}}}$ scaling relation and on the characterization of its intrinsic scatter.\\
\indent One of the goals of the NIKA2 tSZ large program is also to calibrate the scaling relation linking the value of $\rm{Y_{500}^{\rm{meas.}}}$, estimated by low angular resolution instruments using a universal pressure profile template, and the value of $\rm{Y_{500}^{\rm{HSE}}}$ measured at high angular resolution under the hydrostatic equilibrium assumption by excluding potential disturbed ICM regions. This additional scaling relation will enable the characterization of both the bias and scatter induced by disturbed ICM morphologies on the measurement of $\rm{Y_{500}}$ for the whole cluster population in cosmological analyses based on tSZ cluster statistics.\\
\indent As shown in the left panel of Fig. \ref{fig:Nonparam_profiles}, the NIKA2 SZ large program will also allow us to precisely constrain the universal pressure profile of galaxy clusters at high redshift and to compare it with the one obtained from the X-ray observations of a representative sample of galaxy clusters at low redshift \citep[see, e.g.][]{arn10}. This will enable us to characterize potential variations of the universal pressure profile parameters due to redshift evolution, projection effects that affect differently the tSZ and X-ray data, and possible instrumental calibration biases and systematic effects. The high angular resolution and large FOV of NIKA2 will therefore enable us to study the impact of hitherto unidentified substructures within the ICM of galaxy clusters on the characterization of both the scaling relation and the universal pressure profile for a representative cluster sample at intermediate and high redshift.

\section{Conclusions and perspectives}\label{sec:conclusions}
The first NIKA2 tSZ-mapping of a galaxy cluster has been made on \psz\ simultaneously at 150 and 260 GHz for a total of 11.3 hours. Despite the rather poor weather conditions, significant tSZ signal is recovered at 150~GHz from the cluster center up to an angular distance of 1.4~arcmin on the map at an angular resolution of 17.7~arcsec. This is comparable to the ICM extension constrained in X-ray by \xmm\ for a similar angular resolution. The NIKA2 260 GHz map has lead to the identification of a high-redshift submillimeter point source. The SED of this source is constrained by using jointly the \emph{Herschel} and NIKA2 260 GHz data.

The potential of NIKA2 to study the impact of the ICM dynamics at high redshift on the characterization of the tSZ-mass scaling relation and the cluster pressure profile is highlighted by the discovery of a thermal pressure excess in the south-west region of \psz. We characterize the radial extension of the over-pressure region by comparing the tSZ surface brightness profiles computed within circular sectors of the NIKA2 150 GHz map.

We performed a standard analysis based only on NIKA2 and \planck\ data and masking the over-pressure region in the NIKA2 150~GHz map. We emphasize that the NIKA2 mapping of the tSZ signal at 150~GHz enables the setting of significant constraints on the radial pressure profile up to $\rm{R>R_{500}}$. Furthermore, the NIKA2 data enable us to constrain the value of $\rm{Y_{500}}$ with a relative uncertainty of 7\%. These estimates on both the radial pressure profile and the cluster integrated Compton parameter show the typical constraints that will be achieved for all the clusters of the NIKA2 SZ large program.

We study the impact of the over-pressure region on the ICM characterization by performing a non-parametric pressure profile deprojection using jointly the tSZ data obtained by MUSTANG, NIKA2, Bolocam, and \planck\ with and without masking the thermal pressure excess. This enables us to constrain the cluster pressure profile from its core up to $3\rm{R_{500}}$.
We show that excluding the cluster over-pressure region leads to a significant drop in the thermal pressure recovered between radial distances of 500 and 900~kpc from the cluster center. This leads to a variation of 65\% of the integrated Compton parameter $\rm{Y_{500}}$ computed from the integration of the pressure profiles obtained by the two analyses. Both the pressure profile constrained from the tSZ data sets and the density profile deprojected from the \xmm\ observations of \psz\ are jointly used to further constrain the ICM thermodynamic properties. We show that the exclusion of the cluster over-pressure region decreases the hydrostatic mass estimate significantly compared to the constraint that would be obtained by a $>1~\rm{arcmin}$ tSZ instrument, which could not resolve the thermal pressure excess. This shows the potential of NIKA2 to constrain the impact of the ICM dynamics of high-redshift clusters on the characterization of the intrinsic scatter of the $\rm{Y_{500}} - \rm{M_{500}}$ scaling relation.\\

The NIKA2 camera is now installed at the focal plane of the IRAM 30-m telescope and commissioned. Open-time observation projects as well as guaranteed time programs such as the NIKA2 SZ large program have already begun. The latter focuses on the ICM characterization of a representative sample of 50 tSZ-discovered clusters with redshift spanning from 0.5 to 0.9. The work presented in this paper shows the typical products that will be delivered by this large program. The NIKA2 high angular resolution tSZ mapping  will enable the identification of yet-unresolved thermal pressure substructures and the study of the impact of disturbed ICM regions on the pressure profile characterization. The NIKA2 data will be combined with \planck\ and ACT observations to constrain the pressure profile of each cluster from the core to the outskirts. \xmm\ data will be used along with that from NIKA2  to constrain all the ICM thermodynamic properties and characterize the origin of a possible redshift evolution of the tSZ-mass scaling relation and pressure profile parameters. 

\begin{acknowledgements}
We would like to thank the IRAM staff for their support during the campaigns and the anonymous referee for helpful comments. 
The NIKA dilution cryostat has been designed and built at the Institut N\'eel. 
In particular, we acknowledge the crucial contribution of the Cryogenics Group, and 
in particular Gregory Garde, Henri Rodenas, Jean Paul Leggeri, and Philippe Camus. 
This work has been partially funded by the Foundation Nanoscience Grenoble, the LabEx FOCUS ANR-11-LABX-0013, and 
the ANR under the contracts "MKIDS", "NIKA", and ANR-15-CE31-0017. This work has benefited from the support of the European Research Council Advanced Grants ORISTARS and M2C under the European Union's Seventh Framework Programme (Grant Agreement Nos. 291294 and 340519). We acknowledge funding from the ENIGMASS French LabEx (R. A. and F. R.), 
the CNES post-doctoral fellowship program (R. A.),  the CNES doctoral fellowship program (A. R.), and 
the FOCUS French LabEx doctoral fellowship program (A. R.). R.A. acknowledges support from the Spanish Ministerio de Econom\'ia and Competitividad (MINECO) through grant number AYA2015-66211-C2-2.
\end{acknowledgements}

\end{document}